\documentstyle[aps,epsf]{revtex}

\begin{document}

\title{Theory of Bose-Einstein condensation in trapped gases} 
\author{Franco Dalfovo,$^1$ Stefano Giorgini,$^{1}$
Lev P. Pitaevskii,$^{1,2,3}$ and Sandro Stringari$^1$}

\address{$^1$ Dipartimento di Fisica, Universit\`a di Trento, and \\
Istituto Nazionale per la Fisica della Materia,
I-38050 Povo, Italy}

\address{$^2$ Department of Physics, TECHNION, Haifa 32000, Israel}

\address{$^3$ Kapitza Institute for Physical Problems,
ul. Kosygina 2, 117334 Moscow}

\maketitle

\date{  } 

\begin{abstract}
The phenomenon of Bose-Einstein condensation of dilute gases in traps 
is reviewed from a theoretical perspective. Mean-field theory provides 
a framework to understand the main features of the condensation and the 
role of interactions between particles. Various properties of these 
systems are discussed, including the density profiles and the energy of 
the ground state configurations, the collective oscillations and the 
dynamics of the expansion, the condensate fraction and the thermodynamic 
functions. The thermodynamic limit exhibits a scaling behavior in the 
relevant length and energy scales. Despite the dilute nature of the gases, 
interactions profoundly modify the static as well as the dynamic properties 
of the system; the predictions of mean-field theory are in excellent 
agreement with available experimental results. Effects of superfluidity 
including the existence of quantized vortices and the reduction of the 
moment of inertia are discussed, as well as the consequences of coherence 
such as the Josephson effect and interference phenomena. The review also 
assesses the accuracy and limitations of the mean-field approach.

\smallskip
\noindent
{\it Preprint, October 6, 1998. For publication in Reviews of Modern 
Physics.  }

\end{abstract}





\section{Introduction}
\label{sec:intro}

\begin{figure}[t]
\epsfysize=7cm
\hspace{3cm}
\epsffile{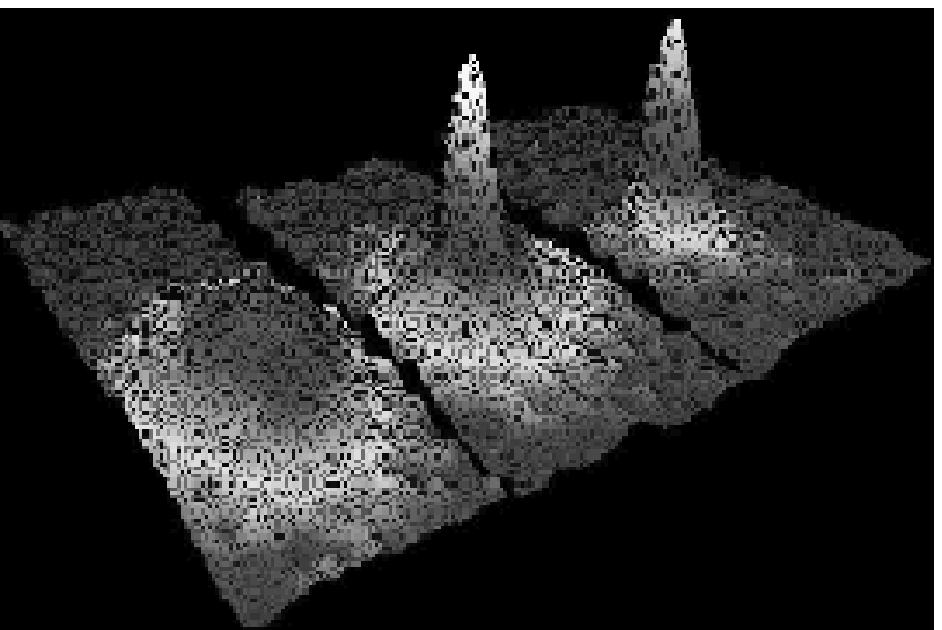}
\caption{ Images of the velocity distribution of rubidium atoms in the
experiment by Anderson {\it et al.} (1995), taken by means of the 
expansion method. The left frame corresponds to a 
gas at a temperature just above condensation; the center frame, just 
after the appearance of the condensate; the right frame, after further 
evaporation leaves a sample of nearly pure condensate. The field of 
view is $200 \mu$m $\times 270 \mu$m, and corresponds to the distance the
atoms have moved in about $1/20$ s. The color corresponds to the number of 
atoms at each velocity, with red being the fewest and white being the
most. From Cornell (1996). [note: this is B/W version of reduced quality
for e-archive only; the original is a color jpeg file]  } 
\label{fig:threepeaks}
\end{figure}

Bose-Einstein condensation (BEC) (Bose, 1924; Einstein, 1924) was observed
in 1995 in a remarkable series of experiments on vapors of rubidium
(Anderson {\it et al.}, 1995) and sodium (Davis {\it et al.}, 1995) in
which the atoms were confined in magnetic traps and cooled down to extremely
low  temperatures, of the order of fractions of microkelvins.  The first
evidence for condensation emerged from time of flight measurements.
The atoms were left to expand by switching off the confining trap and then
imaged with optical methods.  A sharp peak in the velocity distribution
was then  observed  below a certain critical temperature, providing a clear
signature for BEC. In Fig.~\ref{fig:threepeaks}, we show one of the first
pictures of the atomic clouds of rubidium.  In the same year, first 
signatures of the occurrence of BEC in vapors of lithium were also 
reported (Bradley {\it et al.}, 1995). 

Though the experiments of 1995 on the alkalis should be considered a
milestone in the history of BEC,  the experimental and theoretical
research on this unique phenomenon predicted by quantum statistical
mechanics is much older and has  involved  different areas of physics
(for an interdisciplinary review of BEC see Griffin, Snoke and
Stringari, 1995). In particular, from the very beginning, superfluidity 
in helium was considered  by London (1938) as a possible manifestation 
of BEC.  Evidences for BEC in helium have later emerged from
the analysis of the momentum distribution of the atoms measured in 
neutron scattering experiments (Sokol, 1995). In recent years, BEC 
has been also investigated in the gas of paraexcitons in 
semiconductors (see Wolfe, Lin and Snoke, 1995, and references 
therein), but an unambiguous signature for BEC in this system has 
proven difficult to find. 

Efforts to Bose condense atomic gases began with hydrogen more than 15 
years ago.  In a series of experiments hydrogen atoms were first cooled 
in a dilution refrigerator, then trapped by a magnetic field and further 
cooled by evaporation. This approach has come very close to observing BEC, 
but is still limited by recombination of individual atoms to form molecules
(Silvera and Walraven, 1980 and 1986; Greytak and Kleppner, 1984; 
Greytak, 1995; Silvera, 1995). At the time of this review, first 
observations of BEC in spin polarized hydrogen have been reported
(Fried {\it et al.}, 1998). In the '80s laser-based techniques,
such as laser cooling and magneto-optical trapping, were developed
to cool and trap neutral atoms [for recent reviews, see Chu (1998), 
Cohen-Tannoudji (1998) and Phillips (1998)].  Alkali atoms are well 
suited to laser-based methods because their optical transitions can be 
excited by available lasers and because they have a favourable internal 
energy-level structure for cooling to very low temperatures. Once they are 
trapped, their temperature can be lowered further by evaporative cooling 
[this technique has been recently reviewed by Ketterle and van 
Druten (1996a) and by Walraven (1996)].  By combining laser and 
evaporative cooling for alkali atoms, experimentalists 
eventually succeeded in reaching the temperatures and densities required to 
observe BEC. It is worth noticing that, in these conditions, the equilibrium 
configuration of the system  would be the solid phase. Thus, in order to 
observe BEC, one has to preserve the system in a metastable gas phase 
for a sufficiently long time. This is possible because three-body 
collisions are rare events in dilute and cold gases, whose lifetime 
is hence long enough to carry out experiments. So far BEC has 
been realized in $^{87}$Rb (Anderson {\it et al.}, 1995; Han {\it et al.}, 
1998; Kasevich, 1997; Ernst {\it et al.}, 1998a; Esslinger {\it et al.}, 
1998; Dalibard {\it et al.}, 1998), in $^{23}$Na (Davis {\it et al.}, 
1995; Hau, 1997 and 1998;  Lutwak {\it et al.}, 1998) and in $^7$Li 
(Bradley {\it et al.}, 1995 and 1997). The number of experiments on BEC 
in vapors of rubidium and sodium is now growing fast. In the
meanwhile, intense experimental research is currently carried out also 
on vapors of caesium, potassium and metastable helium.   

\begin{figure}[t]
\epsfysize=6cm
\hspace{3cm}
\epsffile{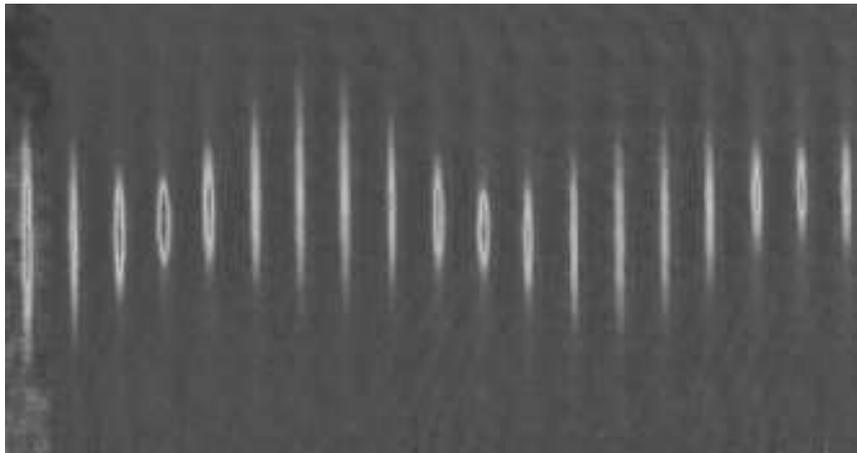}
\caption{Collective excitations of a Bose-Einstein condensate.  Shown 
are in-situ repeated phase-contrast images taken of a ``pure" condensate.  
The excitations were produced by modulating the magnetic fields which 
confine the condensate, and then letting the condensate evolve freely.  
Both the center-of-mass and the shape oscillations are visible, and the 
ratio of their oscillation frequencies can be accurately measured.
The field of view in the vertical direction is about $620 \mu$m, 
corresponding to a condensate width of the order of $200$-$300 \mu$m.
The time step is $5$ ms per frame.  From Stamper-Kurn and Ketterle (1998). 
[note: this is B/W version of reduced quality
for e-archive only; the original is a color jpeg file] }
\label{fig:insitu}
\end{figure}

One of the most relevant features of these trapped Bose gases is that they 
are inhomogeneous and finite-sized systems, the number of atoms ranging 
typically from a few thousands to several millions. In most cases, the 
confining traps are well approximated by harmonic potentials. The trapping
frequency, $\omega_{\rm ho}$, provides also a characteristic length scale 
for the system, $a_{\rm ho} = [\hbar / ( m \omega_{\rm ho} ) ]^{1/2}$, of 
the order of a few microns in the available samples. Density variations 
occur on this scale. This is a major difference with respect to other 
systems, like for instance superfluid helium, where the effects of 
inhomogeneity take place on a microscopic scale fixed by the interatomic 
distance.  In the case of $^{87}$Rb  and $^{23}$Na, the size of the system 
is enlarged as an effect of repulsive two-body forces and  the trapped gases 
can become almost macroscopic objects, directly measurable with optical 
methods. As an example, we show in Fig.~\ref{fig:insitu} a sequence of 
``in situ" images of an oscillating condensate of sodium atoms taken 
at the Massachusetts Institute of Technology (MIT), where the mean axial 
extent is of the order of $0.3$ mm. 

The fact that these gases are highly inhomogeneous has several
important consequences. First BEC shows up not only  in { \it momentum}
space, as happens in superfluid helium,  but also in {\it co-ordinate}
space. This double possibility of investigating  the effects of
condensation is very interesting from both the theoretical  and
experimental viewpoints and provides novel methods of investigation
for relevant quantities, like the temperature dependence 
of the condensate, energy and density distributions, interference 
phenomena, frequencies of collective excitations, and so on.

\begin{figure}[t]
\epsfysize=8cm
\hspace{3cm}
\epsffile{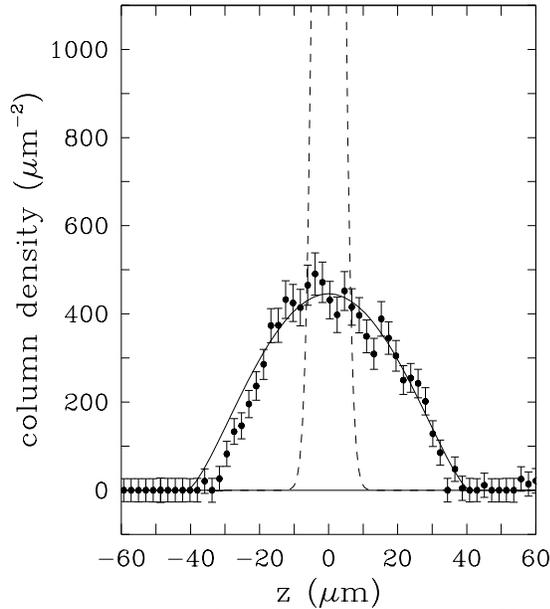}
\caption{Density distribution of $80000$ sodium atoms in the trap of Hau 
{\it et al.} (1998) as a function of the axial co-ordinate. The 
experimental points correspond to the measured optical density,
which is proportional to the column density of the atom cloud along 
the path of the light beam. The data well agree with the prediction 
of mean-field theory for interacting atoms (solid line) 
discussed in  Sec.~\protect\ref{sec:groundstate}. Conversely, 
a noninteracting gas in the same trap would have a much sharper Gaussian 
distribution (dashed line). The same normalization is used for the 
three density profiles. The central peak of the Gaussian is 
found at about $5500 \mu$m$^{-2}$. The figure points out the role 
of atom-atom interaction in reducing the central density and enlarging 
the size of the cloud. }
\label{fig:hau}
\end{figure}

Another important consequence of the inhomogeneity of these systems
is the role played by two-body interactions. This aspect 
will be extensively discussed in the present review. The main point is 
that, despite the very dilute nature of these gases (typically the 
average distance between atoms is more than ten times  the range of
interatomic forces), the combination of BEC and harmonic trapping
greatly enhances the effects of the atom-atom interactions on important 
measurable quantities. For instance, the central density of the 
interacting gas at very low temperature can be easily one or two 
orders of magnitude smaller than the density  predicted for an ideal 
gas in the same trap, as shown in Fig.~\ref{fig:hau}. Despite the 
inhomogeneity of these systems, which makes the solution of the 
many-body problem nontrivial, the dilute nature of the gas allows 
one to describe the effects of the interaction in a rather fundamental 
way. In practice a single physical parameter, the $s$-wave scattering  
length, is sufficient to obtain an accurate description.

The recent experimental achievements of BEC in alkali vapors have renewed  
a great interest in the theoretical studies of Bose gases. A rather massive 
amount of work has been done in the last couple of years, both to 
interpret the initial observations and to predict new phenomena. 
In the presence of harmonic confinement, the many-body theory of
interacting Bose gases gives rise to several unexpected features.
This opens new theoretical perspectives in this interdisciplinary 
field, where useful concepts coming from different areas of 
physics (atomic physics, quantum optics, statistical mechanics and
condensed matter physics) are now merging together.  

The natural starting point for studying the behavior of these 
systems is the theory of weakly interacting bosons which, for inhomogeneous
systems, takes the form of the Gross-Pitaevskii theory. This
is a mean-field approach for the order parameter associated with the
condensate. It provides closed and relatively simple equations
for describing the relevant  phenomena associated with BEC. In 
particular, it reproduces typical properties exhibited by superfluid 
systems, like the propagation of collective excitations and the 
interference effects originating from the phase of the order parameter. 
The theory is well suited to describing most of the effects of two-body 
interactions in these dilute gases at zero temperature and can be 
naturally generalized to explore also thermal effects. 

An extensive discussion of the application of mean-field theory to 
these systems is the main basis of the present review article.
We  also give, whenever possible, simple arguments based on scales of 
length, energy and density, in order to point out the relevant parameters 
for the description of the various phenomena. 

There are several topics which are only marginally discussed in our 
paper. These include, among others, collisional and thermalization 
processes, phase diffusion phenomena, light scattering from the condensate 
and analogies with systems of coherent photons.  In this sense our work 
is complementary to other recent review articles (Burnett, 1996; Parkins 
and Walls, 1998). Furthermore in our paper we do not  discuss the physics 
of ultracold collisions and the determination of the scattering length 
which have been recently the object of important experimental and 
theoretical studies in the alkalis (Heinzen, 1997; Weiner {\it et al.}, 
1998).

The plan of the paper is the following:

In Sec.~\ref{sec:theidealgas}
we summarize the basic features of the noninteracting Bose gas in 
harmonic traps and we introduce the first relevant length and energy 
scales,  like the oscillator length and the critical temperature. 
We also comment on finite  size effects, on the role of dimensionality
and on the possible relevance of anharmonic traps.

In Sec.~\ref{sec:groundstate}
we discuss the effects of the interaction on the ground state. We 
develop the formalism of mean-field theory, based on the Gross-Pitaevskii 
equation. We consider the case of gases interacting  with  both repulsive 
and attractive forces. We then discuss in detail the large $N$ limit for 
systems interacting with repulsive forces, leading to the so called 
Thomas-Fermi approximation, where the ground state properties 
can be calculated in analytic form.  In the last part,  we discuss the
validity of the mean-field approach and give explicit results for 
the first corrections, beyond mean-field, to the ground state properties, 
including the quantum depletion of the condensate, {\it i.e.}, the
decrease in the condensate fraction produced by the interaction. 

In Sec.~\ref{sec:dynamics}
we investigate the dynamic behavior of the condensate using the time 
dependent Gross-Pitaevskii equation. The equations of motion for the
density and the velocity field of the condensate in the large $N$ limit,
where the Thomas-Fermi approximation is valid, are shown to have the 
form of the hydrodynamic equations of superfluids. 
We also discuss the dynamic behavior in the nonlinear regime (large
amplitude oscillations and free expansion), the collective modes in  
the case of attractive forces and the transition from collective to 
single-particle states in the spectrum of excitations. 

In Sec.~\ref{sec:thermodynamics}
we discuss thermal effects. We show how one can define the thermodynamic 
limit in these inhomogeneous systems and how interactions modify the 
behavior compared to the noninteracting case. We extensively discuss
the occurrence of scaling properties in the thermodynamic limit. We 
review several results for the shift of the 
critical temperature and for the temperature dependence of  
thermodynamic functions, like the condensate fraction, the chemical 
potential and the release energy. We also discuss the behavior of 
the excitations at finite temperature.

In Sec.~\ref{sec:superfluidity}
we illustrate some features of these trapped Bose gases in connection
with superfluidity and phase coherence. We discuss in particular
the structure of quantized vortices and the behavior of the moment of
inertia, as well as interference phenomena and quantum effects beyond
mean-field theory, like the collapse-revival of collective 
oscillations.

In Sec.~\ref{sec:conclusions}
we draw our conclusions and we discuss some further future perspectives
in the field.

The overlap between current theoretical and experimental 
investigations of BEC in trapped alkalis is already wide and rich. 
Various theoretical predictions, concerning the ground state, dynamics 
and thermodynamics are found to agree very well with observations;  
others are stimulating new experiments. The comparison between theory and
experiments then represents an exciting feature of these novel 
systems, which will be frequently emphasized in the present review.


\section{The ideal Bose gas in a harmonic trap}
\label{sec:theidealgas}

\subsection{The condensate of noninteracting bosons}
\label{sec:noninteracting}

An important feature characterizing the available magnetic
traps for alkali atoms is that the confining potential can be safely 
approximated with the quadratic form 
\begin{equation}
V_{\rm ext}({\bf r}) = {m\over 2} ( \omega_x^2 x^2 + \omega_y^2 y^2
+ \omega_z^2 z^2) \; .
\label{eq:uext}
\end{equation}
Thus the investigation of these systems starts as a
textbook application of nonrelativistic quantum mechanics for
identical point-like particles in a harmonic potential.

The first step consists in neglecting the atom-atom interaction.
In this case, almost all predictions are analytical and relatively
simple. The many-body Hamiltonian is the sum of single-particle
Hamiltonians whose eigenvalues have the form
\begin{equation}
\varepsilon_{n_x n_y n_z} = \bigg({n_x + {1\over 2}}\bigg)
\hbar \omega_x  + \bigg({n_y + {1\over 2}}\bigg) \hbar \omega_y
+ \bigg({n_z + {1\over 2}}\bigg) \hbar \omega_z \; ,
\label{eq:spectrumho}
\end{equation}
where $\{ n_x, n_y, n_z \} $ are non-negative integers. The ground
state $\phi({\bf r}_1,..,{\bf r}_N)$ of $N$ noninteracting bosons 
confined by the potential (\ref{eq:uext}) is obtained by putting 
all the particles in the lowest single-particle state 
($n_x=n_y=n_z=0$), namely $\phi({\bf r}_1,..,{\bf r}_N)=
\prod_i\varphi_0({\bf r}_i)$,
where $\varphi_0({\bf r})$ is given by 
\begin{equation}  
\varphi_0 ({\bf r}) = \left( { m \omega_{\rm ho}  \over
\pi \hbar } \right)^{3/4} \exp \left[ - {m \over
2 \hbar } (\omega_x x^2 + \omega_y y^2 + \omega_z z^2)
\right]  \; ,
\label{eq:gaussian}
\end{equation}
and we have introduced the geometric average of the oscillator
frequencies:
\begin{equation}
\omega_{\rm ho} = (\omega_x \omega_y \omega_z)^{1/3} \;.
\label{omegaho}
\end{equation}
The density distribution then becomes $n({\bf r})= N|\varphi_0({\bf r})|^2$
and its value grows with $N$. The size of the cloud is instead independent 
of $N$ and is fixed by the harmonic oscillator length
\begin{equation}
a_{\rm ho} = \left( {\hbar \over m \omega_{\rm ho} } \right)^{1/2}
\label{eq:aho}
\end{equation}
which corresponds to the average width of the Gaussian
(\ref{eq:gaussian}). This is the first important length scale
of the system.  In the available experiments, it is typically
of the order of $a_{\rm ho} \approx 1$ $\mu$m.  At finite temperature 
only part of the atoms occupy the lowest state, the others
being thermally distributed in the excited states at higher
energy. The radius of the thermal cloud is larger than $a_{\rm ho}$.
A rough estimate can be obtained by assuming $k_B T \gg \hbar
\omega_{\rm ho}$ and approximating the density of the thermal cloud
with a classical Boltzmann distribution $n_{\rm cl} (r) \propto
\exp[-V_{\rm ext}(r)/k_BT]$. If $V_{\rm ext}(r)=(1/2) 
m\omega_{\rm ho}^2 r^2$, the width of the Gaussian is 
$R_T = a_{\rm ho} (k_BT/\hbar \omega_{\rm ho})^{1/2}$, 
and hence larger than $a_{\rm ho}$. The use of a Bose
distribution function does not change significantly this
estimate. 

\begin{figure}[t]
\epsfysize=8cm
\hspace{3cm}
\epsffile{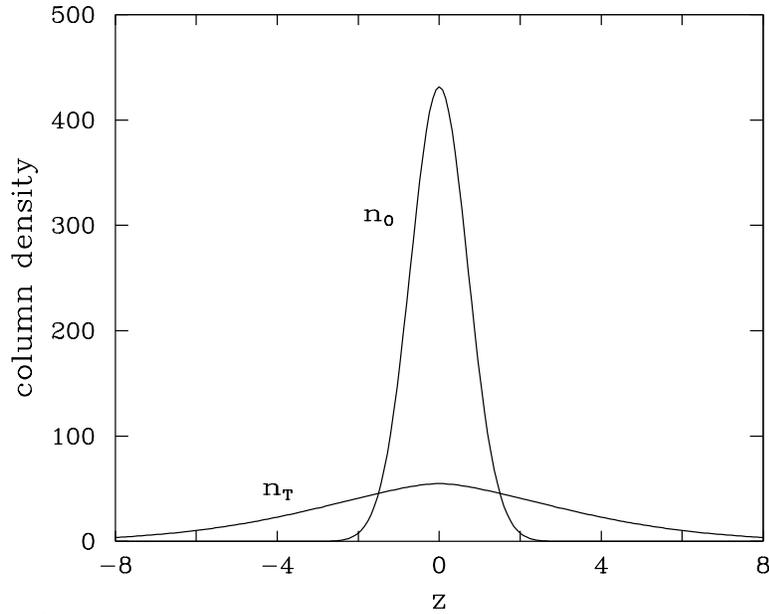}
\caption{Column density for $5000$ noninteracting bosons in 
a spherical trap at temperature $T= 0.9 T_c^0$. The central peak
is the condensate, superimposed on the broader thermal distribution. 
Distance and density are in units of $a_{\rm ho}$ and $a_{\rm ho}^{-2}$,
respectively. The density  is normalized to the number of atoms. The 
same curves can be identified with the momentum distribution of the 
condensed and noncondensed particles, provided the abscissa and the 
ordinate are replaced with $p_z$, in units of $a_{\rm ho}^{-1}$, and 
the momentum distribution, in units of $a_{\rm ho}^2$, respectively. } 
\label{fig:s-peak}
\end{figure}

The above discussion reveals that Bose-Einstein condensation in
harmonic traps shows up with the appearance of a sharp peak in the
central region  of the density distribution. An example is
shown in Fig.~\ref{fig:s-peak}, where we plot the prediction
for the condensate and thermal densities of $5000$ noninteracting
particles in a spherical trap at a temperature $T=0.9 T_c^0$,
where $T_c^0$ is the temperature at which condensation occurs (see
discussion in the next section). The curves correspond to the
column density, namely the particle density integrated along one
direction, $n(z) = \int dx \ n (x,0,z)$; this is a typical measured
quantity, the $x$ direction being the direction of the light beam
used to image the atomic cloud. By plotting directly the
density $n({\bf r})$, the ratio of the condensed and noncondensed 
densities at the center would be even larger. 

By taking the Fourier transform of the ground state wave function,
one can also calculate the momentum distribution of the atoms in the 
condensate. For the ideal gas, it is given by a Gaussian centered at
zero momentum and having a width proportional to $a_{\rm ho}^{-1}$.
The distribution of the thermal cloud is, also in momentum space, 
broader.  Using a classical distribution function one
finds that the width is proportional to $(k_B T)^{1/2}$.  Actually, the 
momentum distributions of the condensed and noncondensed particles 
of an ideal gas in harmonic traps have exactly the same form as the 
density distributions $n_0$ and $n_T$ shown in Fig.~\ref{fig:s-peak}. 

The appearence of the condensate as a narrow peak in both co-ordinate 
and momentum space is a peculiar feature of trapped Bose gases
having important consequences in both the experimental and theoretical 
analysis. This is different from the case of a uniform gas where the 
particles condense into a state of zero momentum,
but BEC cannot be revealed in co-ordinate space, since the condensed
and noncondensed particles fill the same volume.

Indeed, the condensate has been detected experimentally as the
occurrence of a sharp peak over a broader distribution, in both
the  velocity and spatial distributions.  In the
first case, one lets the condensate expand freely, by switching-off
the trap, and measures the density of the expanded cloud with
light absorption (Anderson {\it et al.}, 1995). If the particles
do not interact, the expansion is ballistic and the imaged spatial
distribution of the expanding cloud can be directly related to
the initial momentum distribution.  In the second case, one
measures directly the density of the atoms in the trap by means
of dispersive light scattering (Andrews {\it et al.}, 1996).
In both cases, the appearence of a sharp peak is the main
signature of Bose-Einstein condensation. An important theoretical
task consists of predicting how the shape of these peaks is
modified by the inclusion of two-body interactions. As anticipated
in Fig.~\ref{fig:hau}, the interactions can change the picture 
drastically. This effect will be deeply discussed in 
Sec.~\ref{sec:groundstate}. 

The shape of the confining field fixes also the symmetry of the
problem. One can use spherical or axially symmetric
traps, for instance. The first experiments on rubidium and sodium
were carried out with axial symmetry. In this case one can
define an axial co-ordinate $z$ and a radial co-ordinate
$r_\perp = (x^2 +y^2)^{1/2}$ and the corresponding frequencies,
$\omega_z$ and $\omega_\perp=\omega_x=\omega_y$. The ratio between 
the axial and radial frequencies, $\lambda = \omega_z / \omega_\perp$, 
fixes the asymmetry of the trap. For $\lambda < 1$ the trap is
cigar-shaped while for $\lambda > 1$  is disk-shaped. In terms of $\lambda$ 
the ground state (\ref{eq:gaussian}) for noninteracting bosons
can be rewritten as
\begin{equation}
\varphi_0 ({\bf r}) = { \lambda^{1/4}
\over \pi^{3/4} a_\perp^{3/2} }
\exp \left[ - {1 \over  2 a_\perp^2 }(r_\perp^2 +
\lambda z^2) \right]
\; .
\label{eq:gaussian-lambda}
\end{equation}
Here $a_\perp =(\hbar /m\omega_\perp)^{1/2}$ is the harmonic
oscillator length  in the $x$-$y$ plane and, since $\omega_\perp = 
\lambda^{-1/3} \omega_{\rm ho}$, one has also $a_\perp = 
\lambda^{1/6} a_{\rm ho}$. 

The choice of an axially symmetric trap has proven useful
for providing further evidence of Bose-Einstein condensation
from the analysis of the momentum distribution. To understand 
this point, let us take the Fourier transform of the wave function
(\ref{eq:gaussian-lambda}): $\tilde \varphi_0 ({\bf p}) \propto  
\exp [ -a^2_\perp (p^2_\perp + \lambda^{-1} p^2_z )/2\hbar^2]$. From 
this one can calculate the average axial and radial widths. Their ratio,
\begin{equation}
\sqrt{ \langle p^2_z \rangle /  \langle p^2_\perp \rangle }
=  \sqrt{\lambda} \; , 
\label{eq:arho}
\end{equation}
is fixed by the asymmetry parameter of the trap. Thus,
the shape of the expanded cloud in the $x$-$z$ plane is 
an ellipse,  the ratio between the two axis ({\it aspect ratio})
being equal to $\sqrt{\lambda}$. If the particles, instead of being
in the lowest state (condensate), were thermally distributed among many
eigenstates at higher energy, their distribution function would be
isotropic in momentum space, according to the equipartition principle,
and the aspect ratio would be equal to $1$. Indeed, the occurrence
of anisotropy in the condensate peak has been  interpreted 
from the very beginning as an important signature of BEC (Anderson
{\it et al.}, 1995; Davis {\it et al.}, 1995; Mewes {\it et al.}, 1996a).
In the case of the experiment at the Joint Institute for Laboratory 
Astrophysics (JILA) in Boulder, the trap is disk-shaped with
$\lambda=\sqrt{8}$. The first measured value of the aspect ratio was
about $50$\% \  larger than the prediction, $\sqrt{\lambda}$,  of the
noninteracting model  (Anderson {\it et al.}, 1995).  Of course, a 
quantitative comparison can be obtained only including the atom-atom
interaction, which affects the dynamics of the expansion (Holland and
Cooper, 1996; Dalfovo and Stringari, 1996; Holland {\it et al.}, 1997;
Dalfovo {\it et al.}, 1997c). However, the noninteracting model already
points out this interesting effect due to anisotropy.

\subsection{Trapped bosons at finite temperature: thermodynamic limit}
\label{sec:thermodynamiclimit}

At temperature $T$, the total number of particles is given, in the
grand-canonical ensemble,  by the sum
\begin{equation}
N= \sum_{n_x,n_y,n_z} \left\{ \exp [ \beta
(\varepsilon_{n_x n_y n_z} - \mu )] -1 \right\}^{-1} \;,
\label{eq:bosedistribution}
\end{equation}
while the total energy is given by
\begin{equation}
E=\sum_{n_x,n_y,n_z} \varepsilon_{n_x n_y n_z} \left\{ \exp [ \beta
(\varepsilon_{n_x n_y n_z} - \mu )] -1 \right\}^{-1} \;,
\label{eq:Etotho}
\end{equation}
where $\mu$ is the chemical potential and $\beta= (k_BT)^{-1}$.
Below a given temperature the population of the lowest state becomes
macroscopic and this corresponds to the onset of  Bose-Einstein
condensation. The calculation of the critical temperature, the fraction
of particles in the  lowest state (condensate fraction) and the other
thermodynamic quantities, starts from  Eqs.~(\ref{eq:bosedistribution}) 
and (\ref{eq:Etotho}) with the appropriate spectrum 
$\varepsilon_{n_x n_y n_z}$ (de Groot, Hooman and Ten Seldam, 1950; 
Bagnato, Pritchard and Kleppner, 1987).  Indeed the
statistical mechanics of these trapped gases is less trivial
than expected at first sight. Several interesting problems
arise from the fact that these systems have a finite size and
are inhomogeneous. For example, the usual definition 
of thermodynamic limit (increasing $N$ and volume with the average
density kept constant) is not appropriate for trapped gases. 
Moreover the traps can be made very anisotropic, reaching
the limit of  quasi-2D and quasi-1D systems, so that interesting
effects of reduced dimensionality can be also investigated.

As in the case of a uniform Bose gas, it is convenient to separate
out the lowest eigenvalue $\varepsilon_{000}$ from the sum
(\ref{eq:bosedistribution}) and call $N_0$ the number of particles
in this state. This number can be macroscopic, i.e., of the order
of $N$, when the chemical potential becomes equal to the energy
of the lowest state,
\begin{equation}
\mu \to \mu_c \ = \
{3 \over 2 } \ \hbar \bar{\omega}  \; ,
\label{eq:mucritical}
\end{equation}
where $\bar{\omega} = (\omega_x+\omega_y+\omega_z)/3$ is
the arithmetic average of the trapping frequencies.
Inserting this value in the rest of the sum, one can write
\begin{equation}
N-N_0 = \sum_{n_x,n_y,n_z \neq 0} { 1 \over \exp [ \beta
\hbar (\omega_x n_x  + \omega_y n_y + \omega_z n_z)] -1 } \; .
\label{eq:nminusn0}
\end{equation}
In order to evaluate this sum explicitly, one usually assumes
that the level spacing becomes smaller and smaller when $N \to
\infty$, so that the sum can be replaced by an integral:
\begin{equation}
N-N_0 = \int_0^{\infty} { dn_x dn_y dn_z \over
\exp [ \beta  \hbar (\omega_x n_x  + \omega_y n_y + \omega_z n_z)]
-1 } \; .
\label{eq:nminusn0semiclassical}
\end{equation}
This assumption corresponds to a semiclassical description
of the excited states. Its validity implies that the relevant
excitation energies, contributing to the sum (\ref{eq:nminusn0}),
are much larger than the level spacing fixed by the
oscillator frequencies. The accuracy of the semiclassical
approximation (\ref{eq:nminusn0semiclassical}) is expected to be
good  if the number of trapped atoms is large and $k_BT \gg \hbar
\omega_{\rm ho}$. It can be tested {\it a  posteriori}  by
comparing the integral (\ref{eq:nminusn0semiclassical})
with the numerical summation (\ref{eq:nminusn0}).

The integral (\ref{eq:nminusn0semiclassical}) can be easily
calculated by changing variables ($\beta  \hbar \omega_x n_x
= \tilde{n}_x$, etc.).  One finds
\begin{equation}
N-N_0 =  \zeta(3) \bigg( {k_BT\over \hbar \omega_{\rm ho} }\bigg)^3 \; ,
\label{eq:tcube}
\end{equation}
where $\zeta(n)$ is the Riemann $\zeta$-function and $\omega_{\rm ho}$ is
the geometric average (\ref{omegaho}). From this
result one can also obtain the transition temperature for
Bose-Einstein condensation. In fact, by imposing that $N_0 \to 0$
at the transition, one gets
\begin{equation}
k_B T_c^0 = \hbar \omega_{\rm ho} \left( {N \over \zeta(3) } \right)^{1/3}
= 0.94 \ \hbar \omega_{\rm ho} \ N^{1/3} \; .
\label{eq:tcho}
\end{equation}
For temperatures higher than $T_c^0$ the chemical potential is less
than $\mu_c$ and becomes $N$-dependent, while the population of the lowest
state is of the order of $1$ instead of $N$.  The proper thermodynamic 
limit for these systems is obtained by letting $N\to\infty$ and 
$\omega_{\rm ho}\to 0$, while keeping the product $N \omega_{\rm ho}^3$ 
constant. With this definition the transition temperature (\ref{eq:tcho}) 
is well defined in the thermodynamic limit. Inserting the above expression 
for $T_c^0$ into Eq.~(\ref{eq:tcube}) one gets the $T$-dependence of 
the condensate fraction for $T < T_c^0$: 
\begin{equation}
{ N_0 \over N } = 1 - \left( { T \over T_c^0 } \right)^3 \; .
\label{eq:condfractioho}
\end{equation}

The same result can be also obtained by rewriting 
(\ref{eq:nminusn0semiclassical}) as an integral over the energy,
in the form
\begin{equation}
N - N_0 = \int_0^\infty {  \rho (\varepsilon) \ d\varepsilon 
\over \exp (\beta\varepsilon) - 1 }
\label{eq:nminusn0rho}
\end{equation}
where $\rho(\varepsilon)$ is the density of states. The latter
can be calculated by using the spectrum (\ref{eq:spectrumho}) and
turns out to be quadratic in $\varepsilon$:
$\rho (\varepsilon) =  (1/2) (\hbar \omega_{\rm ho})^{-3}
\varepsilon^2$. Inserting this value into (\ref{eq:nminusn0rho}),
one finds again result (\ref{eq:tcube}). The integral
$E=\int_0^\infty d\varepsilon\rho(\varepsilon)\varepsilon/[\exp(\beta
\varepsilon)-1]$ gives instead the total energy of the system
(\ref{eq:Etotho}) for which one finds
the result
\begin{equation}
\frac{E}{Nk_BT_c^0}=\frac{3\zeta(4)}{\zeta(3)}
\left(\frac{T}{T_c^0}\right)^4
\; .
\label{eq:Etothoperpart}
\end{equation}
Starting from the energy one can calculate specific heat, entropy 
and the other thermodynamic quantities.  

These results can be compared with the well known theory of uniform
Bose gases (see, for example, Huang, 1987). In this case, the 
eigenstates of the Hamiltonian are plane waves of energy $\varepsilon =
p^2 / (2m)$, with the density of states given by $\rho (\varepsilon) =
(2\pi)^{-2} V (2m/\hbar^2)^{3/2} \sqrt{\varepsilon}$, where $V$ is
the volume. The sum (\ref{eq:bosedistribution}) gives
$N_0/N= 1 -(T/T_c^0)^{3/2}$  and $k_B T_c^0 = (2\pi \hbar^2 /m)
[n/\zeta(3/2)]^{2/3}$, with $n=N/V$, while the energy is given by 
$E/(Nk_BT_c^0)=3\zeta(5/2)/[2\zeta(3/2)](T/T_c^0)^{5/2}$. 

Another quantity of interest, which can be easily calculated using the
semiclassical approximation, is the density of thermal particles
$n_T({\bf r})$. The sum of $n_T({\bf r})$ and the condensate density,
$n_0({\bf r})=N_0|\varphi_0({\bf r})|^2$,   gives the total density
$n({\bf r})=n_0({\bf r})+n_T({\bf r})$. At $T<T_c^0$ and in the
thermodynamic limit, the thermal density is given by the integral
over momentum space $n_T({\bf r})=\int d{\bf p}(2\pi\hbar)^{-3}
[\exp(\beta \varepsilon({\bf p},{\bf r}))-1]^{-1}$, where
$\varepsilon({\bf p},{\bf r})= p^2/2m +V_{\rm ext}({\bf r})$ is the
semiclassical energy in phase space. The result is
\begin{equation}
n_T({\bf r}) = \lambda_T^{-3}\; g_{3/2}
\left( e^{-\beta V_{\rm ext}({\bf r})}
\right) \;,
\label{eq:nThor}
\end{equation}
where $\lambda_T=[2\pi\hbar^2/(mk_BT)]^{1/2}$ is the thermal wavelength.
The function $g_{3/2}({\bf r})$ belongs to the class of functions
$g_{\alpha}(z)=\sum_{n=1}^\infty z^n/n^\alpha$ [see, for example, 
Huang (1987)]. 
By integrating $n_T({\bf r})$ over space one gets again the
number of thermally depleted atoms $N-N_0=N(T/T_c^0)^3$, consistently with  
Eq.~(\ref{eq:condfractioho}). In a similar way one can obtain the
distribution of thermal particles in momentum space:
$n_T({\bf p}) = (\lambda_T m \omega_{\rm ho})^{-3} g_{3/2}
(\exp(-\beta p^2/2m))$. 

The above analysis points out the existence of two relevant 
scales of energy for the ideal gas: the transition temperature,
$k_BT_c^0$, and the average level spacing, $\hbar \omega_{\rm ho}$. From
expression (\ref{eq:tcho}), one  clearly
sees that $k_BT_c^0$ can be much larger than $\hbar \omega_{\rm ho}$.
In the available traps, with $N$ ranging from a few
thousand to several millions, the transition temperature is
20 to 200 times larger than $\hbar \omega_{\rm ho}$. This also means
that the semiclassical approximation is expected to work well in these
systems on a wide and useful range of temperatures. The frequency
$\omega_{\rm ho}/(2\pi)$ is fixed by the trapping potential and 
ranges typically from tens to hundreds of Hertz. This  gives
$\hbar \omega_{\rm ho}$ of the order of a few nK.  In one of the
first experiments  at JILA (Ensher {\it et al.}, 1996) for
example, the  average level spacing  was about $9$ nK, corresponding 
to a critical temperature [see Eq.~(\ref{eq:tcho})] of about 
$300$ nK with $40000$ atoms in the trap. 
We also note that, for the ideal gas, the  chemical potential
is of the same  order of $\hbar \omega_{\rm ho}$, as shown by
Eq.~(\ref{eq:mucritical}). However, as we will see later on, its value
depends significantly on the atom-atom  interaction and shall
consequently provide a third important scale of energy.  

\begin{figure}[t]
\epsfysize=8cm
\hspace{3cm}
\epsffile{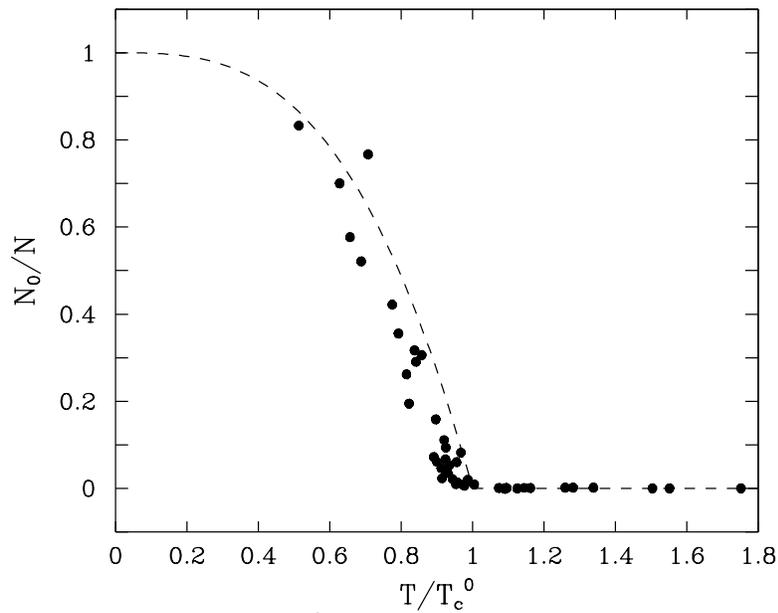}
\caption{Condensate fraction as a function of $T/T_c^0$. Circles are 
the experimental results of Ensher {\it et al.} (1996), while
the dashed line is the law (\protect\ref{eq:condfractioho})}  
\label{fig:condfrac-exp}
\end{figure}

The noninteracting harmonic oscillator model has guided 
experimentalists to the proper value of the critical temperature. 
In fact, the measured transition temperature was found to be very close 
to the ideal gas value (\ref{eq:tcho}), the occupation  of the condensate 
becoming macroscopically large below the critical temperature as   
predicted by (\ref{eq:condfractioho}). As an example, in  
Fig.~\ref{fig:condfrac-exp} we show the first experimental results 
obtained at JILA (Ensher {\it et al.}, 1996). The occurrence of a 
sudden transition at $T/T_c^0 \sim 1$ is evident. Similar results have 
been obtained also at MIT (Mewes {\it et al.},  1996a).  Apart from 
problems  related to temperature calibration,  a more
quantitative comparison between theory and experiments requires
the inclusion of two main effects: the fact that these gases have a  
finite number of particles and that they are interacting.  The role 
of interactions will be analysed extensively in the next
sections. Here we briefly discuss the relevance of finite 
size corrections.

\subsection{Finite size effects}
\label{sec:finitesize}

The number of atoms that can be put into the traps is not truly 
macroscopic. So far experiments have been carried out with a maximum 
of about $10^7$ atoms. As a consequence, the thermodynamic limit is 
never reached exactly. A first effect is the lack of discontinuities 
in the thermodynamic functions. Hence Bose-Einstein condensation in 
these trapped gases is not, strictly speaking, a phase transition. 
In practice, however, the macroscopic  occupation of the lowest state  
occurs rather abruptly as temperature is lowered and can be observed,
as clearly shown in Fig.~\ref{fig:condfrac-exp}.  The transition is 
actually rounded with respect to the  predictions of the $N \to \infty$ 
limit, but this effect, though interesting,  is small enough to 
make the words {\it transition} and {\it critical temperature} 
meaningful even for finite-sized systems. It is also worth noticing 
that, instead of being a limitation, the fact that $N$ is finite  
makes the system potentially richer, because new interesting regimes
can be explored  even in cases where there is no real phase transition 
in the thermodynamic limit. An example is BEC in 1D, as we will see
in Sec.~\ref{sec:dimensionality}. 

\begin{figure}[t]
\epsfysize=8cm
\hspace{3cm}
\epsffile{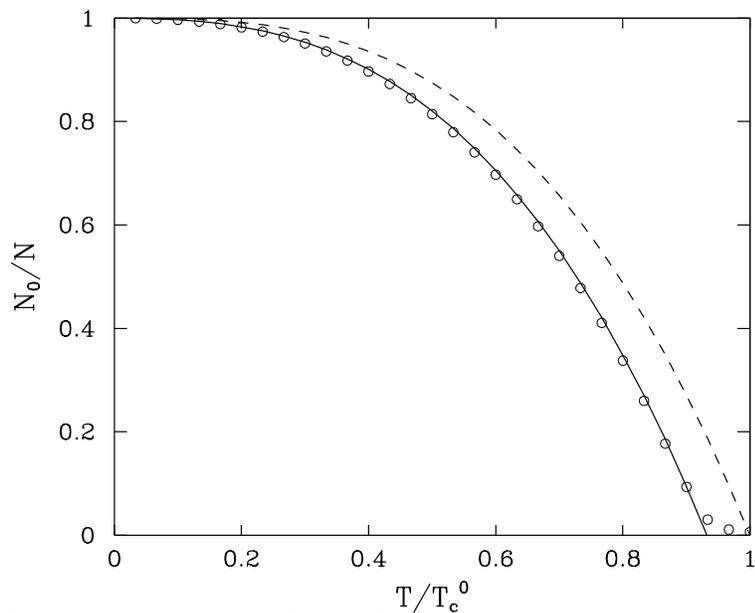}
\caption{Condensate fraction {\it vs.} temperature for an ideal gas in 
a trap. The circles correspond to the exact quantum calculation for $N=1000$ 
atoms in a trap with spherical symmetry and the solid line to the prediction  
(\protect\ref{eq:N0finitesize}). The dashed line refers to the thermodynamic 
limit  (\protect\ref{eq:condfractioho}). } 
\label{fig:s-condfrac}
\end{figure}

In order to work out the thermodynamics of a noninteracting Bose gas, 
all one needs is the spectrum of single particle levels entering the 
Bose distribution function.  Working in the grand-canonical ensemble 
for instance, the average number of atoms is given by the sum 
(\ref{eq:bosedistribution}) and it is not necessary to take the 
$N \to \infty$ limit. In fact, the explicit summation
can be carried out numerically (Ketterle and van Druten, 1996b)
for a fixed number  of particles and a given temperature, the chemical
potential being a function of $N$ and $T$. The condensate fraction
$N_0(T)/N$, obtained in this way, turns out to be smaller than the
thermodynamic limit prediction (\ref{eq:condfractioho}) and, as expected,
the transition is rounded off. An example of an exact calculation of
the condensate fraction for $1000$ noninteracting particles is shown
in Fig.~\ref{fig:s-condfrac} (circles).  With their numerical calculation,
Ketterle and van Druten (1996b) found that  finite size effects are
significant only for rather small values of $N$, less than about $10^4$. 
They calculated  also the occupation  of the first excited levels,
finding that the fraction of atoms in these states vanishes for
$N \to \infty$ and is very small already for $N$ of the order of $100$.

The first finite size correction to the law (\ref{eq:condfractioho}) for
the condensate fraction can be evaluated analytically by studying the large
$N$ limit of the sum (\ref{eq:bosedistribution}) (Grossmann and Holthaus, 
1995; Ketterle and van Druten, 1996b; Kirsten and Toms 1996; Haugerud, 
Haugset and Ravndal, 1997). The result for $N_0(T)/N$ is given by
\begin{equation}
{ N_0 \over N } = 1 - \left( {T \over T_c^0} \right)^3
- { 3 \bar{\omega} \zeta(2) \over 2 \omega_{\rm ho}
[\zeta(3)]^{2/3} } \left( {T \over T_c^0} \right)^2 
N^{-1/3}  \;.
\label{eq:N0finitesize}
\end{equation}
To the lowest order, finite size effects decrease as $N^{-1/3}$ 
and depend on the ratio of the arithmetic ($\bar{\omega}$) and
geometric ($\omega_{\rm ho}$) averages of the oscillator frequencies.
For axially symmetric traps this ratio depends on the deformation parameter
$\lambda=\omega_z/\omega_\perp$ as $\bar{\omega}/\omega_{\rm ho}=(\lambda+2)/
(3\lambda^{1/3})$.  For $N=1000$ prediction (\ref{eq:N0finitesize})
is  already indistinguishable from the exact result obtained by summing
explicitly over the excited states of the  harmonic oscillator
Hamiltonian, apart from a narrow region near $T_c^0$ where higher
order corrections should be included  to get 
the exact result. This is well illustrated in Fig.~\ref{fig:s-condfrac},
where we plot the prediction (\ref{eq:N0finitesize}) (solid line)
together with the exact calculation obtained directly from 
(\ref{eq:bosedistribution}) (circles). Both predictions
are also compared with the thermodynamic limit, $N_0/N=1-(T/T_c^0)^3$.

Finite size effects reduce the condensate fraction and thus result in a
lowering of the transition temperature as compared to the $N\to\infty$
limit.   By setting the left hand
side of Eq.~(\ref{eq:N0finitesize}) equal to zero one can estimate the  
shift of the critical temperature to order $N^{-1/3}$ 
(Grossmann and Holthaus, 1995; Ketterle and van Druten, 1996b; 
Kirsten and Toms, 1996):
\begin{equation}
{ \delta T_c^0 \over T_c^0 } = - { \bar{\omega}  \zeta(2) \over
2 \omega_{\rm ho} [\zeta(3)]^{2/3} } N^{-1/3} \simeq
- \ 0.73 \  { \bar{\omega} \over \omega_{\rm ho} }
\ N^{-1/3} \; .
\label{eq:dtc}
\end{equation}

Another problem, which deserves to be mentioned in connection
with the finite size of the system, is the equivalence between 
different statistical ensembles and the problem of fluctuations. 
In the thermodynamic limit the grand canonical, canonical and 
microcanonical ensembles are expected to provide the same results. 
However, their equivalence is no longer ensured when $N$ is finite. 
Rigorous results concerning the ideal Bose gas in a box and, in 
particular, the behavior of fluctuations, can be found in Ziff 
{\it et al.} (1977), and Angelescu {\it et al.} (1996). In the case 
of a trapped gas, Gajda and  Rz\c{a}\.{z}ewski (1997) have  
shown that the differences between the predictions of the micro-
and grand canonical ensembles for the temperature dependence of 
the condensate fraction are small already at $N \sim 1000$.  
The fluctuations of the number of atoms in 
the condensate are instead much more sensitive to the choice of 
the ensemble (Navez {\it et al.}, 1997; Wilkens and Weiss, 1997; see 
also  Holthaus, Kalinowski and Kirsten, 1998, and references therein). 
Inclusion of two-body interactions can, however, change the scenario 
significantly (Giorgini, Pitaevskii and Stringari, 1998).

\subsection{Role of dimensionality}
\label{sec:dimensionality}

So far we have discussed the properties of the ideal Bose gas in 
three-dimensional space.  Though the trapping frequencies in each
direction can be quite different, nevertheless the relevant results 
for the temperature dependence of the condensate have been obtained
assuming that $k_BT$ is much larger than all the oscillator energies
$\hbar\omega_x,  \hbar\omega_y, \hbar\omega_z$. In order to
observe effects of reduced dimensionality,  one should remove
such a condition in one or two directions.

The statistical behavior of 2D and 1D Bose gases  exhibits very 
peculiar features. Let us first recall that in a uniform gas
Bose-Einstein condensation cannot occur in 2D and 1D at finite
temperature because thermal fluctuations destabilize the
condensate. This can be seen by noting that, for an ideal gas in 
the presence of BEC,  the chemical potential vanishes and the 
momentum distribution, $n(p) \propto [\exp(\beta p^2/2m)-1]^{-1}$, 
exhibits an infrared $1/p^2$ divergence. In the thermodynamic limit, 
this yields a divergent contribution to the integral $\int \! d{\bf p} 
\ n(p)$ in 2D and 1D, thereby violating the normalization condition. 
The absence of BEC in 1D and 2D can be also proven for interacting 
uniform systems, as shown by Hohenberg (1967). 

In the presence of harmonic trapping, the effects of thermal
fluctuations are  strongly quenched due to the different behavior
exhibited by  the density of states $\rho(\varepsilon)$.
In fact, while in the uniform gas $\rho(\varepsilon)$ behaves as
$ \varepsilon^{(d-2)/2}$, where $d$ is the dimensionality of
space, in the presence of an harmonic potential one has instead the law
$\rho(\varepsilon) \sim \varepsilon^{d-1}$ and, consequently, the
integral (16) converges also in 2D. The corresponding value of the
critical temperature is given by
\begin{equation}
k_B T_{2D}= \hbar\omega_{2D} \left(\frac{N}{\zeta(2)}\right)^{1/2} \; ,
\label{eq:t2d}
\end{equation}
where $\omega_{2D}=(\omega_x\omega_y)^{1/2}$ (see, for example, Mullin, 
1997, and references therein).  One notes first that in 2D the 
thermodynamic limit corresponds to taking $N \to \infty$
and $\omega_{2D}\to 0$ with the product $N\omega_{2D}^2$ kept constant.
In order to achieve 2D Bose-Einstein condensation in real 3D traps, one 
should choose the frequency $\omega_z$ in the third direction large 
enough to satisfy the condition  $\hbar\omega_{2D}\ll k_BT_{2D} < 
\hbar \omega_z$; this implies rather severe conditions on the deformation 
of the trap. The main features of BEC in 2D gases confined in harmonic 
traps and, in particular, the applicability of the Hohenberg theorem and
of its extensions to nonuniform gases, have been discussed in details
by Mullin (1997).  

\begin{figure}[t]
\epsfysize=8cm
\hspace{3cm}
\epsffile{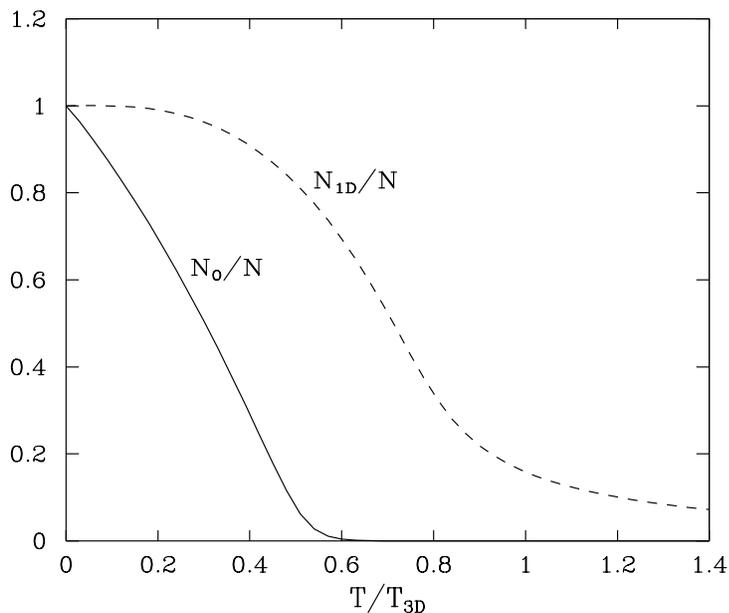}
\caption{Behavior of an ideal gas with $N=10^6$ particles in a highly 
anisotropic trap: $\omega_\perp=5.6 \times 10^4 \omega_z$, corresponding 
to $T_{3D}=2T_{1D}$. Solid line: fraction of atoms in the ground state 
($n_x=0,n_y=0,n_z=0$), dashed line: fraction of atoms in the lowest radial 
state ($n_x=0,n_y=0$). } 
\label{fig:twostep}
\end{figure}

In 1D the situation is also very interesting. In this case,
Bose-Einstein condensation cannot occur even in the presence of 
harmonic confinement 
because of the logarithmic  divergence in the integral
(\ref{eq:nminusn0rho}).  This means that the
critical temperature  for 1D Bose-Einstein condensation
tends to zero in  the thermodynamic limit if one keeps the product
$N\omega_{1D}$ fixed. In fact, in 1D the critical temperature for the
ideal  Bose gas can be estimated to be (Ketterle and van Druten, 1996b)
\begin{equation}
k_B T_{1D} = \hbar\omega_{1D} \frac{N}{\ln (2 N)}  
\label{eq:t1d}
\end{equation}
with $\omega_{1D}\equiv \omega_z$. 
Despite the fact that one cannot have BEC in the thermodynamic limit,
nevertheless for finite values of $N$ the system  can exhibit a
large occupation of the lowest single-particle state in a useful
interval of temperatures. Furthermore, if the value of $N$ and the 
parameters of the trap are chosen in a proper way, one observes a new 
interesting phenomenon  associated with the macroscopic occupation of 
the lowest energy state, taking place in two distinct steps (van 
Druten and Ketterle, 1997).  This happens when the relevant
parameters of the trap satisfy simultaneously the conditions
$T_{1D} < T_{3D}$ and $\hbar \omega_{\perp} < k_B T_{3D}$, where
$T_{3D}$ coincides with the usual critical temperature  given
in Eq.~(\ref{eq:tcho}) and $\omega_\perp$ is the frequency of the trap
in the  $x$-$y$ plane.  In the interval $T_{1D} < T < T_{3D}$, only 
the radial degrees of freedom are frozen,
while no condensation occurs in the axial degrees of freedom. 
At lower temperatures, below $T_{1D}$, also the axial variables
start being frozen and the overall ground state is  occupied in
a macroscopic way.  An example of this two-step BEC is shown in
Fig.~\ref{fig:twostep}. It is also interesting to  notice that the
conditions for the occurrence of two-step condensation in harmonic
potentials are peculiar of the 1D geometry. In fact, it is easy to
check that the corresponding conditions $T_{2D} < T_{3D}$ and
$\hbar \omega_{z} <  k_B T_{3D}$, which would yield two-step BEC
in 2D, cannot be easily satisfied because of the absence
of the $\ln N$ factor.

It is finally worth pointing out that the above discussion
concerns the behavior of the ideal Bose gas. Effects of two-body
interactions are expected to modify in a deep way the nature of
the phase transition in reduced dimensionality. In particular,
interacting Bose systems exhibit the well known 
Berezinsky-Kosterlitz-Thouless transition in 2D (Berezinsky,
1971;  Kosterlitz and Thouless, 1973).  The case of trapped gases in 
2D has been recently discussed by Mullin (1998) and is expected to 
become an important issue in future investigations.

\subsection{Non harmonic traps and adiabatic transformations}
\label{sec:nonharmonic}

A crucial step to reach the low temperatures needed for BEC in the
experiments realized so far is evaporative cooling. This
technique is intrinsically irreversible since it is based on the loss
of hot particles from the trap. New interesting perspectives would
open if one could adiabatically cool the system in a reversible way
(Ketterle and Pritchard, 1992; Pinkse {\it et al.}, 1997).
Reversible cooling  of the gas is achieved by adiabatically
changing the shape of the trap at a rate slow compared to the internal
equilibration rate.

An important class of trapping potentials for studying the effects
of adiabatic changes is provided by power-law potentials of the form
\begin{equation}
V_{\rm ext}({\bf r}) = A \; r^\alpha \;,
\label{eq:Upl}
\end{equation}
where, for simplicity, we assume spherical symmetry. The critical
temperature for Bose-Einstein condensation in the trap (\ref{eq:Upl})
has been calculated by Bagnato, Pritchard and Kleppner (1987) and is 
given by
\begin{equation}
k_BT_c^0 = \left[ \frac{N\hbar^3}{(2m)^{3/2}}
\frac{6\sqrt{\pi}A^\delta} {\Gamma(1+\delta)\zeta(3/2+\delta)}
\right]^{1 \over (3/2+\delta) } \;.
\label{eq:Tcpl}
\end{equation}
Here we have introduced the parameter $\delta=3/\alpha$, while
$\Gamma(x)$ is the usual gamma function.  By setting $\delta=3/2$ and 
$A=m\omega_{\rm ho}^2/2$, one recovers the  result for the transition
temperature in an isotropic harmonic trap. The result
for a rigid box is instead obtained  by letting $\delta\to 0$.

It is straightforward  to work out the thermodynamics of a
noninteracting gas in the confining potential (\ref{eq:Upl}) (Bagnato, 
Pritchard and Kleppner, 1987;  Pinkse {\it et al.} 1997). For example,  
for the condensate fraction one finds: $N_0/N=1-(T/T_c^0)^{3/2+\delta}$.
More relevant to the discussion of reversible processes is the 
entropy which remains constant during the adiabatic change.
Above $T_c$ the system can be approximated by a classical
Maxwell-Boltzmann gas and the entropy per particle takes the simple
form
\begin{equation}
\frac{S}{Nk_B} = \left( \frac{5}{2}+\delta-\ln \zeta(3/2+\delta) \right)
+ \left(\frac{3}{2}+\delta\right) \ln \left(\frac{T}{T_c^0}\right) \;.
\label{Spl}
\end{equation}
\noindent From this equation one sees that the entropy depends on the
parameter $A$ of the external potential (\ref{eq:Upl}) only through the
ratio $T/T_c^0$. Thus, for a fixed power-law dependence of the trapping
potential ($\delta$ fixed), an adiabatic change of $A$, like for example
an adiabatic expansion of the harmonic trap, does not bring us closer
to the transition, since the ratio $T/T_c^0$ remains constant. A reduction 
of the ratio $T/T_c^0$ is instead obtained by increasing
adiabatically $\delta$, that is,  changing the power-law dependence
of the trapping potential (Pinkse {\it et al.} 1997). For example, in
going from a harmonic ($\delta_1=3/2$) to a linear trap ($\delta_2=3$), 
one gets the relation $t_2\simeq 0.7 t_1^{2/3}$ between the initial and
final reduced temperature $t=T/T_c^0$. In this case a system at twice
the critical temperature ($t_1=2$) can be cooled down to nearly the
critical point ($t_2\simeq 1.1$).  Using this technique it should be
possible, by a proper change of $\delta$, to cool adiabatically the
system from the high temperature phase without condensate down to
temperatures below $T_c$ with a large fraction of atoms in the condensate
state. 

The possibility of reaching BEC using adiabatic transformations has 
been recently successfully explored in an experiment carried out at MIT 
(Stamper-Kurn {\it et al.}, 1998b).

\section{Effects of interactions: ground state}
\label{sec:groundstate}

\subsection{Order parameter and mean-field theory} 
\label{sec:manybody}

The many body Hamiltonian describing  $N$ interacting bosons confined by
an external potential $V_{\rm ext}$ is given, in  second quantization, by:
\begin{equation}
\hat{H}  =  \int \! d{\bf r}  \ \hat\Psi^{\dagger} ({\bf r})
\left[ - {\hbar^2 \over 2m} \nabla^2 + V_{\rm ext}({\bf r}) \right]
\hat\Psi ({\bf r})
 +  {1\over 2} \int \! d{\bf r}d{\bf r}' \
\hat\Psi^{\dagger} ({\bf r}) \hat\Psi^{\dagger} ({\bf r}')
V({\bf r} - {\bf r}')
\hat{\Psi} ({\bf r}') \hat{\Psi} ({\bf r})
\label{eq:manybodyhamiltonian}
\end{equation}
where $\hat{\Psi}({\bf r})$ and $\hat\Psi^{\dagger}({\bf r})$ are
the boson field operators that annihilate and create a particle
at the position ${\bf r}$, respectively, and $V({\bf r} - {\bf r}')$
is the two-body interatomic potential.

The ground state of the system as well as its thermodynamic properties
can be directly calculated starting from the Hamiltonian
(\ref{eq:manybodyhamiltonian}). For instance, Krauth (1996) has 
used a Path Integral Monte Carlo method to calculate the thermodynamic
behavior of  $10^4$ atoms interacting with a repulsive ``hard-sphere''
potential. In principle, this procedure gives exact results within
statistical errors. However, the calculation can be heavy or even
impracticable for systems with much  larger values of  $N$.
Mean-field approaches are commonly developed for interacting
systems in order to overcome the problem of solving
exactly the full many-body Schr\"odinger equation. Apart from
the convenience of avoiding heavy numerical work, mean-field
theories  allow one to understand the behavior of a system in terms
of a set of parameters having a clear physical meaning.  This is
particularly true in the case of trapped bosons. Actually most
of the results reviewed in this paper show that the mean-field
approach is very effective in providing quantitative predictions
for the  static, dynamic and thermodynamic properties of these
trapped gases.

The basic idea for a mean-field description of a dilute Bose
gas was formulated by Bogoliubov (1947). The key point
consists in separating out the condensate
contribution to the bosonic field operator. In general, the field
operator can be written as $\hat\Psi ({\bf r}) = \sum_\alpha
\Psi_\alpha ({\bf r})  a_\alpha$,  where $\Psi_\alpha ({\bf r})$
are single-particle wave functions and $a_\alpha$ are the
corresponding annihilation operators. The bosonic creation
and annihilation operators $a^{\dagger}_\alpha$ and $a_\alpha$ are
defined in Fock space through the relations
\begin{eqnarray}
a_{\alpha}^{\dagger} \mid n_0, n_1, \dots, n_{\alpha}, \dots
\rangle   & = &  \sqrt{n_{\alpha} + 1} \mid n_0, n_1, \dots,
n_{\alpha} +1,  \dots \rangle  \\ 
a_{\alpha} \mid n_0, n_1, \dots, n_{\alpha}, \dots \rangle & = & 
\sqrt{n_{\alpha}} \mid n_0, n_1, 
\dots, n_{\alpha} - 1, \dots \rangle 
\label{eq:aadagoperators}
\end{eqnarray}
where $n_{\alpha}$ are the eigenvalues of the operator
$\hat n_{\alpha}= a^{\dagger}_{\alpha}a_{\alpha}$ giving the
number of atoms in the single-particle $\alpha$-state. They 
obey the usual  commutation rules:
\begin{equation}
\big[ a_{\alpha}, a_{\beta}^{\dagger} \big] =  \delta_{\alpha, \beta}
\; , \;  \big[ a_{\alpha}, a_{\beta} \big]  =  0  \; , \; 
\big[ a_{\alpha}^{\dagger}, a_{\beta}^{\dagger} \big]  =  0 \; .
\label{eq:commrules }
\end{equation}
Bose-Einstein  condensation occurs when the number of atoms $n_0$ of a
particular single-particle state becomes very large: $n_0 \equiv N_0
\gg 1$ and the ratio $N_0/N$ remains finite in the thermodynamic limit
$N\to\infty$. In this limit the states with $N_0$ and $N_0 \pm 1\simeq
N_0$ correspond to the same physical configuration and, consequently, the
operators  $a_0$ and  $a_0^{\dagger}$ can be treated like $c$-numbers:
$a_0=a_0^{\dagger}=\sqrt{N_0}$. For a uniform gas in a volume $V$, where
BEC occurs in the single-particle state $\Psi_0= 1/\sqrt{V}$ having zero
momentum, this means that the field operator $\hat \Psi({\bf r})$ can
be decomposed in the form $\hat\Psi ({\bf r}) =  \sqrt{N_0/V} +
\hat\Psi' ({\bf r})$. By treating  the operator $\hat\Psi'$ as a small
perturbation, Bogoliubov developed the ``first-order" theory for the
excitations of interacting Bose gases.

The generalization of the Bogoliubov prescription to the case of  
nonuniform and time dependent configurations is given by
\begin{equation}
\hat \Psi ({\bf r},t) = \Phi ({\bf r},t) + \hat\Psi' ({\bf r},t) 
\; ,
\label{eq:prescription}
\end{equation}
where we have used the Heisenberg representation for the field operators.
Here $\Phi({\bf r},t)$ is a complex function defined as the expectation
value of the field operator: $\Phi({\bf r},t) \equiv  \langle \hat{\Psi}
({\bf r},t)\rangle$. Its modulus fixes the condensate density through 
$n_0({\bf r},t)=|\Phi({\bf r},t)|^2$. The function $\Phi({\bf r},t)$ 
possesses also a well-defined phase and, similarly to the case of uniform 
gases, this corresponds to assuming the occurrence of a broken gauge 
symmetry in the many-body system.

The function $\Phi({\bf r},t)$ is a classical field having the meaning of 
an order parameter and is often called ``wave function of the condensate". 
It characterizes the off-diagonal long-range behavior of the one-particle 
density matrix $\rho_1({\bf r}',{\bf r},t)= \langle \hat \Psi ^\dagger 
({\bf r}',t) \hat \Psi ({\bf r},t)\rangle $. In fact the decomposition
(\ref{eq:prescription}) implies the following asymptotic behavior
(Ginzburg and Landau, 1950; Penrose, 1951; Penrose and Onsager, 1956):
\begin{equation}
\lim_{|{\bf r}'-{\bf r}|\to\infty}  \rho_1 ({\bf r}',{\bf r},t) =
\Phi ^*({\bf r}',t) \Phi ({\bf r},t) \; .
\label{eq:nondiag}
\end{equation}
Notice that, strictly speaking, in a finite-sized system neither
the concept of broken gauge symmetry, nor the one of off-diagonal
long-range order can be applied. The condensate wave function $\Phi$
has nevertheless still a clear meaning: it can be in fact determined
through  the diagonalization of the one-body density matrix, 
$\int \! d{\bf r}' \rho_1 ({\bf r}',{\bf r}) \Phi_i ({\bf r}') =
N_i \Phi_i ({\bf r})$,  and corresponds to the eigenfunction, $\Phi_i$, 
with the largest eigenvalue, $N_i$. This procedure has been used, for
example, to explore Bose-Einstein condensation in finite drops of
liquid helium by Lewart {\it et al.} (1988).  The connection between 
the condensate wave function, defined through the diagonalization of 
the density matrix and the concept of order parameter commonly used 
in the theory of superfluidity, is an interesting and nontrivial 
problem in itself. Another important question concerns the possible
fragmentation of the condensate, taking place when two or more 
eigenstates of the density matrix $\rho_1 ({\bf r}',{\bf r})$ are
macroscopically occupied. One can show (Nozi\`eres and Saint
James, 1982; Nozi\`eres, 1995) that,  due to exchange effects, in 
uniform gases interacting with repulsive forces the fragmentation costs 
a macroscopic energy. The  behavior can be however different in the 
presence of attractive forces and almost degenerate single-particle 
states (Nozi\`eres and Saint James, 1982; Kagan, Shlyapnikov and 
Walraven, 1996; Wilkin, Gunn and Smith, 1998).

The decomposition (\ref{eq:prescription}) becomes particularly useful
if $\hat\Psi'$ is small, i.e., when the depletion of the condensate
is small. Then, an equation for the order parameter
can be derived  by expanding the theory to  the lowest orders in
$\hat\Psi'$ as in the case of uniform gases. The main difference 
is that here one gets also a nontrivial ``zeroth-order" theory for
$\Phi({\bf r},t)$. 

In order to derive the equation for the condensate wave function
$\Phi({\bf r},t)$, one has to write the time evolution of the field
operator $\hat \Psi({\bf r},t)$ using the Heisenberg equation with
the many-body Hamiltonian (\ref{eq:manybodyhamiltonian}):
\begin{eqnarray}
 i\hbar \frac{\partial}{\partial t} \hat \Psi({\bf r},t)&=&
 [ \hat \Psi, \hat H ]
\nonumber \\
&=& \left[ -\frac{\hbar^2\nabla ^2}{2m} +
V_{\rm ext}({\bf r})  +\int d{\bf r}' \
\hat \Psi ^{\dagger}({\bf r}',t) V({\bf r}'-{\bf r})
\hat \Psi ({\bf r}',t) \right] \hat \Psi ({\bf r},t) \; .
\label{eq:H}
\end{eqnarray}
Then one has to replace the operator $\hat\Psi$ with the classical
field $\Phi$. In the integral containing the atom-atom interaction
$V({\bf r}'-{\bf r})$, this replacement is, in general, a poor 
approximation when short distances $({\bf r}'-{\bf r})$ are involved. 
In a dilute and cold gas, one can nevertheless obtain a proper
expression for the interaction term by observing that, in this
case, only binary collisions at low energy are relevant and these 
collisions are characterized by a single parameter, the $s$-wave 
scattering length, independently of the details of the two-body 
potential. This allows one to replace $V({\bf r}'-{\bf r})$ in 
(\ref{eq:H}) with an effective interaction 
\begin{equation}
V({\bf r}'-{\bf r}) = g \delta ({\bf r}'-{\bf r})
\label{eq:vdelta}
\end{equation}
where the coupling constant $g$ is related to the scattering
length $a$ through
\begin{equation}
g = { 4 \pi \hbar^2 a \over m} \; .
\label{eq:g}
\end{equation}
The use of the effective potential (\ref{eq:vdelta}) in 
(\ref{eq:H}) is compatible with the replacement of $\hat\Psi$ with 
$\Phi$ and yields the following closed equation for the order
parameter: 
\begin{equation}
i \hbar { \partial \over \partial t} \Phi ({\bf r},t) =
 \left( - { \hbar^2 \nabla^2 \over 2m } + V_{\rm ext}({\bf r})
+g |\Phi({\bf r},t)|^2 \right) \Phi ({\bf r},t) \; . 
\label{eq:TDGP}
\end{equation}
This equation, known as Gross-Pitaevskii (GP) equation, was
derived independently by Gross (1961 and 1963) and  Pitaevskii (1961).
Its validity is based on the condition that the $s$-wave scattering
length be much smaller than the average distance between atoms and
that the number of atoms in the condensate be much larger than $1$.
The GP equation can be used, at low temperature, to explore the
macroscopic behavior of the system, characterized by variations of the
order parameter over distances larger than the mean distance between
atoms.

The Gross-Pitaevskii equation (\ref{eq:TDGP}) can also be obtained 
using a variational procedure:
\begin{eqnarray} 
 i\hbar\frac{\partial}{\partial t} \Phi =
\frac{\delta  E}{\delta  \Phi ^*} \;,
\label{eq:varE}
 \end{eqnarray}
 where the energy functional $E$ is given by
\begin{equation}
E[\Phi] = \int d{\bf r} \  \left[ {\hbar^2 \over 2m}
|  \mbox{\boldmath$\nabla$} \Phi |^2
+  V_{\rm ext}({\bf r}) |\Phi|^2  + { g \over 2}
|\Phi|^4 \right]   \; .
\label{eq:energy}
\end{equation}
The first term in the integral (\ref{eq:energy}) is the kinetic energy
of the condensate $E_{\rm kin}$, the second is the harmonic oscillator 
energy $E_{\rm ho}$, while the last one is the mean-field interaction 
energy $E_{\rm int}$. Notice that the mean-field term, $E_{\rm int}$, 
corresponds to the first correction in the virial expansion for the
energy of the gas. In the case of non-negative and finite-range
interatomic potentials, rigorous bounds for this term have been
obtained by Dyson (1967) and Lieb and Yngvason (1998).  

The dimensionless parameter controlling the validity of the dilute-gas 
approximation, required for the derivation of Eq.~(\ref{eq:TDGP}), is 
the number of particles in a  ``scattering volume" $|a|^3$. 
This can be written  as $\bar{n} |a|^3$, 
where $\bar{n}$ is the average density of the gas. Recent determinations  
of the scattering length for the atomic species used in the experiments 
on BEC give: $a=2.75$ nm for $^{23}$Na (Tiesinga {\it et al.}, 1996), 
$a=5.77$ nm for $^{87}$Rb (Boesten {\it et al.}, 1997) and $a=-1.45$ nm 
for $^7$Li (Abraham {\it et al.}, 1995). Typical values of density range 
instead  from $10^{13}$ to $10^{15}$ cm$^{-3}$, so that $\bar{n} |a|^3$ 
is always less than $10^{-3}$. 

When $\bar{n} |a|^3 \ll 1$ the system is said to be dilute or weakly
interacting.  However, one should better clarify the meaning of the 
words ``weakly interacting", since the smallness of  the parameter 
$\bar{n} |a|^3$ does not imply  necessarily that the interaction effects 
are small. These effects, in fact, have to be compared with the kinetic 
energy of the atoms in the trap.  A first estimate can be obtained by 
calculating the  interaction energy, $E_{\rm int}$, on the ground state 
of the harmonic oscillator. This energy is given by $g N \bar{n}$, where 
the average density is of the order of $N / a_{\rm ho}^3$, so that
$E_{\rm int} \propto N^2 |a| / a_{\rm ho}^3$.  On the other hand, the 
kinetic energy is of the order of $N \hbar \omega_{\rm ho}$ and thus 
$E_{\rm kin} \propto N a_{\rm ho}^{-2}$. One finally finds
\begin{equation}
{ E_{\rm int} \over E_{\rm kin} } \propto { N |a| \over a_{\rm ho} } \; .
\label{eq:naoveraho}
\end{equation}
This is the  parameter expressing the importance of the atom-atom
interaction compared to the kinetic energy. It can be easily larger
than $1$ even if $\bar{n} |a|^3 \ll 1$, so that also very dilute gases
can exhibit an important {\it nonideal} behavior, as we will 
discuss in the following sections. In 
the first experiments with rubidium atoms at JILA (Anderson {\it et al.}, 
1995) the ratio $|a| / a_{\rm ho}$ was about $7\times 10^{-3}$, with $N$ of
the order of a few thousands. Thus  $N a / a_{\rm ho}$ is larger
than $1$. In the experiments with $^7$Li at Rice University
(Bradley {\it et al.}, 1997; Sackett {\it et al.}, 1997) the same 
parameter is smaller
than $1$, since the number of particles is of the order of $1000$
and $|a|/a_{\rm ho} \approx 0.5 \times 10^{-3}$. Finally, in the
experiments with sodium at MIT (Davis {\it et al.}, 1995) the number
of atoms in the condensate is very large ($10^6 - 10^7$)
and $N |a| /a_{\rm ho} \sim 10^3 - 10^4$.

Due to the assumption $\hat\Psi' \equiv 0$, the above formalism
is strictly valid only in the limit of zero temperature, when all
the particles are in the condensate. The dynamic behavior and the
generalization to finite temperatures will be discussed in
Secs.~\ref{sec:dynamics} and \ref{sec:thermodynamics}, respectively.
Here we present the results for the stationary solution of the
Gross-Pitaevskii (GP) equation at zero temperature.

\subsection{Ground state}
\label{sec:stationaryGP}

For a system of noninteracting bosons in a harmonic trap, the condensate 
has the form of a Gaussian of average width $a_{\rm ho}$ [see 
Eq.~(\ref{eq:gaussian})], and the central density is proportional to $N$.
If the atoms are interacting, the shape of the condensate can change
significantly with respect to the Gaussian.  The scattering length
entering the Gross-Pitaevskii equation can be positive or negative, its
sign and magnitude depending crucially on the details of the  atom-atom
potential. Positive and negative values of $a$ correspond to
an effective repulsion and attraction between the atoms, respectively.
The change can be dramatic when the interaction energy is much
greater than the kinetic energy, that is, when  $N |a| /a_{\rm ho} \gg 1$.
The central density is  lowered  (raised)  by a repulsive (attractive)
interaction and the radius of the atomic cloud consequently increases
(decreases). This effect of the interaction has important
consequences, not only for the structure of the ground state,
but also for the dynamics and thermodynamics of the system, as
we will see later on. 

The ground state can be easily obtained within
the formalism of mean-field theory. For this, one can write the
condensate wave function as $\Phi({\bf r},t)=\phi({\bf r})
\exp(-i \mu t/ \hbar)$,  where $\mu$ is the chemical potential
and $\phi$ is real and normalized to the  total number of particles,
$\int d{\bf r} \ \phi^2 = N_0 = N$.  Then the Gross-Pitaevskii
equation (\ref{eq:TDGP}) becomes
\begin{equation}
\left( - { \hbar^2\nabla^2 \over 2m } + V_{\rm ext}({\bf r})  +
g \phi^2 ({\bf r}) \right) \phi({\bf r}) = \mu  \phi({\bf r}) \; .
\label{eq:GP}
\end{equation}
This has  the form of a ``nonlinear Schr\"odinger equation", the
nonlinearity coming from the mean-field term, proportional to the
particle density $n({\bf r})=\phi^2({\bf r})$. In the absence of 
interactions ($g=0$), this equation reduces to the usual 
Schr\"odinger equation for the single-particle Hamiltonian 
$-\hbar^2/(2m)\nabla^2+V_{\rm ext}({\bf r})$ and, for harmonic 
confinement, the ground state solution coincides, apart from a 
normalization factor, with the Gaussian function (\ref{eq:gaussian}):
$\phi({\bf r})=\sqrt{N}\varphi_0({\bf r})$. We note, in passing, 
that a similar nonlinear equation for the order parameter has been 
also considered in connection with the theory of superfluid helium 
near the $\lambda$-point (Ginzburg and Pitaevskii, 1958); in that
case, however, the ingredients of the equation have a different 
physical meaning. 

The numerical solution of the GP equation (\ref{eq:GP}) is relatively
easy to obtain (Edwards and Burnett, 1995; Ruprecht {\it et al.}, 1995;
Edwards {\it et al.}, 1996b; Dalfovo and Stringari, 1996; Holland and
Cooper, 1996). Typical wave functions $\phi$, calculated 
from Eq.~(\ref{eq:GP}) with different values of the  parameter
$N |a| /a_{\rm ho}$, are shown in Figs.~\ref{fig:aneg} and \ref{fig:apos}
for attractive and repulsive interaction, respectively. The effects
of the interaction are revealed by the deviations from the Gaussian 
profile (\ref{eq:gaussian}) predicted by the noninteracting model. 
Excellent agreement has been found by comparing the solution of the 
GP equation with the experimental density profiles obtained at low 
temperature (Hau {\it et al.}, 1998), as shown in Fig.~\ref{fig:hau}.
The condensate wave function obtained with the stationary GP equation
has been also compared with the results of an {\it ab initio} Monte Carlo
simulation starting from Hamiltonian (\ref{eq:manybodyhamiltonian}),
finding a very good agreement (Krauth, 1996).  

The role of the parameter $N |a| /a_{\rm ho}$, already
discussed in the previous section, can be easily pointed out, in the
Gross-Pitaevskii equation, by using rescaled dimensionless variables.
Let us consider a spherical trap with frequency $\omega_{\rm ho}$
and use $a_{\rm ho}$,  $a_{\rm ho}^{-3}$ and $\hbar\omega_{\rm ho}$ as 
units of length,  density and energy, respectively. By putting a tilde
over the rescaled quantities, Eq.~(\ref{eq:GP}) becomes
\begin{equation}
\left[ - \tilde \nabla^2 + \tilde r^2 
+ 8 \pi ( N a/ a_{\rm ho})  \tilde \phi^2(\tilde{\bf r})
\right]  \tilde \phi(\tilde{\bf r}) = 2 \tilde \mu
\tilde \phi(\tilde{\bf r}) \;.
\label{eq:dimensionlessGP}
\end{equation}
In these new units the order parameter satisfies the normalization
condition $\int d\tilde{\bf r} |\tilde\phi|^2=1$.
It is now evident that the importance of the atom-atom interaction
is completely fixed by the parameter $N a /a_{\rm ho}$. 

It is worth noticing that the solution of the stationary GP equation 
(\ref{eq:GP}) minimizes the energy functional (\ref{eq:energy}) for 
a fixed number of particles. Since the ground state has no currents, 
the energy is a functional of the density only, which can be written
in the form
\begin{equation}
E[n] \  = \  \int d{\bf r} \  \left[ {\hbar^2 \over 2m}
| \mbox{\boldmath$\nabla$} \sqrt{n} |^2
+  n V_{\rm ext}({\bf r})  + { g n^2 \over 2} \right]  
\  = \  E_{\rm kin} + E_{\rm ho} + E_{\rm int}  \; .
\label{eq:funcofdensity}
\end{equation}  
The first term corresponds to the quantum kinetic energy coming from 
the uncertainty principle; it is usually named ``quantum pressure" and 
vanishes for uniform systems.  In general, for a nonstationary order 
parameter, the kinetic energy in  (\ref{eq:energy}) includes also 
the contribution of currents in the form of an additional term 
containing  the gradient of the phase of $\Phi$. 

By direct integration of the GP equation (\ref{eq:GP}) one finds
the useful expression
\begin{equation}
\mu = ( E_{\rm kin} + E_{\rm ho} + 2 E_{\rm int} ) /N
\label{eq:newmu}
\end{equation}
for the chemical potential in terms of the different contributions
to the energy functional (\ref{eq:funcofdensity}). Further important
relationships can be also found by means of the virial theorem. In
fact, since the energy (\ref{eq:energy}) is stationary  for any variation
of $\phi$ around the exact solution of the GP equation, one can choose
scaling transformations of  the form $\phi(x,y,z) \to (1+\nu)^{1/2}
\phi ( (1+\nu)x, y,z)$, and  insert them in (\ref{eq:energy}).
By imposing the energy variation to vanish at first order
in $\nu$, one finally gets 
\begin{equation}
(E_{\rm kin})_x - (E_{\rm ho})_x + {1\over 2} E_{\rm int} = 0 \; ,      
\label{eq:virial}
\end{equation}
where $(E_{\rm kin})_x= \langle \sum_i p_{ix}^2 \rangle /2m$
and $(E_{\rm ho})_x= (m/ 2) \omega_x^2 \langle \sum_i x_i^2 \rangle$. 
Analogous expressions are found by choosing similar scaling 
transformations for the $y$ and $z$ co-ordinates. By summing over 
the three directions one finally finds the virial relation:
\begin{equation} 
2 E_{\rm kin} - 2 E_{\rm ho}+ 3 E_{\rm int}=0 \; .
\label{eq:virial1}
\end{equation}
The above results are exact  within  Gross-Pitaevskii theory and can 
be used, for instance, to check the numerical solutions of
Eq.~(\ref{eq:GP}).

\begin{figure}[t]
\epsfysize=8cm
\hspace{3cm}
\epsffile{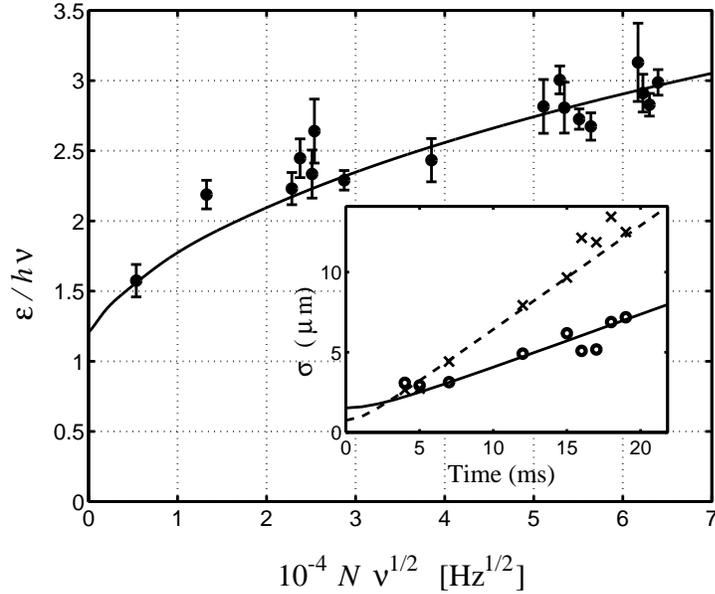}
\caption{Comparison of the release energy as a function of
interaction strength from the stationary GP equation (solid line) 
and  the experimental measurements (solid circles). Inset shows the
expansion  of widths of the condensate  in the horizontal (empty 
circles) and vertical (crosses) directions against the predictions
of the time dependent GP equation (dashed and solid lines) for the 
data point at $10^{-4} N \nu^{1/2} = 0.53$. Here $\nu$ is the 
frequency of the trapping potential and the trapped gas is 
rubidium. From Holland {\it et al.} (1997).  } 
\label{fig:releasejila}
\end{figure}

\begin{figure}[t]
\epsfysize=8cm
\hspace{3cm}
\epsffile{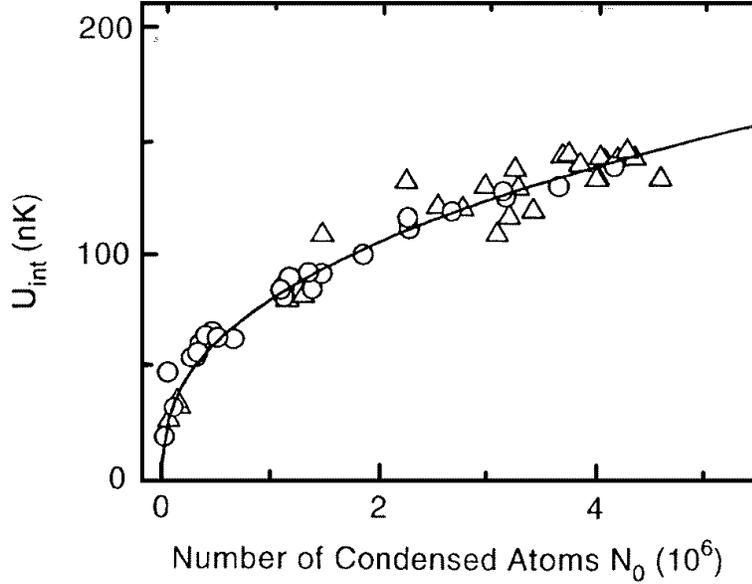}
\caption{Release energy of the condensate as a function of the number 
of condensed atoms in the MIT trap with sodium atoms. For these 
condensates the initial kinetic energy is negligible and the release
energy coincides with the mean-field energy. The symbol $U_{int}$
is here used for the mean-field energy per particle.
Triangles: clouds with no visible thermal component. Circles: clouds
with both thermal and condensed fractions visible. The solid line is 
a fit proportional to $N_0^{2/5}$ (see discussion in 
Sec.~\protect\ref{sec:TF}). From Mewes {\it et al.} (1996a).  } 
\label{fig:releasemit}
\end{figure}

In a series of experiments the gas has been imaged after a sudden
switching-off of the trap and the kinetic energy of the atoms has
been measured by integrating over the observed velocity distribution.
This energy, which is also called release energy, coincides with the
sum of the kinetic and interaction energies of the atoms at the
beginning of the expansion:
\begin{equation}
E_{\rm rel}=E_{\rm kin}+E_{\rm int} \; .
\label{release}
\end{equation}
During the first phase of the expansion both the quantum kinetic
energy (quantum pressure) and the interaction energy are rapidly 
converted into kinetic energy of motion. Then the atoms expand at 
constant velocity. Since energy is conserved during the expansion,
its initial value (\ref{release}), calculated with the stationary  
GP equation, can be directly compared with experiments. This 
comparison provides  clean evidences for the crucial role 
played by two-body interactions. In fact, the noninteracting model
predicts a release energy per particle given by $E_{\rm rel} /N = 
(1/2) (1+ \lambda/2) \hbar \omega_{\rm ho}$, independent of $N$. 
Conversely, the observed release energy per particle 
depends rather strongly on $N$, in good agreement with the 
theoretical predictions for the interacting gas. In 
Figs.~\ref{fig:releasejila} and \ref{fig:releasemit}, we show 
the experimental data obtained at JILA (Holland {\it et al.},1997) 
and MIT (Mewes {\it et al.}, 1996a), respectively. 

Finally, we notice that the balance between the quantum pressure and the  
interaction energy of the condensate fixes a typical length scale, 
called the healing length, $\xi$.  This is the minimum distance over 
which the order parameter can heal. If the condensate density grows 
from $0$ to $n$ within a distance $\xi$, the two terms in 
Eq.~(\ref{eq:GP}) coming from the quantum pressure and the interaction 
energy are $\sim \hbar^2/(2m\xi^2)$ and $\sim  4 \pi \hbar^2 a n/m$, 
respectively. By equating them, one finds the following expression 
for the healing length:
\begin{equation}
\xi = (8 \pi n a )^{-1/2} \; .
\label{eq:healinglength}
\end{equation}
This is a well known result for weakly interacting Bose gases. In the 
case of trapped bosons, one can use the central  density, or the average 
density,  to get an order of magnitude  of the healing length. This 
quantity is relevant for  superfluid effects. For instance, it provides
the typical size of the core of quantized vortices (Gross, 1961; 
Pitaevskii, 1961). Note that in condensed matter physics the
same quantity is often named ``coherence length", but the name  
``healing length" is preferable here in order to avoid confusion with 
different definitions of coherence length used in atomic physics and
optics.

\subsection{Collapse for attractive forces}
\label{sec:attractiveforces}

\begin{figure}[t]
\epsfysize=8cm
\hspace{3cm}
\epsffile{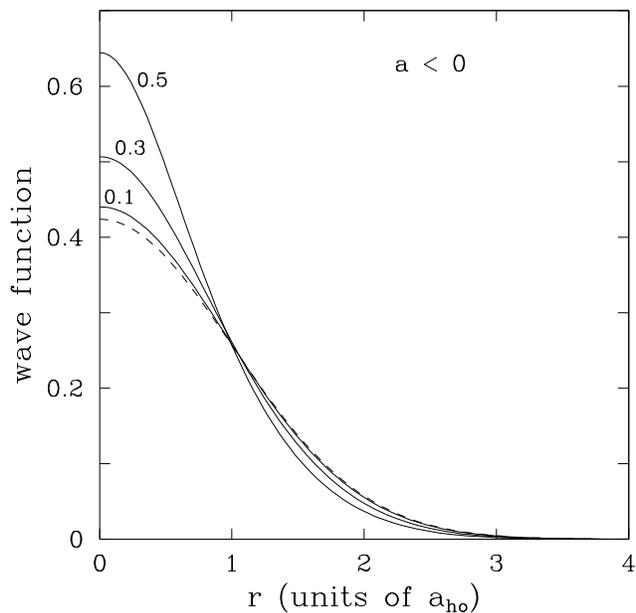}
\caption{Condensate wave function, at $T=0$, obtained by solving 
numerically the stationary GP equation (\protect\ref{eq:GP}) 
in a spherical trap and with attractive interaction among the atoms
($a<0$). The  three solid lines correspond to $N|a|/a_{\rm ho}= 0.1, 0.3, 
0.5$. The dashed line is the prediction for the ideal gas. Here 
the radius, $r$, is in units of the oscillator length $a_{\rm ho}$ 
and we plot $(a_{\rm ho}^3/N)^{1/2} \phi(r)$, so that the curves are 
normalized to $1$ [see also 
Eq.~(\protect\ref{eq:dimensionlessGP})]. }
\label{fig:aneg}
\end{figure}

\begin{figure}[t]
\epsfysize=8cm
\hspace{3cm}
\epsffile{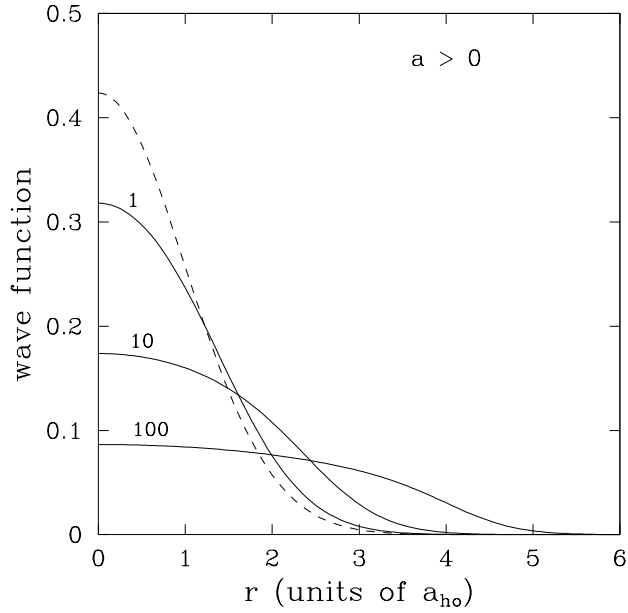}
\caption{Same as in Fig.~\protect\ref{fig:aneg}, but 
for repulsive interaction ($a>0$) and $Na/a_{\rm ho}= 1, 10, 100$.   } 
\label{fig:apos}
\end{figure}

If forces are attractive ($a<0$), the gas tends to increase
its density in the center of the trap in order to lower the 
interaction energy, as seen in  Fig.~\ref{fig:aneg}. 
This tendency is contrasted by the zero point kinetic energy
which can stabilise the system. However, if the central density
grows too much, the kinetic energy is no longer able to avoid the
collapse of the gas. For a given atomic species in a given trap,
the collapse is expected to occur when the number of particles in 
the condensate exceeds a critical value $N_{\rm cr}$, of the order 
of $a_{\rm ho}/|a|$. It is worth stressing that in a uniform gas, 
where quantum pressure is absent, the condensate is always unstable.

The critical number $N_{\rm cr}$ can be calculated at zero temperature
by means of the Gross-Pitaevskii equation.  The condensates shown in 
Fig.~\ref{fig:aneg} are metastable, corresponding to local minima of the
energy functional (\ref{eq:energy}) for different $N$. When $N$
increases, the depth of the local minimum decreases.  Above $N_{\rm cr}$
the minimum no longer exists and the Gross-Pitaesvkii equation
has no solution. For a spherical trap this happens at  (Ruprecht
{\it et al.}, 1995)
\begin{equation}
{ N_{\rm cr} |a| \over a_{\rm ho} } = 0.575 \;.
\label{eq:Ncr}
\end{equation}
For the axially symmetric trap with $^7$Li used in the experiments at
Rice University  (Bradley {\it et al.}, 1995 and 1997; Sackett {\it et 
al.}, 1997), the GP equation predicts $N_{\rm cr} \simeq 1400$ (Dalfovo 
and Stringari, 1996; Dodd {\it et al.}, 1996);  this value is consistent 
with recent experimental measurements  (Bradley {\it et al.}, 1997; 
Sackett {\it et al.}, 1997). The same problem has been investigated 
theoretically by several authors (Kagan, Shlyapnikov and Walraven, 1996; 
Houbiers and Stoof, 1996; Shuryak, 1996; Pitaevskii, 1996; Bergeman,1997).  

\begin{figure}[t]
\epsfysize=8cm
\hspace{3cm}
\epsffile{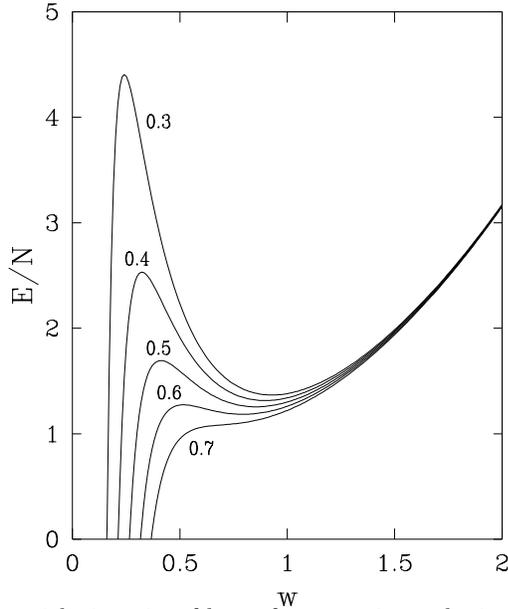}
\caption{Energy per particle, in units of $\hbar \omega_{\rm ho}$, for
atoms in a spherical trap interacting with attractive forces, as
a function of the effective width $w$ in the Gaussian model of 
Eqs.~(\protect\ref{eq:gaussianapprox})-(\protect\ref{eq:eofd}). 
Curves are plotted for several values of the parameter 
$N|a|/a_{\rm ho}$. The local minimum disappears at $N=N_{\rm cr}$.} 
\label{fig:gauss}
\end{figure}

A direct insight into the behavior of the gas with attractive forces
can be obtained by means of a variational approach based on Gaussian
functions (Baym and Pethick, 1996).  For a spherical trap one can 
minimize the energy (\ref{eq:energy}) using the {\it ansatz}
\begin{equation}
\phi (r) = \left( {N \over w^3 a_{\rm ho}^3  \pi ^{3/2} }
\right)^{1/2} \exp \left( - { r^2 \over 2 w^2 a_{\rm ho}^2 } \right)  \; ,
\label{eq:gaussianapprox}
\end{equation}
where $w$ is a dimensionless variational parameter which fixes the
width of the condensate. One gets
\begin{equation}
{ E (w) \over N \hbar \omega_{\rm ho} }  =
{3 \over 4} ( w^{-2} + w^2 ) - (2\pi)^{-1/2}
{ N |a| \over a_{\rm ho} } \ w^{-3} \; .
\label{eq:eofd}
\end{equation}
This energy is plotted in Fig.~\ref{fig:gauss} as
a function of $w$, for several values of the parameter
$N|a|/a_{\rm ho}$. One clearly sees that the local minimum disappears
when this parameter exceeds a critical value. This can
be calculated by requiring that the first and second derivative
of $E(w)$ vanish at the critical point ($w=w_{\rm cr}$ and $N=N_{\rm cr}$).
One finds
$w_{\rm cr} = 5^{-1/4} \approx 0.669$ and
$N_{\rm cr}|a|/a_{\rm ho}
\approx 0.671$.
The last formula provides an estimate of the critical
number of atoms, for given trap and atomic species, reasonably
close to the  value (\ref{eq:Ncr}) obtained by solving exactly the GP
equation. The Gaussian {\it ansatz} has been used by several authors
in order to explore both static and dynamic properties of the trapped
gases. The stability of a gas with $a<0$ has been explored in details,
for instance, by  Stoof (1997), P\'erez-Garc\'\i a {\it et al.} 
(1997), Shi and Zheng (1997a), Parola, Salasnich and Reatto (1998). 
The variational function proposed by Fetter (1997), which interpolates 
smoothly  between the ideal gas and the Thomas-Fermi limit for 
positive $a$, also reduces to a Gaussian for $a<0$. 

The behavior of the gas close to collapse could be significantly 
affected by mechanisms not included in the Gross-Pitaevskii theory. 
Among them, inelastic two- and three-body collisions can cause a 
loss of atoms from the condensate through, for instance, spin 
exchange or recombination (Hijmans {\it et al.}, 1993; 
Edwards {\it et al.}, 1996b;  Moerdijk {\it et al.}, 1996; 
Fedichev {\it et al.}, 1996).  This is an important problem not 
only for attractive forces but also for repulsive forces when 
the density of the system becomes large. 

Recent discussions about the collapse, including quantum tunneling
phenomena, can be found, for instance, in Sackett, Stoof and Hulet (1998), 
Kagan, Muryshev and Shlyapnikov (1998), Ueda and Leggett (1998), Ueda and 
Huang (1998). 

\subsection{\bf Large $N$ limit for repulsive forces}
\label{sec:TF}

In the case of atoms with repulsive interaction ($a>0$), the limit
$Na/a_{\rm ho} \gg 1$ is particularly interesting, since this condition
is well satisfied by the parameters $N$, $a$ and $a_{\rm ho}$ used in
most of current experiments. Moreover, in this limit the
predictions of mean-field theory take a rather simple analytic
form (Edwards and Burnett 1995; Baym and Pethick 1996).

As regards the ground state, the effect of increasing the parameter
$Na/a_{\rm ho}$ is clearly seen in Fig.~\ref{fig:apos}: the
atoms are pushed outwards, the central density becomes rather flat
and the radius grows. As a consequence, the quantum
pressure term in the Gross-Pitaevskii equation (\ref{eq:GP}),
proportional to $\nabla^2 \sqrt{n({\bf r})}$, takes a significant
contribution only near the boundary and becomes less and less
important with respect to  the interaction energy. If one neglects
completely the quantum pressure in (\ref{eq:GP}), one gets the
density profile in the form
\begin{equation}
n({\bf r}) = \phi^2({\bf r}) = g^{-1} [ \mu - V_{\rm ext}({\bf r})]
\label{eq:rhoTF}
\end{equation}
in the region where $\mu > V_{\rm ext}({\bf r})$, and $n=0$ outside.
This is often referred to as Thomas-Fermi (TF) approximation.

The normalization condition on $n({\bf r})$ provides the relation
between chemical potential and number of particles:
\begin{equation}
\mu =   {\hbar \omega_{\rm ho} \over 2}
\left( { 15 N a \over a_{\rm ho} } \right)^{2/5} \; .
\label{eq:muTF}
\end{equation}
Note that the chemical potential depends on the trapping frequencies,
entering the potential $V_{\rm ext}$ given in (\ref{eq:uext}), only
through the geometric average $\omega_{\rm ho}$ [see Eq.~(\ref{omegaho})]. 
Moreover,  since $\mu = \partial E/ \partial N$, the energy 
per particle turns out to be
$E/N = (5/7) \mu$. This energy is the sum of the interaction
and oscillator energies, since the kinetic energy gives a vanishing
contribution for large $N$.  Finally, in the same limit, the release 
energy (\ref{release}) coincides with the interaction energy: 
$E_{\rm rel}/N =(2/7)\mu$.

The chemical potential, as well as the interaction and oscillator
energies obtained by solving numerically the GP equation (\ref{eq:GP})
become closer and closer to the Thomas-Fermi values when $N$
increases (see for instance, Dalfovo and Stringari, 1996). For 
sodium atoms in the MIT traps, where $N$ is larger than $10^6$,
the Thomas-Fermi approximation is practically indistinguishable
from the solution of the GP equation. The release energy per 
particle measured by Mewes {\it et al.} (1996a) is indeed well 
fitted with a $N^{2/5}$ law, as shown in Fig.~\ref{fig:releasemit}. 
The same agreement is expected to occur for rubidium atoms in the 
most recent JILA traps, having $N$ larger than $10^5$ (Matthews 
{\it et al.}, 1998). 

The density profile (\ref{eq:rhoTF}) has the form of an inverted 
parabola, which vanishes at the classical turning point ${\bf R}$
defined by the condition $\mu=V_{\rm ext}({\bf R})$. For a spherical
trap, this implies $\mu=m \omega_{\rm ho}^2 R^2/2$ and, using result
(\ref{eq:muTF}) for $\mu$, one finds the
following expression for the radius of the condensate
\begin{equation}
R = a_{\rm ho} \left( { 15 N a \over a_{\rm ho} } \right)^{1/5}
\label{eq:capitalr}
\end{equation}
which grows with $N$. For an axially symmetric trap, the widths in
the radial and axial directions are fixed by the conditions
$\mu=m\omega_\perp^2 R_\perp^2/2=m\omega_z^2 Z^2/2$. It is worth
mentioning that, in the case of the cigar-shaped trap used at MIT,
with a condensate of about  $10^7$ sodium atoms, the axial width
becomes  macroscopically large ($Z\sim 0.3$ mm), allowing
for direct {\it in situ} measurements.  

The value of the density (\ref{eq:rhoTF}) in the center of the trap
is $n_{\rm TF} (0) = \mu/g$. It is worth stressing that this density is much
lower than the one predicted for noninteracting particles. In the
latter case, using Eq.~(\ref{eq:gaussian}) one gets $n_{\rm ho}(0)=
N/(\pi^{3/2} a_{\rm ho}^3)$. The ratio between the central densities in
the two cases is then
\begin{equation}
{ n_{\rm TF} (0) \over n_{\rm ho} (0) } = {15^{2/5} \pi^{1/2} \over 8 }
\left( {N a \over a_{\rm ho} } \right)^{-3/5} \; ,
\label{eq:ration0}
\end{equation}
and decreases with $N$. For the available traps
with $^{23}$Na and $^{87}$Rb, where $Na/a_{\rm ho}$ ranges from about
$10$ to $10^4$, the atom-atom repulsion reduces the density by
one or two orders of magnitude, which is a quite remarkable effect
for such a dilute systems. An example was already shown in
Fig.~\ref{fig:hau}; in that case, the number of particles is 
about $80000$ and $Na/a_{\rm ho} \sim 300$.  

\begin{figure}[t]
\epsfysize=10cm
\hspace{3cm}
\epsffile{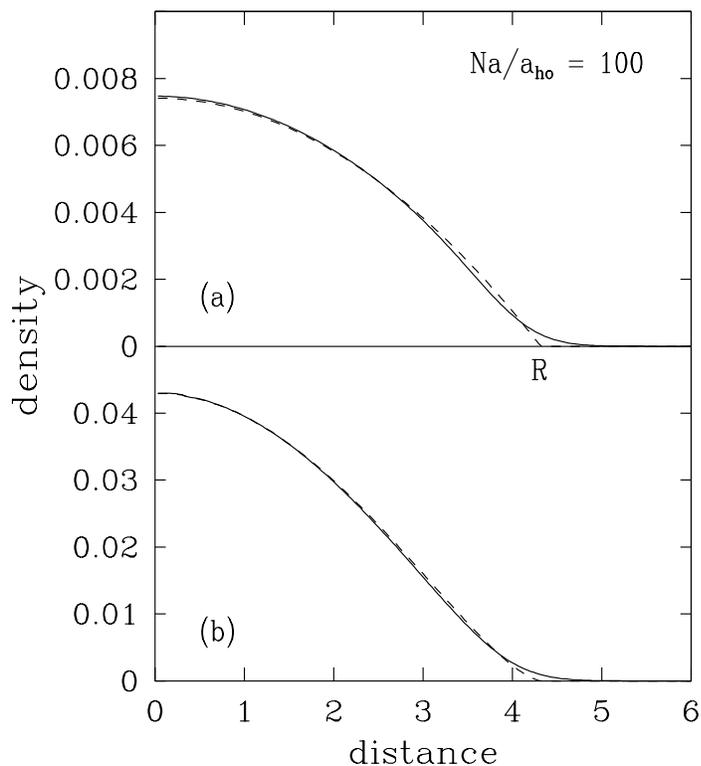}
\caption{Density profile for atoms interacting with repulsive 
forces in a spherical trap, with $Na/a_{\rm ho}=100$. Solid line: 
solution of the stationary GP equation (\protect\ref{eq:GP}). 
Dashed line: Thomas-Fermi approximation (\protect\ref{eq:rhoTF}).
In the upper part the atom density is plotted in arbitrary units,
while the distance from the center of the trap is in units of 
$a_{\rm ho}$. The classical turning point is at $R \simeq 4.31 a_{\rm ho}$. 
In the lower part the column density for the same system is 
reported.  } 
\label{fig:tf-column}
\end{figure}

In  Fig.~\ref{fig:tf-column}a we show the density profile for a
gas in a spherical trap with  $N a / a_{\rm ho} =  100$. The comparison 
with the exact solution of the GP equation (\ref{eq:GP}) shows that the
TF approximation is very accurate except in the surface region close 
to $R$. In part $b$ of the same figure, we plot the column density,
$n(z) = \int dx \ n (x,0,z)$, which is the measured quantity when 
the atomic cloud is imaged by light absorption or dispersive light
scattering.  Using the TF density (\ref{eq:rhoTF}) with 
$V_{\rm ext}=(1/2)m\omega_{\rm ho}^2 r^2$, 
one  finds  $n(z) = (4/3)  [2/(m\omega_{\rm ho}^2)]^{1/2} g^{-1}
[\mu-(1/2)m\omega_{\rm ho}^2 z^2]^{3/2}$. One notes that the accuracy of the 
Thomas-Fermi approximation is even better in the case of the column 
density, because the extra integration makes the cusp in the outer 
part of the condensate smoother. 

The only region where the Thomas-Fermi density (\ref{eq:rhoTF})
is inadequate is close to the classical turning point. This
region plays a crucial role for the calculation of the kinetic energy
of the condensate. The shape of the outer part of the condensate
is fixed by the balance of the zero point kinetic energy and the
external potential. In particular, this balance can be used to define
an effective surface thickness, $d$. For a spherical trap, for
instance, one can assume the two energies  to have the form
$\hbar^2/(2md^2)$ and $m\omega_{\rm ho}^2 Rd$, respectively. One then
gets (Baym and Pethick, 1996)
\begin{equation}
{ d \over R} 
= 2^{-1/3} \left( {a_{\rm ho} \over R} \right)^{4/3}  \; ;
\label{eq:surfacethickness}
\end{equation}
this ratio is small when TF approximation is valid, i.e., when
$R \gg a_{\rm ho}$. It is interesting to compare the surface thickness $d$ 
with the healing length (\ref{eq:healinglength}). In terms of the ratio 
$a_{\rm ho}/R$ one can write $\xi/R=(a_{\rm ho}/R)^2$, 
showing that the healing length decreases with $N$ more rapidly than 
the surface thickness $d$.  

A good approximation for the density in the region close to the classical 
turning point, can be obtained by a suitable
expansion of the GP equation (\ref{eq:GP}). In fact, when $|r-R|\ll R$, 
the trapping potential $V_{\rm ext}(r)$ can be replaced with a linear ramp,
$m \omega_{\rm ho}^2 R (r-R)$, and the GP equation takes a universal form
(Dalfovo, Pitaevskii and Stringari, 1996; Lundh, Pethick and 
Smith, 1997),  yielding the rounding of
the surface profile. 

Using the above procedure it is possible to calculate the kinetic energy  
which, in the case of a spherical trap, is found to follow the 
asymptotic law 
\begin{equation}
{ E_{\rm kin} \over N } \simeq { 5 \hbar^2 \over 2mR^2 } 
\ln \left( { R  \over C a_{\rm ho} } \right)
\label{eq:ekinlog}
\end{equation}
where $C\simeq 1.3$ is a numerical factor.
Analogous expansions can be derived for the harmonic potential energy,
$E_{\rm ho}$,  and interaction energy, $E_{\rm int}$, in the same large $N$
limit (Fetter and Feder, 1997). A straightforward derivation is obtained 
by using nontrivial relationships among the various energy components 
$E_{\rm kin}$, $E_{\rm ho}$ 
and $E_{\rm int}$ of Eq.~(\ref{eq:funcofdensity}).  A first relation is 
given by the virial theorem (\ref{eq:virial1}). A second one is obtained 
by using expression (\ref{eq:newmu}) for the chemical potential and the 
thermodynamic definition $\mu=\partial E/\partial N$. These two 
relationships, together with the asymptotic law (\ref{eq:ekinlog}) 
for the kinetic energy, allow one to obtain the expansions 
$E_{\rm ho}/N =(3/7)\mu_{\rm TF} +  \hbar^2/(mR^2) \ln [R/(C a_{\rm ho})]$
and $E_{\rm int}/N=(2/7)\mu_{\rm TF} - \hbar^2/(mR^2) \ln [R/(C a_{\rm ho})]$. 
From them one gets the useful results 
\begin{equation}
\mu =  \mu_{\rm TF} \left[ 1+ 3\frac{a_{\rm ho}^4}{R^4}
\ln \left(\frac{R}{Ca_{\rm ho}} \right) \right]
\label{eq:expansionmu}
\end{equation}
and
\begin{equation}
E =  {5 \over 7} N \mu_{\rm TF} \left[ 1+ 7\frac{a_{\rm ho}^4}{R^4}
\ln \left(\frac{R}{Ca_{\rm ho}}\right) \right]
\label{eq:expansionenergy}
\end{equation}
for the chemical potential and the total energy, respectively. In
these equations $\mu_{\rm TF}$ and $R$ are the Thomas-Fermi values
(\ref{eq:muTF}) and (\ref{eq:capitalr}) of the chemical potential and
the radius of the condensate. Equations 
(\ref{eq:ekinlog})-(\ref{eq:expansionenergy}), which apply to spherical 
traps, clearly show that the relevant small parameter in the 
large $N$ expansion is $a_{\rm ho}/R = (15 Na / a_{\rm ho})^{-1/5}$. 

The Thomas-Fermi approximation (\ref{eq:rhoTF}) for the ground
state density of trapped Bose gases is very useful not only for
determining the static properties of the system, but also for dynamics and
thermodynamics, as we will see in Secs.~\ref{sec:dynamics} and
\ref{sec:thermodynamics}. It is worth noticing that this approximation 
can be derived more directly using  local density theory
as we are going to discuss in the next section.

\subsection{Beyond mean-field theory}
\label{sec:corrections}

Before closing this discussion about the effect of interactions
on the ground state properties,  we wish to come back to the basic
question of the validity of the Gross-Pitaevskii theory. All
the results so far presented are expected to be valid if the system
is dilute, that is,  if ${n}|a|^3 \ll 1$. In order to estimate 
the accuracy of this approach we will now calculate the first
corrections to the mean-field approximation. Such corrections have
been recently investigated in several papers as, for instance, by
Timmermans, Tommasini and Huang (1997) and by Braaten and Nieto (1997).
Here we limit the discussion to the case of repulsive interactions
and large $N$, where analytic results can be found. In fact, in 
this limit the
solution of the stationary GP equation (\ref{eq:GP}) for the
ground state density can be  safely replaced with the Thomas-Fermi
expression (\ref{eq:rhoTF}) and the energy of the system is
given by $E/N=(5/7) \mu_{\rm TF}$, where $\mu_{\rm TF}$ is the TF chemical
potential (\ref{eq:muTF}).

Let us first discuss the behavior of the ground state density. For
large $N$ one can use the local density approximation for the chemical
potential:
\begin{equation}
\mu = \mu_{\rm local} [n({\bf r})] + V_{\rm ext}({\bf r}) \; .
\label{eq:local}
\end{equation}
The use of the local density approximation for $\mu$ is well justified 
in the thermodynamic limit $N\to \infty$, $\omega\to 0$  where the
profile of the density distribution is very smooth.  Equation
(\ref{eq:local}) fixes the density profile $n({\bf r})$ of the ground 
state once the thermodynamic relation $\mu_{\rm local}(n)$ for the uniform 
fluid is known, the parameter $\mu$ in the l.h.s. of Eq.~(\ref{eq:local})
being fixed by the normalization of the density. For example, in a very 
dilute Bose gas at $T=0$, one has $\mu_{\rm local}(n) =g n$ and immediately
finds the mean-field Thomas-Fermi result (\ref{eq:rhoTF}).  The first 
correction to the Bogoliubov equation of state is given by the law 
(Lee and Yang, 1957; Lee, Huang and Yang, 1957)
\begin{equation}
\mu_{\rm local}(n) = gn \left[ 1+\frac{32}{3\sqrt{\pi }}(na^3)^{1/2} \right]
\; , 
\label{eq:mu1}
\end{equation}
which includes nontrivial effects associated with the renormalization 
of the scattering length. Using expression (\ref{eq:mu1}) for 
$\mu_{\rm local}$, one can solve equation (\ref{eq:local}) by iteration.
The result is  
\begin{equation}
n({\bf r}) = g^{-1} \left[ \mu - V_{\rm ext}({\bf r}) \right] -
\frac{4m^{3/2}}{3\pi^2\hbar^3} 
\left[ \mu-V_{\rm ext}({\bf r})\right]^{3/2} \; , 
\label{eq:n1}
\end{equation}
with $\mu$ given by 
\begin{equation}
\mu = \mu_{\rm TF} \left( 1 + \sqrt{\pi a^3 n(0)} \right) \; . 
\label{eq:mubeyond}
\end{equation}
Then the energy can be also evaluated through the thermodynamic 
relation $\mu=\partial E/\partial N$, and one finds
\begin{equation}
E = \frac{5}{7}N\mu_{\rm TF} \left(1+\frac{7}{8}
\sqrt{\pi a^3n(0)}\right)
\label{eq:Ebeyond}
\end{equation}
where, in the second term, we have safely used the lowest order 
relation $\mu_{\rm TF}=gn(0)$. In an equivalent way, results 
(\ref{eq:n1})-(\ref{eq:Ebeyond}) can be derived  using a variational 
procedure by writing the energy functional of the system in the local 
density  approximation.  

Equations (\ref{eq:mubeyond})-(\ref{eq:Ebeyond}) show that, as expected, 
the corrections to the mean-field results are fixed by the gas parameter 
$a^3n$ evaluated at the center of the trap. This quantity can be 
directly expressed in terms of the relevant parameters of the system:
\begin{equation}
a^3n(0) = \frac{15^{2/5}}{8\pi}
 \left(N^{1/6}\frac{a}{a_{\rm ho}}\right)^{12/5} \; .
\label{eq:gasparameter}
\end{equation}
Inserting typical values for the available experiments, the corrections 
to the chemical potential and the energy turn out to be of the order 
of $1$\%. These corrections to the mean-field predictions should be 
compared with the ones due to finite size effects (quantum pressure) in 
the solution of the Gross-Pitaevskii equation 
[see Eqs.~(\ref{eq:expansionmu}) and (\ref{eq:expansionenergy})], 
which have a different dependence on the parameters $N$ and $a/a_{\rm ho}$. 
One finds that finite size effects become smaller than the corrections
given by Eqs.~(\ref{eq:mubeyond})-(\ref{eq:Ebeyond}) when $N$ is larger 
than about $10^6$.

Another important quantity to discuss is the quantum depletion of the
condensate. This gives the fraction of atoms which do not occupy the
condensate at zero temperature, because of correlation effects. The quantum
depletion is ignored in the derivation of the Gross-Pitaevskii equation.
It is consequently useful to have a reliable estimate of its value in order
to check the validity of the theory. Also in this case we can use local
density approximation (Timmermans, Tommasini and Huang, 1997) and write 
the density of atoms out of the condensate, $n_{\rm out} ({\bf r})$,
using Bogoliubov's theory for uniform gases at density $n=n({\bf r})$ 
(see for example Huang, 1987). One
gets $n_{\rm out} ({\bf r}) = (8/3) [ n({\bf r}) a^3/ \pi ]^{1/2}$. 
Integration of  $n_{\rm out}$ yields the result:
\begin{equation}
{ N_{\rm out} \over N } = { 5 \sqrt{\pi} \over 8 } \sqrt{a^3n(0)} \; .
\label{eq:depletion}
\end{equation}
for the quantum depletion of the condensate. Similarly to the correction 
to the mean-field energy (\ref{eq:Ebeyond}), this effect is very small
(less than 1\%) in the presently available experimental conditions.

The above results justify {\it a posteriori} the use of the Bogoliubov
prescription for the Bose field operators and the perturbative treatment
of the noncondensed part at zero temperature. We recall that this situation
is completely different from the one of superfluid $^4$He where quantum
depletion amounts to  about 90\% \ (Griffin, 1993; Sokol, 1995).

\section{Effects of interactions: dynamics}
\label{sec:dynamics}

\subsection{Excitations of the condensate and time dependent GP equation}
\label{sec:TDGP}

The study of elementary excitations is a task of primary importance
of quantum many-body theories. In the case of Bose fluids,
in particular, it plays a crucial role in the understanding
of the properties of superfluid liquid helium and was the subject
of pioneering work by Landau, Bogoliubov and Feynman
(for a recent discussion on the dynamic behavior of interacting 
Bose superfluids see, for instance, Griffin, 1993).

After the experimental realization of BEC in  trapped Bose gases,
there has been an intensive study of the excitations
in these systems. Measurements of the frequency of the lowest
modes have soon become available and the direct observation of
the propagation of wave packets has been also obtained. In the
meanwhile, on the theoretical side, a variety of papers has
been written to explore several interesting features exhibited by
the dynamic behavior of trapped Bose gases.

Let us start our discussion recalling that for dilute Bose gases
an appropriate description of the  excitations can be
obtained from the time dependent GP equation (\ref{eq:TDGP}) for
the order parameter. This equation  has been already used in
Sec.~\ref{sec:groundstate} for  evaluating the stationary
solution  $\phi({\bf r})$ characterizing the ground state. 
In the low temperature limit, where the
properties of the excitations do not depend on temperature, the
excited states can be found from the ``classical" frequencies
$\omega$ of the linearized GP equation. Namely, one can look for
solutions of the form
\begin{equation}
\Phi ({\bf r},t) = e^{-i\mu t/\hbar} \left[ \phi ({\bf r}) + 
u({\bf r}) e^{-i \omega t} + v^*({\bf r}) e^{i \omega t} \right]
\label{eq:linearized}
\end{equation}
corresponding to small oscillations of the order parameter around
the ground state value. By keeping  terms linear in the complex
functions $u$ and $v$, Eq.~(\ref{eq:TDGP}) becomes 
\begin{eqnarray}
 \hbar \omega u({\bf r}) &=& [ H_0 - \mu + 2 g \phi^2({\bf r})]
u ({\bf r}) + g  \phi^2({\bf r}) v ({\bf r})
\label{eq:coupled1}
\\
- \hbar \omega v({\bf r}) &=& [ H_0 - \mu + 2 g \phi^2({\bf r})] 
v ({\bf r}) + g  \phi^2({\bf r}) u ({\bf r}) \; .
\label{eq:coupled2}
\end{eqnarray}
where $H_0= - (\hbar^2/2m) \nabla^2 +  V_{\rm ext}({\bf r})$. These 
coupled equations allow one to calculate the eigenfrequencies $\omega$ 
and hence the energies $\varepsilon= \hbar \omega$ of the excitations.
This formalism was introduced by Pitaevskii (1961), in order to 
investigate the excitations of vortex lines in a uniform Bose gas.

This procedure is also equivalent to the diagonalization of the
Hamiltonian in Bogoliubov approximation, in which one expresses
the field operator $\hat\Psi'$ in terms of the quasiparticle
operators $\alpha_j$ and $\alpha_j^\dagger$ through (Fetter, 1972
and 1996)
\begin{equation}
\hat\Psi' ({\bf r})  = \sum_j [ u_j({\bf r}) \alpha_j(t)
+ v_j^*({\bf r}) \alpha_j^\dagger(t) ] \; .
\label{eq:tildepsilinearized}
\end{equation}
By imposing the Bose commutation rules to the operators $\alpha_j$
and $\alpha_j^\dagger$,  one finds that the quasiparticle amplitudes
$u$ and $v$ must obey the normalization condition
\begin{equation}
\int d{\bf r} \ [ u_i^*({\bf r}) u_j({\bf r}) -
v_i^*({\bf r}) v_j({\bf r}) ] \ = \ \delta_{ij} \; .
\label{eq:normalizationuandv}
\end{equation}

In a uniform gas, the amplitudes $u$ and $v$ are plane waves and
the resulting dispersion law takes the most famous Bogoliubov form
(Bogoliubov, 1947)
\begin{equation}
(\hbar \omega)^2 = \left( {\hbar ^2 q^2 \over 2m} \right) \left(
{ \hbar^2 q^2 \over 2 m} + 2 g n \right)
\label{eq:bogoliubovspectrum}
\end{equation}
where ${\bf q}$ is the wavevector of the excitation and
$n=|\phi |^2$ is the  density of the gas. For large momenta the spectrum
coincides with the free-particle energy $\hbar^2q^2/2m$.  At low momenta
Eq.~(\ref{eq:bogoliubovspectrum}) instead yields the phonon dispersion
$\omega=cq$, where
\begin{equation}
c = \sqrt{\frac{g n}{m}}
\label{eq:phonons}
\end{equation}
is the sound velocity. It is worth noticing that this velocity coincides 
with the hydrodynamic expression $c= [(1/m) \partial P/\partial n]^{1/2}$ 
for a gas with equation of state $P = (1/2) gn^2$ [see also the discussion 
after Eq.~(\ref{eq:eulerHD})].

In the case of harmonic trapping, an important role is played by the ratio
$Na/a_{\rm ho}$,  and one expects different behaviors in the two opposite
limits $Na/a_{\rm ho} \ll 1$ and $Na/a_{\rm ho} \gg 1$. In the first case, 
one recovers the excitation spectrum   $\omega= n_x \omega_x + n_y \omega_y
+ n_z \omega_z$ of the noninteracting harmonic potential [see
Eq.~(\ref{eq:spectrumho})]. In the second case, one obtains a different
dispersion law for the excitations of the system which are the analog
of phonons [see Eq.~(\ref{eq:hdsdispersion}) below].

The coupled  equations (\ref{eq:coupled1})-(\ref{eq:coupled2}) were
first used to calculate numerically the excitations of trapped gases 
by Burnett and co-workers (Ruprecht {\it et al.}, 1996; 
Edwards {\it et al}, 1996a and 1996c). Similar calculations have been 
also performed by other authors, for both spherical  and anisotropic
configurations (Singh and Rokhsar, 1996; Esry, 1997; Hutchinson, 
Zaremba and Griffin, 1997; Hutchinson and Zaremba, 1997; You, Hoston 
and Lewenstein,1997; Dalfovo {\it et al.}, 1997a).  

\begin{figure}[t]
\epsfysize=8cm
\hspace{3cm}
\epsffile{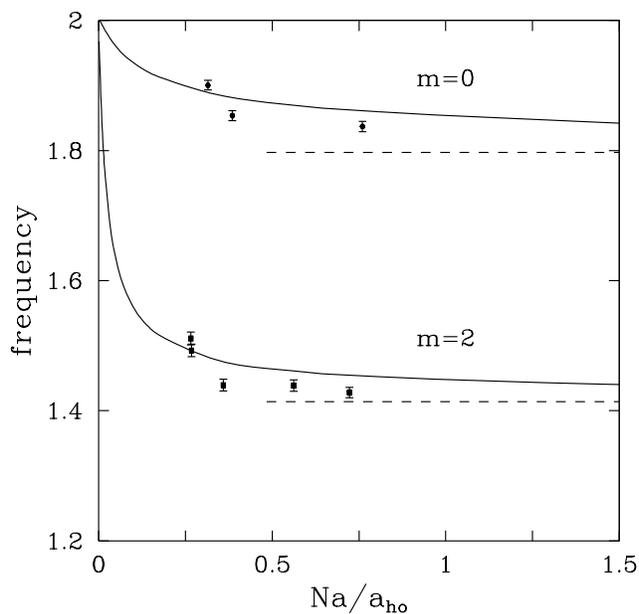}
\caption{Frequency of the lowest collective modes of even parity,
$m=0$ and $m=2$, for rubidium atoms in the JILA trap ($\lambda=
\protect\sqrt{8}$). The abscissa is the dimensionless parameter 
$Na/a_{\perp}$, with $a_\perp=[ \hbar / (m \omega_\perp)]^{1/2}$, 
while the frequency is given in units of $\omega_\perp$.  Points are 
taken from  the experimental data of Jin {\it et al.} (1996). Solid 
lines  are the predictions of the mean-field equations 
(\protect\ref{eq:coupled1})-(\protect\ref{eq:coupled2}) [see, for
instance, Edwards {\it et al.} (1996c); Esry, (1997); You, Hoston
and Lewenstein (1997)]. Dashed lines are the asymptotic results 
for $ Na/a_{\perp} \to \infty $ (Stringari, 1996b), as discussed in 
Sec.~\protect\ref{sec:hydrodynamics}.  } 
\label{fig:m0m2}
\end{figure}

For spherical traps, the solutions of 
Eqs.~(\ref{eq:coupled1})-(\ref{eq:coupled2}) are 
characterized by the quantum  numbers $n_r$, $\ell$ and $m$, where $n_r$ 
is the number of radial nodes, $\ell$ is the angular momentum of the 
excitation and $m$  its $z$ component. For axially symmetric 
traps the third  component $m$ of angular momentum is still a good quantum
number. In  Fig.~\ref{fig:m0m2} we report  the lowest solutions
of even parity with $m=0$ and $m=2$, obtained  for a gas of rubidium
atoms confined in an axially symmetric trap ($\omega_x=\omega_y=
\omega_{\perp}$).  The asymmetry parameter of the trap ($\lambda
= \omega_z / \omega_{\perp} = \sqrt{8}$) corresponds to the experimental
conditions of Jin {\it et al.} (1996) and values of $N$ up to $10^4$
are  considered. Actually the results are reported as a function of
the dimensionless parameter $N a/ a_\perp$ where $a_\perp=
\sqrt{\hbar/(m\omega_\perp)}$.  The theoretical predictions are compared
with   the experimental results. In the experiments these oscillations
are observed by shaking the condensate through the modulation of the
trapping magnetic  fields.  The general  agreement between theory and
experiments is good and reveals the  important role played by two-body
interactions.  In fact, in the absence  of interactions, the
eigenfrequencies would be  the ones predicted by  the ideal harmonic
oscillator, which gives  $\omega = 2 \omega_\perp$ for both modes.

Among the various excitations exhibited by these trapped gases,
special attention should be devoted to the dipole mode. This oscillation
corresponds to the motion of the center of mass of the system which, due
to the harmonic confinement, oscillates with the frequency of the harmonic
trap (this frequency can of course be different in the three directions).
Two-body interactions cannot affect this mode because, in the presence of
harmonic trapping, the motion of the center of mass is exactly decoupled 
from the internal degrees of freedom of the system.  This is best 
understood by considering Eq.~(\ref{eq:TDGP}) and looking for solutions 
of the form 
\begin{equation}
e^{iz\beta(t)} \ \Phi(x,y,z+\alpha(t))\; ,
\label{eq:dipole}
\end{equation}
where, for simplicity, we have considered only oscillations along the
$z$-axis. By a proper change of variables,  $z \to z+ \alpha$, one finds
that (\ref{eq:dipole}) corresponds to an exact solution of
the time dependent equation (\ref{eq:TDGP}) oscillating with frequency
$\omega_z$. This property holds not only in the context of
the Gross-Pitaevskii equation, but is valid for any interacting system
confined in a harmonic potential at zero as well as finite temperature,
and is independent of statistics (Fermi or Bose). For example, such a 
decoupling is a well known property of shell model theory in nuclear 
physics (Elliott and Skyrme, 1955; Brink, 1957). It also exhibits 
interesting analogies with Kohn's theorem for electrons in a static 
magnetic field, stating that the cyclotron frequency is not affected by 
interactions (Kohn, 1961) [see also Dobson (1994) and references 
therein for discussions about the generalization of Kohn's theorem to 
the case of electrons confined in harmonic traps].   

The fact that the dipole frequency is not affected by two-body interactions 
offers a direct test on the numerical accuracy of the various methods 
used to solve the equations of motion.  On the other hand
the experimental determination of the dipole frequency turns out to
be a  very useful procedure to check the harmonicity of the
trap and  to determine accurately the value of the trapping frequencies.
The properties of the dipole excitation in the framework of Bogoliubov 
theory have been discussed in detail by Fetter and Rokhsar (1998) [see
also Kimura and Ueda (1998)]. 

Of course the coupled equations (\ref{eq:coupled1})-(\ref{eq:coupled2})
provide a full series of solutions, with different values of the
corresponding quantum numbers. So far experiments have provided
direct information only on the low energy modes which can  be
directly excited by suitable modulation of the harmonic trap. These
excitations will be further discussed in the next section using the
formalism of collisionless hydrodynamic equations. States at higher 
energy and multipolarity are also important, since they characterize 
the thermodynamic behavior of the system, as we will see later on.

Finally, we note that, starting from the solutions of 
Eqs.~(\ref{eq:coupled1})-(\ref{eq:coupled2}), one can also evaluate 
the density of particles out of the condensate at zero temperature 
(quantum  depletion) by summing the 
square modulus  of the ``hole" amplitude $v$ over all the 
excited states: $n_{\rm out}({\bf r}) = \sum_j |v_j({\bf r})|^2$ (Fetter, 
1972). The results (Hutchinson, Zaremba and Griffin, 1997; Dalfovo 
{\it et al.}, 1997a) are in agreement with the local density estimate 
(\ref{eq:depletion}).

\subsection{ Large $Na/a_{\rm ho}$ limit and collisionless hydrodynamics}
\label{sec:hydrodynamics}

When the number of atoms in the trap increases, the eigenfrequencies
of the coupled equations (\ref{eq:coupled1})-(\ref{eq:coupled2}) approach
an asymptotic value. The new regime is achieved when the condition
$Na/a_{\rm ho} \gg 1$ is satisfied.  In this limit the excitations are
properly described by the hydrodynamic theory of superfluids in the
collisionless regime at zero temperature. 
In a dilute gas this theory can be explicitly  
derived starting from the time dependent GP equation (\ref{eq:TDGP}). 
To this purpose, it is convenient to write the complex order parameter 
$\Phi$ in terms of a modulus and a phase,
as follows:
\begin{equation}
\Phi ({\bf r},t) = \sqrt{n({\bf r},t) }\ e^{i S ({\bf r},t)} \; .
\label{eq:orderparameter}
\end{equation}
The phase fixes the velocity field
\begin{equation}
n({\bf r},t) {\bf v} ({\bf r},t) =  { \hbar \over 2 i m  }  (\Phi^*
\mbox{\boldmath$\nabla$} \Phi -\Phi \mbox{\boldmath$\nabla$} \Phi^*)
\; ,
\label{eq:velocity}
\end{equation}
so that 
\begin{equation}
{\bf v} ({\bf r},t) = { \hbar \over  m }  
\mbox{\boldmath$\nabla$}  S ({\bf r},t) \; . 
\label{eq:gradientS}
\end{equation}
The GP equation (\ref{eq:TDGP}) can hence be  rewritten in the form of
two coupled equations for the density  and the velocity field:
\begin{equation}
{ \partial \over \partial t} n + \mbox{\boldmath$\nabla$}
\cdot ({\bf v}n) = 0
\label{eq:continuity}
\end{equation}
and
\begin{equation}
m  { \partial \over \partial t} {\bf v} +
\mbox{\boldmath$\nabla$} \left( V_{\rm ext} + g n - 
{\hbar^2 \over 2m \sqrt{n} }
\nabla^2 \sqrt{n}
+ { mv^2 \over 2 } \right) = 0 \; .
\label{eq:euler}
\end{equation}
Equation (\ref{eq:continuity}) is the equation of continuity, while
(\ref{eq:euler}) establishes the irrotational nature of the superfluid
motion. It is worth noticing that, at this stage, 
Eqs.~(\ref{eq:continuity}) and (\ref{eq:euler}) do not
involve any approximation with respect to the GP equation (\ref{eq:TDGP})
and can be used in the linear as well as nonlinear regimes.

If the repulsive interaction among atoms is strong enough, then the
density profiles become smooth and one can safely neglect the kinetic pressure 
term, proportional to $\hbar^2$,  in the equation for the velocity field, 
which then takes the form
\begin{equation}
m  { \partial \over \partial t} {\bf v} +
\mbox{\boldmath$\nabla$} \left( V_{\rm ext} + g n
+ { mv^2 \over 2 } \right) = 0 \; .
\label{eq:eulerHD}
\end{equation}
This result corresponds to  the equation of potential flow for a 
fluid whose pressure and density are related by the equation of state 
$P=(1/2)g n^2$. Equations (\ref{eq:continuity}) and (\ref{eq:eulerHD}) 
have the typical structure of the dynamic equations of superfluids at 
zero temperature (see, for example, Pines and Nozieres, 1966, Vol.II) 
and can be viewed as a particular case of the more general Landau's 
theory of superfluidity.  According to this theory, which is valid if 
the relevant physical quantities change slowly on distances larger than 
the healing length, a complete description of the dynamics of the
fluid is obtained by coupling the equation for the superfluid velocity 
field with a Boltzmann-type equation for the distribution function of 
elementary excitations [see, Lifshitz and Pitaevskii, 1981, \S 77].  
At high temperature, when the mean free path of elementary excitations 
is short, one gets a system of two-fluid hydrodynamics equations. 
Conversely, at low temperature, where the role of thermally excited
states is negligible, the same equations reduce to the hydrodynamic-type
equations (\ref{eq:continuity}) and (\ref{eq:eulerHD}) involving only
the superfluid velocity. In this sense, equations  can be referred 
to as the hydrodynamic equations of superfluids. They should not be 
confused with the hydrodynamic equations valid in the collisional 
regime at high temperature.

The stationary solution of Eq.~(\ref{eq:eulerHD}) coincides with the
Thomas-Fermi density (\ref{eq:rhoTF}) while the  time 
dependent equations (\ref{eq:continuity}) and (\ref{eq:eulerHD}),
after linearization, take the following simplified form:
\begin{equation}
{ \partial^2 \over \partial t^2 } \delta n = \mbox{\boldmath$\nabla$}
\cdot \left[ c^2({\bf r}) \mbox{\boldmath$\nabla$} \delta n
\right]
\label{eq:localsound}
\end{equation}
where $m c^2({\bf r}) =\partial P/\partial n =  \mu- V_{\rm ext}({\bf r})$, 
the quantity $c$ having the meaning of a local sound velocity.

The validity of the equation (\ref{eq:localsound}) is based
on the assumption that the  spatial variations of the density are smooth
not only in the ground  state, but also during the oscillation. In a uniform
system ($V_{\rm ext}\equiv 0$) this is equivalent to 
imposing that the collective
frequencies be much smaller than the chemical potential.
In this case, the solutions of (\ref{eq:localsound}) are sound  waves
propagating with the Bogoliubov velocity (\ref{eq:phonons}).
Sound waves can propagate also in  nonuniform media, provided we
look for solutions varying   rapidly with respect to the size of the
system, so that  one  can assume a locally uniform sound velocity (Landau
and Lifshitz, 1987, \S 67). This is  possible if both
the conditions  $qL \gg 1$ and $\hbar q \ll mc$ are satisfied, where $L$ is
the size of  the condensate and $q$ is the wavevector of the sound wave.
Furthermore, if the system is highly deformed and cigar-shaped,
one can simultaneously
satisfy the conditions $qZ \gg 1$ and $qR_{\perp} \ll 1$, characterizing
one-dimensional waves propagating in the $z$ direction. Here $Z$ and
$R_{\perp}$ are the radii of the condensate in the axial and radial
directions, respectively. In this case, one can show (Zaremba, 1998) 
that the sound
velocity in the central region of the trap is given by $\sqrt{\mu/2m}$,
instead of the usual Bogoliubov value   $\sqrt{\mu/m}$, where $\mu =
g n(0)$ and $n(0)$ is the value of the central density. The occurrence
of the extra factor $2$ follows from the fact that, in the
``one-dimensional" geometry, the sound velocity
is fixed by the density averaged over  the radial direction, which is
of course smaller than its central value.

\begin{figure}[t]
\epsfysize=8cm
\hspace{3cm}
\epsffile{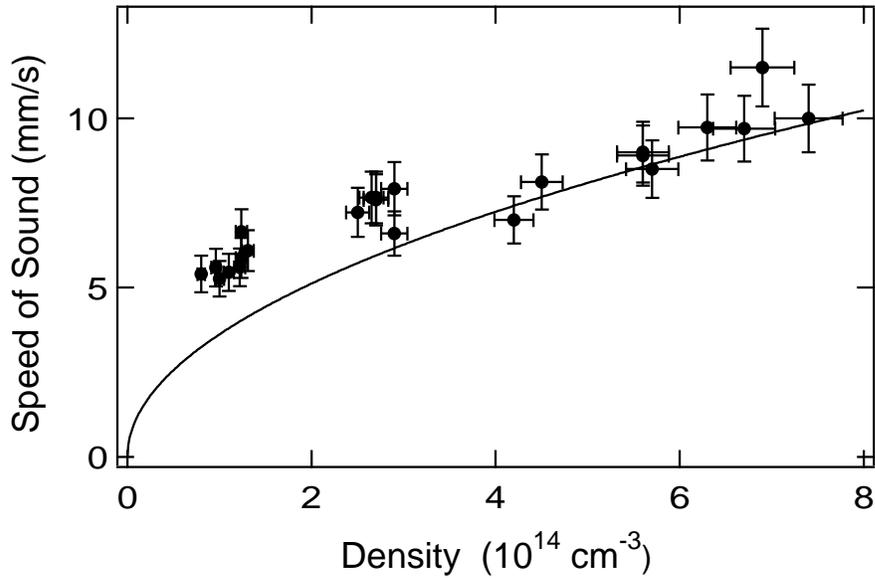}
\caption{Speed of sound, $c$, versus condensate peak density, $n(0)$, 
for waves propagating along the axial direction in the cigar-shaped
condensate at MIT. The experimental points are compared with the 
theoretical prediction $c=[g n(0)/ 2m]^{1/2}$ (solid line). From 
Andrews, Kurn {\it et al.} (1997).  } 
\label{fig:sound}
\end{figure}

In the experiments of Andrews, Kurn {\it et al.} (1997), one-dimensional 
sound waves are generated by focusing a laser pulse in  the center of the 
trap. A wave packet forms in this way, propagating outwards. It is then
imaged at different times so that the value of the sound velocity 
can be directly  measured.  In Fig.~\ref{fig:sound} we show the observed 
values of $c$ at different densities.  The agreement with the theoretical
predictions is reasonably good especially at high density. Possible 
sources of inaccuracy at low density are discussed by the same authors.  
The theoretical analysis of the
propagation of wave packets and sound waves in the elongated geometry has
been the object of several recent works (Zaremba, 1998; Kavoulakis and 
Pethick, 1998; Stringari, 1998).

In these nonuniform condensates, as already said,  oscillations 
having wavelength much smaller than the size of the system or, 
equivalently, frequency much larger than the trapping frequency 
$\omega_{\rm ho}$, propagate as usual sound waves. Conversely, 
solutions of (\ref{eq:localsound}) at lower frequency, of the 
order of $\omega_{\rm ho}$, involve a motion of the whole system 
(Baym and Pethick, 1996). They coincide with the low energy 
solutions of Eqs.~(\ref{eq:coupled1})-(\ref{eq:coupled2}) discussed in 
the previous section.  For a spherical trap these solutions  
are defined   in the interval $0 \le r \le R$ and have the form
$\delta n({\bf r}) = P^{(2n_r)}_{\ell}(r/R)\  r^{\ell} \
Y_{\ell m}(\theta,\phi)$ where $P^{(2n)}_{\ell}$ are  polynomials of
degree $2n$, containing only even powers. The dispersion  law of the
discretized normal modes is given by the formula (Stringari, 1996b)
\begin{equation}
\omega (n_r,\ell) = \omega_{\rm ho} (2n_r^2 + 
2n_r\ell + 3n_r +\ell)^{1/2} \; .
\label{eq:hdsdispersion}
\end{equation}
This result can be compared with the prediction for noninteracting
particles in harmonic potential:
\begin{equation}
\omega (n_r,\ell) = \omega_{\rm ho} (2n_r + l)
\label{eq:frequencyho}
\end{equation}
with $2n_r+l = n_x+n_y+n_z$ [see Eq.~(\ref{eq:spectrumho})]. Of
particular interest is the case of the  so called {\it surface}
excitations ($n_r=0$) for which  Eq.~(\ref{eq:hdsdispersion}) predicts
the dispersion law  $\omega = \sqrt{\ell}\ \omega_{\rm ho}$.
The frequency of these modes is  systematically  smaller than the harmonic
oscillator result $\ell\omega_{\rm ho}$. Notice that in the dipole case
($n_r=0, \ell=1$) the prediction (\ref{eq:hdsdispersion}) coincides with 
the oscillator frequency, in agreement with the general considerations 
discussed in the previous section.

As concerns compressional modes ($n_r \ne 0$), the lowest solution of
(\ref{eq:localsound}) is the monopole oscillation, also called the {\it
breathing  mode}, characterized  by the quantum numbers $n_r=1$ and
$\ell=0$. The  formula  (\ref{eq:hdsdispersion}) gives the
result $\sqrt{5}\ \omega_{\rm ho}$, higher  than the corresponding
prediction of  the noninteracting model, which gives $2\omega_{\rm ho}$.

For a fixed value of $N$ the accuracy of prediction (\ref{eq:hdsdispersion})
is expected to become lower and lower as $n_r$ and $\ell$ increase. In fact,
for large $n_r$ and $\ell$ the oscillations of the density have shorter
wavelength and neglecting the kinetic energy pressure in (\ref{eq:euler})
is no longer justified.  In analogy with the  case of uniform Bose gases,
the condition for the applicability of the hydrodynamic theory of superfluids
is expected to be $\hbar \omega < \mu$. However, as discussed in 
Sec.~\ref{sec:collectivevssingleparticle}, more severe restrictions are
imposed when one considers surface excitations.

The result (\ref{eq:hdsdispersion}) reveals that, in the Thomas-Fermi 
limit $Na/a_{\rm ho}\gg 1$, the dispersion relation of the normal modes
of the condensate has changed significantly from the noninteracting 
behavior, as a consequence of two-body interactions. However it might 
appear surprising that in this limit the dispersion does not depend any 
more on the value of the interaction parameter $a$. This  differs from the 
uniform case where the dispersion law, in the corresponding phonon regime,
is given by $\omega =cq$ and depends explicitly on the interaction
through the  velocity of sound. The behavior exibited in  the harmonic
trap is well understood if  one notes that the values of $q$ are fixed by the
boundary and vary as $1/L$  where $L$ is the size of the system. While in
the box this size is fixed, in the case of harmonic confinement it increases
with $N$ due to the repulsive effect of two-body interactions: $L\sim
(Na/a_{\rm ho})^{2/5}(m\omega_{\rm ho})^{-1/2}$. On the other hand the 
value of the sound velocity, calculated at the center of the trap, is 
given by  $c= (Na/a_{\rm ho})^{2/5}(\omega_{\rm ho}/m)^{1/2}$ and
also increases with $N$. One finally finds that  in the product $cq$ both
the interaction parameter and the number of  atoms in the trap cancel out, 
so that the collective frequency is  proportional to  the oscillator 
frequency $\omega_{\rm ho}$.

The results  for the spherical trap can be generalized to the case of
anisotropic configurations.   Let us consider the case of a harmonic
oscillator trap with axial symmetry along the $z$ axis. In this case the
differential equation (\ref{eq:localsound}) takes the form
\begin{equation}
m { \partial^2 \over \partial t^2 } \delta n =
\mbox{\boldmath$\nabla$} \cdot \left\{
\left[ \mu - {m\over2} \left( \omega_\perp^2 r_\perp^2 
+  \omega_z^2 z^2  \right) \right] \mbox{\boldmath$\nabla$} 
\delta n \right\}
\label{eq:hddeformed}
\end{equation}
where we have used $mc^2({\bf r})=\mu-V_{\rm ext}({\bf r})$. [We notice,
in passing, that the corresponding Eq.~(21) in Stringari (1996b) was
misprinted, since it contains the chemical potential counted twice.] 

Because of the axial symmetry of the trap the third component $m$ of
the angular momentum is a good quantum number. However, in contrast to
the spherical case, the
dispersion law   depends on $m$. Explicit results are available
in some particular cases. For example, quadrupole solutions of the form
$\delta n = r^2 Y_{2 m}(\theta,\phi)$ satisfy Eq.~(\ref{eq:hddeformed})
for $m=\pm 2$ and $m=\pm 1$. The  resulting dispersion laws are:
\begin{equation}
\omega^2(\ell=2, m=\pm 2) = 2\omega_{\perp}^2
\label{eq:pml}
\end{equation}
and
\begin{equation}
\omega^2(\ell=2,m=\pm 1) = \omega_{\perp}^2 +\omega_z^2 \, .
\label{eq:pml-1}
\end{equation}
Conversely the $\ell=2, m=0$ mode is  coupled to the monopole $\ell=0$
excitation and  the  dispersion law of the two decoupled modes is given by
(Stringari, 1996b)
\begin{equation}
\omega^2(m=0) = 2\omega_{\perp}^2  + \frac{3}{2}\omega_z^2
\mp\frac{1}{2}
\sqrt{9\omega_z^4-16\omega_z^2\omega_{\perp}^2+16\omega_{\perp}^4} \; .
\label{eq:pml-2}
\end{equation}
 When $\omega_z=\omega_{\perp}$
one recovers the solutions for the quadrupole and monopole excitations
in the spherical trap. The occurrence of analytic solutions for the 
excitation spectrum, like Eqs.~(\ref{eq:hdsdispersion}) and 
(\ref{eq:hddeformed})-(\ref{eq:pml-2}), is the result of nontrivial 
underlying symmetries of the Hamiltonian that have been exploited
by Fliesser {\it et al.} (1997). The result (\ref{eq:pml-2}) can be 
generalized to a triaxially deformed trap of the form (\ref{eq:uext}).
In this case, the collective frequencies are given by the solution of
the equation
\begin{equation}
\omega^6 - 3 \omega^4 (\omega_x^2 + \omega_y^2 +\omega_z^2) 
+ 8 \omega^2 (\omega_x^2 \omega_y^2 +\omega_y^2 \omega_z^2 
+ \omega_z^2 \omega_x^2 ) - 20 \omega_z^2 \omega_y^2 \omega_z^2
= 0 \; . 
\label{eq:triaxial}
\end{equation} 

\begin{figure}[t]
\epsfysize=8cm
\hspace{3cm}
\epsffile{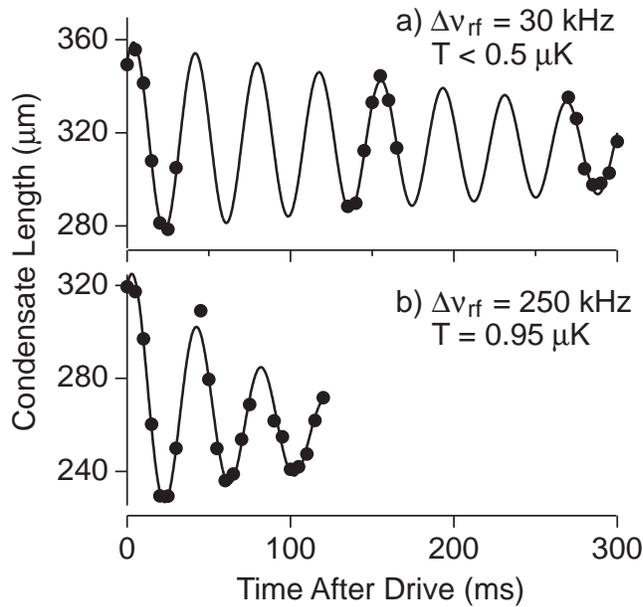}
\caption{Oscillations of the axial width of the condensate in the 
cigar-shaped trap at MIT. The excited collective motion is the 
low-lying $m=0$ mode. Oscillations are shown at low (a) and 
high (b) temperature. Points show the axial width determined from 
fits to phase-contrast images, similar to the ones in 
Fig.~\protect\ref{fig:insitu}. 
Lines are fits to a damped sinusoidal oscillation. From 
Stamper-Kurn {\it et al.}, (1998c).  } 
\label{fig:oscillationsMIT}
\end{figure}

 From Fig.~\ref{fig:m0m2} one can see that the  experiments at JILA do 
not fully fall in the asymptotic $Na/a_{\rm ho} \gg 1$ regime, where
the frequencies are given by Eqs.~(\ref{eq:pml}-\ref{eq:pml-2}). Conversely
the experimental results obtained  on sodium vapors at MIT
(Stamper-Kurn {\it et al.}, 1998c) represent a very clear example of
excitations belonging to the Thomas-Fermi regime. 
In this experiment the magnetic trap is highly asymmetric,  with 
$\lambda = \omega_z/ \omega_\perp = 17/230$ (cigar-shaped geometry). 
Furthermore the number of atoms is very high, so that the
condition $Na/a_{\rm ho} \gg 1$ is well satisfied  and the energies of the 
collective oscillations along the axial direction  are much smaller than
the chemical potential, $\mu \approx 200 \omega_z$.  This explains 
the excellent  agreement between the observed  frequency
for  the lowest  axial $m=0$ mode of even parity ($\omega/ \omega_z =
1.569(4)$) and the theoretical prediction ($\omega/\omega_z = \sqrt{5/2}
= 1.581$) given by Eq.~(\ref{eq:pml-2}) with $\omega_z \ll \omega_{\perp}$.  
In Fig.~\ref{fig:oscillationsMIT} we show  the  oscillations observed
in the MIT experiment (see also Fig.~\ref{fig:insitu}). These measurements
correspond  to nondestructive {\it in situ} images of the oscillating 
condensate, while the ones at JILA (Jin {\it et al.}, 1996 and 1997), as  
well as the first experiments  carried out at MIT (Mewes {\it et al.}, 
1996b) were taken after switching-off the trap and letting the gas expand.

For highly deformed traps it is possible to obtain simple analytic
results also for the excitations  with higher quantum numbers. For
example, in the case of cigar-shaped traps ($\omega_z \ll
\omega_{\perp}$) one finds the dispersion law (Fliesser {\it et al.}, 1997; 
Stringari, 1998)
\begin{equation}
\omega^2(k) = {1\over 4}k(k+3)\omega_z^2 \, .
\label{eq:highlydeformed}
\end{equation}
where $k$ is the relevant quantum number characterizing the spatial
shape of the density oscillation $\delta n(z) = (z^k + \alpha
z^{k-1} + \dots)$. Equation (\ref{eq:highlydeformed}) is valid if 
$\omega(k) \ll \omega_{\perp}$.  It includes, as special cases,
the dipole ($k=1, \omega= \omega_z)$ and ``quadrupole" ($k=2, \omega
=\sqrt{5/2}\ \omega_z$) modes already discussed. It also permits one to
understand  the transition between the discretized (small $k$) and
``continuum" ($k\gg 1$) regimes, through the identification
$k\equiv qz$ where $q$ is the wave vector of  ``one-dimensional" phonons
propagating with sound velocity $\sqrt{\mu/2m}$. As already discussed,
these phonons can be considered one-dimensional only if the
conditions $qZ \gg 1$ and  $qR_{\perp}\ll 1$ are satisfied. The
first condition implies large values of $k$, the second one is
equivalent to imposing $\omega \ll \omega_{\perp}$.

Analogously for disk-shaped traps ($\omega_{\perp}\ll\omega_z$) the 
dispersion law of the lowest modes takes the analytic form (Stringari, 1998)
\begin{equation}
\omega^2(n_r,m) = \left( \frac{4}{3}n_r^2+\frac{4}{3}n_rm+2n_r+m \right)
\omega_{\perp}^2
\, ,
\label{eq:omegadisk}
\end{equation}
where $n_r=0,1,..$ is the number of radial nodes and $m$ is the $z$-component 
of the angular momentum.

\subsection{Sum rules and collective excitations}
\label{sec:collectivewithanegative}

In the previous section we have discussed the excitations of the 
condensate when atoms interact with repulsive forces ($a>0$). 
In the opposite case of attractive interactions ($a<0$), one expects 
a different behavior. For example, interesting effects can originate
from the fact that the system becomes more and more compressible
when approaching the critical number, $N_{\rm cr}$, for collapse. 
In terms of the excitation spectrum, this means a lowering of the
frequency of the monopole oscillation. For repulsive forces, we
have previously discussed the Thomas-Fermi $Na/a_{\rm ho}\gg 1$ 
limit, in which the time dependent GP equation takes the form 
of the zero temperature hydrodynamic theory of superfluids. When the 
interaction is attractive, the large $N$ limit is never reached, 
since the collapse occurs at $Na/a_{\rm ho}$ of the order of $1$,
and one has to solve numerically the GP equation (see, for example, 
Dodd {\it et al.}, 1996) or use different theoretical schemes,
as shown in the following.

A useful physical insight on the behavior of collective oscillations
for both positive and negative $a$ can be obtained using the 
formalism of linear response and sum rules [see, for instance,
Bohigas, Lane and Martorell (1979), Lipparini and Stringari (1989)]. 
This approach allows one to evaluate the energy weighted moments, 
$m_p= \int_0^{\infty} S_F(E)E^p dE$, of the strength distribution 
function (dynamic form factor) associated with a given operator $F$:
\begin{equation}
S_F(E) = \sum_j | \langle j | F |0 \rangle |^2
\delta(E-E_{j0})
\label{eq:SF}
\end{equation}
where the quantity $E_{j0}= (E_j-E_0)$ is the excitation energy of the
eigenstate $|j\rangle$ of the Hamiltonian. Consequently, the method 
provides information on the dynamic behavior of the system. Quantities
like $m_{p+1}/m_p$ or $(m_{p+2}/m_p)^{1/2}$ correspond to rigorous 
upper bounds for the energy of the lowest state excited by the 
operator $F$. They are close to the exact energy when this state
is highly collective, that is, when the strength distribution is
almost exhausted by a single mode. This is often true in the case 
of trapped gases, as we will see below.  

A major advantage of sum rules is that they can be often evaluated in 
a direct way, avoiding the full solution of the Schr\"odinger equation 
for the eigenstates of the Hamiltonian. For example, using the 
completeness of the eigenstates $|j \rangle$, the  energy weighted 
moment, $m_1$, can be easily transformed into the calculation of 
commutators involving the operator $F$ and the Hamiltonian:
\begin{equation}
m_1 = {1\over 2}  \langle 0 | [F^{\dagger},[H,F] ] | 0 \rangle  \; .
\label{eq:m_1}
\end{equation}
Furthermore, if the operator $F$ depends only on spatial co-ordinates
then only the kinetic energy  gives a contribution to $m_1$, whose 
calculation becomes straightforward. In a similar way, one can write 
the cubic energy weighted moment, $m_3$, in the form $m_3={1\over 2} 
\langle [[F^{\dagger},H],[H,[H,F]]] \rangle$. Unlike $m_1$ and $m_3$,
the inverse energy weighted moment, $m_{-1}$, cannot be expressed in 
terms of commutators; it can be however written in the useful form
\begin{equation}
m_{-1}= {1\over 2}\ \chi \; ,
\label{eq:mminus1}
\end{equation}
where $\chi$ is the linear static response of the system. 

Let us first consider the case of compressional modes. The natural 
monopole operator is given by the choice $F=\sum_i^Nr^2_i$ and, from 
(\ref{eq:m_1}), one gets the result $m_1= 2N \hbar^2 \langle r^2 
\rangle/m$ for the energy weighted sum rule.  Furthermore, in the 
monopole case one can easily evaluate also the inverse energy weighted
sum rule through Eq.~(\ref{eq:mminus1}). In fact, the static response
$\chi_M$ (monopole compressibility) is fixed by the linear change 
$\delta\langle r^2\rangle = \epsilon \chi_M$ of the mean square radius 
induced by the external field $-\epsilon r^2$. Adding this field to 
the Hamiltonian is equivalent to renormalizing the trapping harmonic 
potential and hence, for isotropic confinement, one can write
\begin{equation}
\chi_M = -{2N\over m} {\partial \langle r^2 \rangle \over
\partial \omega_{\rm ho}^2} \; .
\label{eq:chim}
\end{equation}
where $\omega_{\rm ho}$ is the frequency of the harmonic oscillator.
Using the properties of  the Gross-Pitaevskii equation
(\ref{eq:TDGP}), one can express exactly the derivative $\partial 
\langle r^2 \rangle / \partial \omega_{\rm ho}^2$ in terms of the square 
radius, $\langle r^2 \rangle$,  and its derivative with respect to $N$.
One finds (Zambelli, 1998)
\begin{equation}
\chi_M =  { N \over m \omega_{\rm ho}^2} \left[ \langle r^2 \rangle
- {N \over 2} {\partial \over \partial N} \langle r^2 \rangle \right]
\label{eq:chiexplicit}
\end{equation}
where the term depending on the derivative arises from two-body
interactions; in the case of an ideal gas, this term vanishes and 
the mean square radius, $\langle r^2 \rangle = 
(3/2) a^2_{\rm ho}$, is independent of $N$.

Using the  moments $m_1$ and $m_{-1}$ one can define an average 
excitation energy, $\hbar\omega$, through the ratio
\begin{equation}
(\hbar\omega)^2 = {m_1\over m_{-1}} \; ,
\label{eq:omega1-1}
\end{equation}
yielding the useful result (Zambelli, 1998)
\begin{equation}
\omega^2_M = 4 \omega^2_{\rm ho}{\langle r^2 \rangle \over
\langle r^2 \rangle- {N\over 2}
{\partial \over  \partial N }  \langle r^2 \rangle } \; .
\label{eq:omegasumrule}
\end{equation}
In the noninteracting case,  
one  recovers  $\omega_M= 2\omega_{\rm ho}$. When $Na/a_{\rm ho}$ 
is large and  positive, the Thomas-Fermi approximation (\ref{eq:rhoTF}) 
for the density provides the analytic behavior of the radius, $\langle r^2
\rangle \propto N^{2/5}$, and hence the result $\omega_M
=\sqrt{5}\ \omega_{\rm ho}$ already discussed in the previous section.  
For negative $a$ and close to the critical size $N_{\rm cr}$,  the monopole
frequency goes to zero  because the compressibility of the system becomes
larger and larger.   Actually the $N$-dependence of $\omega_M$ near
$N_{\rm cr}$ can be determined analytically. For example, using the Gaussian
variational procedure developed in  Sec.~\ref{sec:attractiveforces},
one finds the result  $(\langle r^2 \rangle - \langle r^2
\rangle_{\rm cr})= \langle r^2 \rangle_{\rm cr}
\sqrt{8/5}(1-N/N_{\rm cr})^{1/2}$,
where  $\langle r^2 \rangle_{\rm cr}$ is the square radius of the
condensate at  the critical value $N_{\rm cr}$. As a
consequence of this peculiar $N$ dependence, the monopole
compressibility diverges  near $N_{\rm cr}$ and the monopole frequency
vanishes as (Singh and Rokhsar, 1996; Zambelli, 1998; Ueda and Leggett, 
1998)
\begin{equation}
\omega_M = \omega_{\rm ho}(160)^{1/4}
 \left(1-{N\over N_{\rm cr}}\right)^{1/4} \; .
\label{eq:omegacritical}
\end{equation}
By using the numerical solution of the Gross-Pitaevskii equation to
calculate the $N$-dependence of  the square radius one finds  a  slightly
smaller value for the numerical coefficient in (\ref{eq:omegacritical}),
namely $3.43$ instead of $3.56$. It has been suggested that the behavior
of the monopole frequency near $N_{\rm cr}$ might play an important 
role in the decay mechanism of the condensate for $N$ very close to 
$N_{\rm cr}$, due to quantum tunneling (Ueda and Leggett, 1998).

\begin{figure}[t]
\epsfysize=8cm
\hspace{3cm}
\epsffile{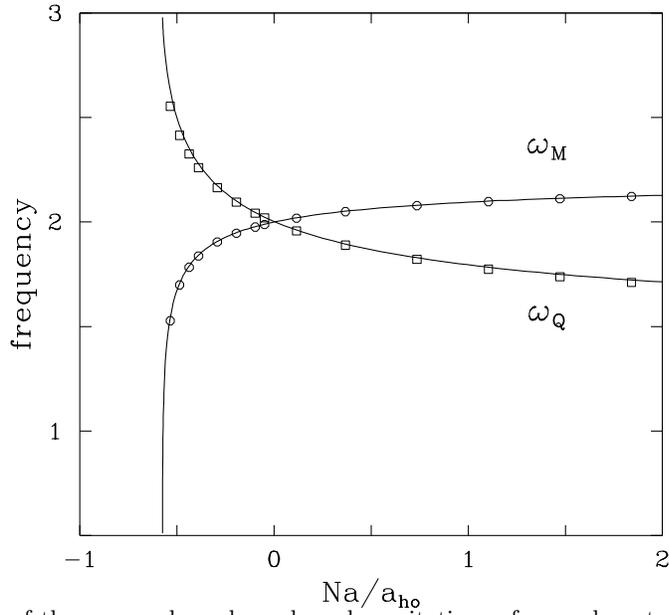}
\caption{Frequencies of the monopole and quadrupole excitations
of a condensate in a spherical trap as a function of the parameter
$Na/a_{\rm ho}$, for positive and negative values of $a$. 
The solid line for the monopole mode is obtained from 
the ratio $(m_1/m_{-1})^{1/2}$, as in  Eq.~(\protect\ref{eq:omegasumrule}). 
For the quadrupole mode it corresponds to the ratio $(m_3/m_1)^{1/2}$, as 
in Eq.~(\protect\ref{eq:omegasumrule2}). Circles and squares are the 
eigenenergies of the linearized time dependent GP equation 
(\protect\ref{eq:coupled1})-(\protect\ref{eq:coupled2}).  } 
\label{fig:sumrules}
\end{figure}

In Fig.~\ref{fig:sumrules} we show the frequency $\omega_M$ obtained 
from Eq.~(\ref{eq:omegasumrule}) as a 
function of the parameter  $Na/a_{\rm ho}$ (solid line). The square radius 
$\langle r^2 \rangle$ has been calculated  by solving numerically the 
stationary GP equation (\ref{eq:GP}) in a spherical trap. As already said,  
the ratio (\ref{eq:omega1-1}) between moments of the strength distribution 
function $S_F(\omega)$ provides a rigorous upper bound to the lowest 
monopole frequency. The comparison with the numerical solutions of the 
time dependent GP equation (circles) shows that the sum rule estimate 
actually gives an excellent approximation to the collective frequency 
for both positive and negative values of $a$, practically indistinguishable 
from the exact result. This means that the strength distribution of the 
monopole operator $F$ almost coincides with a $\delta$-function located 
at the energy of the lowest compressional mode. For the same reason, 
also the ratio $m_3/m_1$ turns out to be very close to $m_{1}/m_{-1}$
(Zambelli, 1998).

Unlike the monopole frequency, the quadrupole frequency increases with 
$N$ when $a<0$, due to the increase of the kinetic energy of the
condensate. This behavior is well understood by calculating the quadrupole
frequency through the ratio $(\hbar\omega)^2 = m_3/m_1$, where
$m_1$ and $m_3$ are the energy and cubic energy weighted moments for
the natural quadrupole operator $F=\sum_{i=1}^Nr^2 Y_{2m}$.
By explicitly working out the commutators of the two sum rules one finds 
the following result for the quadrupole frequency (Stringari, 1996b):
\begin{equation}
\omega_Q^2 = 2\omega_{\rm ho}^2\ \left( 1 + 
{E_{\rm kin} \over E_{\rm ho}} \right) \; .
\label{eq:omegasumrule2}
\end{equation}
In the noninteracting gas one has $E_{\rm kin}=E_{\rm ho}$ and
(\ref{eq:omegasumrule2})  gives the harmonic oscillator result
$\omega_Q=2\omega_{\rm ho}$. In the Thomas-Fermi limit $Na/a_{\rm ho}
\gg 1$, the kinetic energy term is negligible and one
finds the value $\omega_Q= \sqrt{2} \ \omega_{\rm ho}$, while
for negative $a$ the kinetic energy term  is larger than
$E_{\rm ho}$  and  one finds an enhancement of the quadrupole frequency.
The numerical results are reported in Fig.~\ref{fig:sumrules}.
Also in the quadrupole case the sum rule estimate (\ref{eq:omegasumrule2})
turns out to be very close to the exact numerical solution of the linearized 
time dependent GP equations (\ref{eq:coupled1})-(\ref{eq:coupled2}),
indicating that the lowest quadrupole mode almost exhausts the
strength distribution of $F$, as already found for the monopole mode. 

We have here applied the sum rule approach to the case of spherical traps, 
but the same analysis can be easily generalized to more complex geometries, 
the main physical arguments remaining unchanged. Calculations of sum rules
for axially symmetric traps have been carried out by Kimura and Ueda (1998), 
finding accurate predictions for the collective frequencies.

\subsection{Expansion and large amplitude oscillations}
\label{sec:nonlinear}

So far we have discussed the behavior of normal modes of the condensate and 
sound propagation. It is also interesting to investigate nonlinear features
associated, for example,  with the dynamics of the expansion of the
gas, following the switching-off of the trap, as well as with the frequency
shifts of large amplitude oscillations. The dynamics of the expansion
is an important issue because much  information on these Bose condensed
gases is obtained experimentally from images of the expanded atomic cloud.
This includes in particular the temperature of the gas (which is extracted
from the tail of the thermal component), the release energy and the aspect
ratio of the velocity distribution. Nonlinear phenomena are also crucial 
in the analysis  of  the large amplitude oscillations which are produced 
and detected in current experiments. Phenomena like mode coupling, 
harmonic generation, frequency shifts and  stochastic behavior may become 
interesting subjects of research in these systems.

 From the theoretical viewpoint one can again attack the problem starting
from the time dependent Gross-Pitaevskii equation. Indeed, the GP equation
(\ref{eq:TDGP}) for the order parameter of the condensate can be applied
to the nonlinear regime and it is important to check the validity of its 
predictions through a direct comparison with experiments. In Sec.~\ref{sec:TDGP}
we have linearized this equation in order to obtain the coupled equations
(\ref{eq:coupled1})-(\ref{eq:coupled2}) for the  excitations.
The numerical solution in the nonlinear regime is also feasible (Holland and
Cooper, 1996; Holland {\it et al.}, 1997; Ruprecht {\it et al.}, 1996; Smerzi
and Fantoni, 1997; Morgan {\it et al.}, 1998;  Brewczyk {\it et al.}, 1998). 
For instance, Holland {\it et al.} (1997) obtained
results  for the density and energy of the expanding gas of $^{87}$Rb in
good agreement with the first measurements at JILA. In the inset of
Fig.~\ref{fig:releasejila}  their results for the axial and radial widths
are plotted as a function of expansion time. 

When the number of atoms in the trap is large, the time dependent GP 
equation (\ref{eq:TDGP}) reduces to the equations of continuity 
(\ref{eq:continuity}) for the density  and the  Euler equation 
(\ref{eq:eulerHD}) for the velocity field. These equations can be used 
to investigate nonlinear phenomena in a simplified way. Let us write 
the external potential in the form $V_{\rm ext}({\bf r}) = 
(m/2)\sum_i \omega_i^2 r_i^2$, with $r_i\equiv x,y,z$. In general, 
the trapping frequencies can depend on time, $\omega_i=\omega_i(t)$; 
their static values, $\omega_{0i}=\omega_i(0)$, fix the
initial equilibrium configuration of the system, corresponding to the
Thomas-Fermi density (\ref{eq:rhoTF}). One can easily prove that the 
equations of motion admit a class of analytic solutions having the
density in the form
\begin{equation}
n({\bf r},t) = a_0(t) - a_x(t) x^2 - a_y(t) y^2 - a_z(t) z^2   \; ,
\label{eq:quadraticn}
\end{equation}
within the region where $n({\bf r},t)$ is positive and $n({\bf r},t)=0$
elsewhere, and the velocity field as
\begin{equation}
{\bf v} ({\bf r},t)  =  {1\over2} \mbox{\boldmath$\nabla$} [
\alpha_x(t) x^2 + \alpha_y(t) y^2 + \alpha_z(t) z^2] \; .
\label{eq:quadraticv}
\end{equation}
These results, combined with Eq.~(\ref{eq:gradientS}), allow one to obtain 
an explicit expression for the order parameter (\ref{eq:orderparameter}).
In particular, its phase $S$ takes the form 
\begin{equation}
S ({\bf r},t) = { m \over 2 \hbar } [\alpha_x(t) x^2 + 
\alpha_y(t) y^2 + \alpha_z(t) z^2] \; . 
\label{eq:S}
\end{equation}
Notice that, while the velocity field (\ref{eq:quadraticv}) is governed 
by the classical equations (\ref{eq:continuity}) and (\ref{eq:eulerHD}),
the phase of the order parameter depends explicitly on the Planck
constant $\hbar$.  

The results (\ref{eq:quadraticn}) and (\ref{eq:quadraticv}) include
the ground state solution (\ref{eq:rhoTF}) in the Thomas-Fermi limit.
This is recovered by putting 
$\alpha_i \equiv 0$ and $a_i \equiv m\omega_{0i}^2/(2g)$, with 
$i=x,y,z$, while $a_0=\mu/g$. In general, one can insert  
expressions (\ref{eq:quadraticn}) and (\ref{eq:quadraticv}) into
the equations (\ref{eq:continuity})-(\ref{eq:eulerHD}),
getting six coupled differential equations for the time dependent
coefficients $a_i(t)$ and $\alpha_i(t)$, while the relation 
$a_0=(15N/8\pi)^{2/5} (a_x a_y a_z)^{1/5}$ is fixed, at any time, 
by the normalization of the density to the total number of particles. 
Instead of writing these equations, we note that the assumptions 
(\ref{eq:quadraticn}) and (\ref{eq:quadraticv}) for the density and 
velocity distributions correspond to assuming a scaling transformation 
of the order parameter. This means that, at each instant, the parabolic 
shape of the density is preserved, while the classical radii $R_i$, where 
the density (\ref{eq:quadraticn}) vanishes, scale in time as
\begin{equation}
R_i(t) = R_i(0) b_i(t) = \sqrt{ {2 \mu \over m \omega_{0i}^2} }
\  b_i(t) \; . 
\label{eq:scalingr}
\end{equation}
The relation among the coefficients $a_i$ of Eq.~(\ref{eq:quadraticn})
and the variables $b_i$ is found to be $a_i= m \omega_{0i}^2 / ( 2g  
b_x b_y b_z b_i^2)$ and the equations 
(\ref{eq:continuity})-(\ref{eq:eulerHD}) then give $\alpha_i = \dot{b}_i/b_i$ 
and
\begin{equation}
\ddot{b}_i + \omega_i^2 b_i - { \omega_{0i}^2 \over b_ib_xb_yb_z } = 0
\; . 
\label{eq:ddotb}
\end{equation}
These are three  coupled differential equations for the scaling 
parameters $b_i(t)$, which in turn give the time evolution of the 
classical radii, $R_i(t)$, of the order parameter. The second term in 
(\ref{eq:ddotb}) comes from the confining potential, while the third 
one originates from the atom-atom interaction.
Equations (\ref{eq:ddotb}) have been derived and used by different 
authors (Kagan, Surkov and Shlyapnikov, 1996, 1997a and 1997b; Castin 
and Dum, 1996; Dalfovo {\it et al.}, 1997b and 1997c).  Their major
advantage is that they are ordinary differential equations, very easy
to solve, giving results close to the solutions of the time
dependent GP equation in most situations. Kagan, Surkov and 
Shlyapnikov (1996) have shown that in 2D the scaling transformation
of the order parameter, starting from the stationary configuration 
at the initial time, actually corresponds to an exact solution of 
the GP equation. 

\begin{figure}[t]
\epsfysize=8cm
\hspace{3cm}
\epsffile{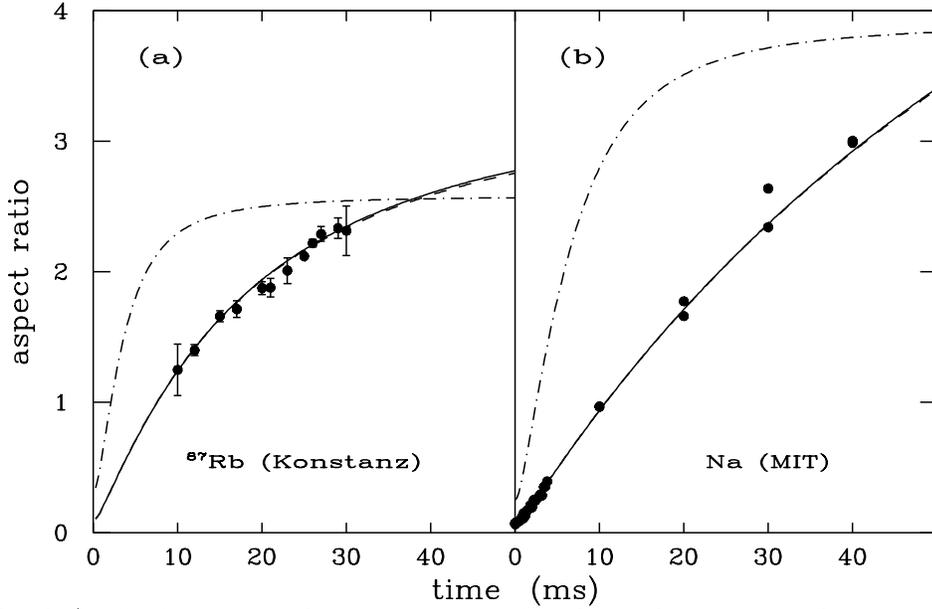}
\caption{ Aspect ratio, $R_\perp /Z$, of a freely expanding condensate 
as a function of time. The experimental points in part (a) correspond to 
$^{87}$Rb atoms initially confined in a trap with $\lambda=0.099$ (Ernst 
{\it et al.}, 1998b). The points in part (b) are measurements on sodium 
atoms, initially in a trap with $\lambda=0.065$ (Stamper-Kurn and 
Ketterle, 1998). The solid lines are obtained by solving 
Eqs.~(\protect\ref{eq:axialexpansion}), which are equivalent to the time 
dependent GP equation in the Thomas-Fermi approximation. The dashed lines 
correspond to the $\lambda \ll 1$ limit of the same equations, that is 
to Eqs.~(\protect\ref{eq:castin1})-(\protect\ref{eq:castin2}), and are almost
indistinguishable from the solid lines. The dot-dashed lines are the 
predictions for noninteracting atoms. Theoretical curves have no fitting 
parameters. In part (a), they have been corrected to include the effect 
of the observation angle, as explained by Ernst {\it et al.} (1998b).   }
\label{fig:expansion}
\end{figure}

Equations (\ref{eq:ddotb}) can be used to simulate the expansion
starting from a gas in equilibrium in the trap, by dropping at
a certain time, $t=0$, the term linear in $b_i$ associated with
the confining potential. For an axially symmetric trap one can 
define $b_\perp \equiv b_x = b_y$ and introduce a dimensionless 
time $\tau = \omega_\perp t$, with $\omega_\perp \equiv \omega_{0x}
=\omega_{0y} = \lambda^{-1} \omega_{0z}$. Then Eqs.~(\ref{eq:ddotb}) 
take the form
\begin{equation}
{d^2 \over d \tau^2} b_\perp = {1 \over b_\perp^3 b_z} \; \; \;
{\rm and}  \; \; \;  
{d^2 \over d \tau^2} b_z = {\lambda^2 \over b_\perp^2 b_z^2} \; . 
\label{eq:axialexpansion}
\end{equation}
By solving these equations, one can look, for instance, at the time 
evolution of the aspect ratio $R_\perp / Z  = \lambda b_\perp/b_z$. 
When $\tau$ is large, both $b_\perp$ and $b_z$ increase linearly 
with $\tau$ and the parameters $\alpha_\perp$ and $\alpha_z$,
characterizing the velocity field (\ref{eq:quadraticv}), behave 
as $1/t$, consistently with the classical equation of motion for 
free particles, ${\bf v}={\bf r}/t$.  In  
Fig.~\ref{fig:expansion} we show the results of this calculation in 
two cases where accurate experimental data are available: atoms of 
$^{87}$Rb released from a trap with $\omega_\perp = 2\pi \times 247$ Hz 
and $\omega_z=2\pi \times 24$ Hz at Konstanz (Ernst, 1998b), and 
sodium atoms released from a trap with $\omega_\perp = 2\pi \times 248$ Hz 
and $\omega_z=2\pi \times 16.23$ Hz at MIT (Stamper-Kurn and 
Ketterle, 1998). Both traps are cigar-shaped and the number of atoms 
is large enough for applying the Thomas-Fermi approximation. The
agreement between theory (solid lines) and experiments (points)
is remarkable.  It is also worth mentioning that, as shown by Castin 
and Dum (1996), the two equations (\ref{eq:axialexpansion}) can be 
solved analitycally for $\lambda \ll 1$, leading to the useful expressions
\begin{eqnarray}
b_\perp (\tau) & = & \sqrt{1+\tau^2} 
\label{eq:castin1} \\
b_z (\tau)     & = & 1+ \lambda^2 [ \tau \arctan \tau - 
\ln \sqrt{ 1+\tau^2 }  ] \; . 
\label{eq:castin2}
\end{eqnarray}
The corresponding aspect ratio is plotted in Fig.~\ref{fig:expansion} 
as a dashed line. As one can see, the analytic small-$\lambda$ limit 
practically coincides with the exact solution of (\ref{eq:axialexpansion})
for the two traps here considered.  Moreover, from expressions
(\ref{eq:castin1})-(\ref{eq:castin2}) one also gets the asymptotic
value $\lim_{\tau \to \infty} (R_\perp / Z)  = 2/(\pi \lambda)$. 
In the case of Fig.~\ref{fig:expansion}, this asymptotic limit 
is approximately $6.5$ and $9.7$ for the Konstanz and MIT data, 
respectively, but is far from being attained even after tens of ms; in fact,
it takes a relatively long time to reach the regime of constant speed
for the motion along the direction of weaker initial confinement, due
to the slow acceleration induced by the mean-field potential.  

The general agreement between theory and experiments
in Fig.~\ref{fig:expansion} is even better appreciated if one considers
the predictions for the expansion of noninteracting particles. The aspect
ratio obtained from the dispersion of a free atomic wave packet is 
represented by the two dot-dashed lines. The asymptotic limit for
$\tau \to \infty$ is $\lambda^{-1/2}$. The comparison with the behavior
of the interacting gas shows, once again, the important role of the 
atom-atom interaction.

The same formalism allows one to calculate the time evolution of the 
various contributions to the release energy of the condensate. In terms of 
the scaling  parameters $b_i$ the release energy takes the form  
\begin{equation}
E_{\rm rel} = {2 \mu \over 7} \left( { 1 \over b_x b_y b_z } 
+ {1\over 2} \sum_i { \dot{b}_i^2 \over \omega_{0i}^2 } 
\right) \; . 
\label{eq:release}
\end{equation}
This quantity is conserved during the expansion. At $t=0$, when $b_i=1$ 
and $\dot{b}_i=0$, the release energy is equal to $(2/7)\mu$. During the 
expansion the mean-field energy  (first term in the bracket) is converted 
into kinetic energy (second term).  After a certain time, which can 
be estimated using Eq.~(\ref{eq:release}), the mean-field energy becomes 
negligible, since the system is more and more dilute, and the 
expansion proceeds at constant speed in each direction. 

\begin{figure}[t]
\epsfysize=8cm
\hspace{3cm}
\epsffile{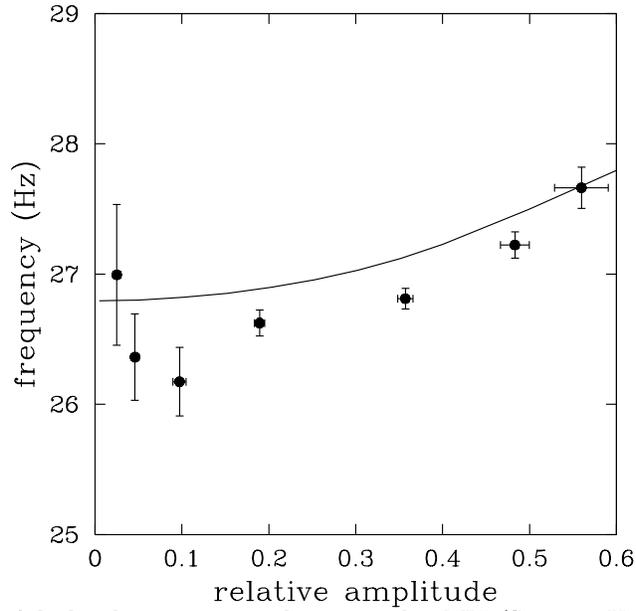}
\caption{ Frequency of the low-lying $m=0$ mode measured at 
MIT (Stamper-Kurn {\it et al.}, 1998, Stamper-Kurn and Ketterle, 1998) 
as a function of the amplitude of the oscillation. 
The solid line is the prediction of 
Eqs.~(\protect\ref{eq:ddotb}). } 
\label{fig:ampshift}
\end{figure}

The same Eqs.~(\ref{eq:ddotb}) allow one also to study the effects of
a sinusoidal driving force which simulates  the modulation of the
confining potential used in experiments to generate collective modes in 
the trap. In the small amplitude limit, Eqs.~(\ref{eq:ddotb}) yields the 
frequencies of the normal modes in the regime of collisionless 
hydrodynamics already discussed in Sec.~\ref{sec:hydrodynamics}. 
In particular for axially
symmetric traps, expression (\ref{eq:quadraticn}) includes the
lowest $m=0$ and $m=2$ modes, which have been measured experimentally,
while other modes could be investigated by adding terms in $xy$, $xz$ and 
$yz$. When the amplitude of the oscillations grows, the frequency of
the modes can shift and the modes themselves can couple. For the
experiments carried out at  JILA and MIT (Jin {\it et al.}, 1996 and
1997; Stamper-Kurn {\it et al.},  1998c) both the frequency shift and
the mode coupling are small. An example is given in Fig.~\ref{fig:ampshift}, 
where we show the frequency of the lowest $m=0$ mode, observed in the 
cigar-shaped MIT trap (Stamper-Kurn {\it et al.}, 1998c; Stamper-Kurn 
and Ketterle, 1998), as a function of its amplitude. The
solid line is the prediction of Eqs.~(\ref{eq:ddotb}). The overall 
agreement is good for both the zero amplitude frequency and its
shift. It is worth recalling that the theory has no fitting
parameters.

An effect which deserves to be mentioned is the large enhancement
of nonlinear effects for special values of the asymmetry parameter
$\lambda$. First, one notes that the frequencies of the collective 
modes depend on the shape of the trapping potential and hence on 
$\lambda$. For certain values of this parameter, it may happen that 
different modes have the same frequency; this has been shown to occur,
in the linear regime,  for $|m|>2$ by \"Ohberg {\it et al.} (1997)
and a systematic investigation of the level crossing has been done
by Hutchinson and Zaremba (1997). In the 
nonlinear regime, one finds strong mode-coupling {\it via} harmonic
generation when the frequency of a mode becomes equal to the one
of the second harmonics of other modes (Dalfovo {\it et al.},
1997b; Graham {\it et al.}, 1998). The conditions for this degeneracy
can be found numerically from Eqs.~(\ref{eq:ddotb}). In the limit
of small amplitude one can also expand the solutions finding
analytical results (Dalfovo {\it et al.}, 1997b). In particular,
one gets a quadratic shift in the form
\begin{equation}
\omega (A) = \omega(0) [ 1 +\delta(\lambda) A^2 ]
\label{eq:quadraticshift}
\end{equation}
where $A$ is the relative amplitude of the oscillation and
$\delta(\lambda)$ is an analytic  coefficient depending on the
anisotropy of the trapping potential and on the mode considered.
For instance, in the case of the $m=2$ mode, one finds:
\begin{equation}
\delta (\lambda) = { (16 - 5 \lambda^2)
\over 4 (16 - 7 \lambda^2) } \; .
\label{eq:delta-2}
\end{equation}
The divergence at $\lambda=\sqrt{16/7}$  is due to the degeneracy
between the frequencies of the high-lying $m=0$ mode and the second
harmonic of the $m=2$ mode.  In this case, it is difficult to drive
the system in a pure mode and, even for relatively small amplitudes, the 
motion is rather complex and the resulting trajectories can exhibit
a chaotic-like behavior. The coefficient $\delta(\lambda)$ can be
calculated also for other modes. For the low-lying $m=0$ mode,
for instance, similar divergences are found when $\lambda= (\sqrt{125}
\pm \sqrt{29}) /\sqrt{72}$ (i.e., $\lambda \approx 0.683$ and
$\lambda \approx 1.952$). They occur because the frequency of the
high-lying mode becomes equal to the second harmonics of the low-lying
mode. It would be very interesting to study experimentally the system
in these conditions. 

As a final remark we note that, by means of a variational
approach based on Gaussian wave functions,
P\'erez-Garc\'\i a {\it et al.} (1996, 1997) have derived equations of
motion of the form (\ref{eq:ddotb}), but with an additional term included,
proportional to $1/b_i^3$, accounting for the quantum pressure
in the Gross-Pitaevskii equation.  Even though the equilibrium
configuration in the Thomas-Fermi regime $Na/a_{\rm ho} \gg1$ is
not exactly recovered, because of the Gaussian {\it ansatz}, these
equations represent a good approximation to the  GP equation for
finite $N$, interpolating  between the noninteracting and strongly
interacting  systems. The frequencies of the lowest $m=0$ and $m=2$
modes, calculated with the parameter of the JILA trap,
differ from the exact solutions of (\ref{eq:coupled1})-(\ref{eq:coupled2})
by less than $1$\% \ over all the relevant range of $N$.  Since this
method includes quantum pressure effects, it can be used to explore
the nonlinear dynamics of the gas also in the case of attractive forces 
(P\'erez-Garc\'\i a {\it et al.}, 1997).

\subsection{ Density of states: collective vs. single-particle excitations}
\label{sec:collectivevssingleparticle}

In the previous sections we have discussed several features of collective
excitations pointing out the crucial role played by two-body interactions.
We may ask whether these collective modes are relevant for the statistical
properties of these many-body systems.  One knows, for instance, that the 
thermodynamic behavior of superfluid $^4$He is dominated by the thermal
excitation of  phonons  and rotons up to the critical temperature. For the
trapped gas the situation is very different. First the system is very
dilute and one expects that the effects of collectivity shall be less
relevant except at very low temperature. Second the harmonic confinement
leaves space for  excitations of single-particle nature which actually
dominate the thermodynamic behavior even at  low temperature.

\begin{figure}[t]
\epsfysize=10cm
\hspace{3cm}
\epsffile{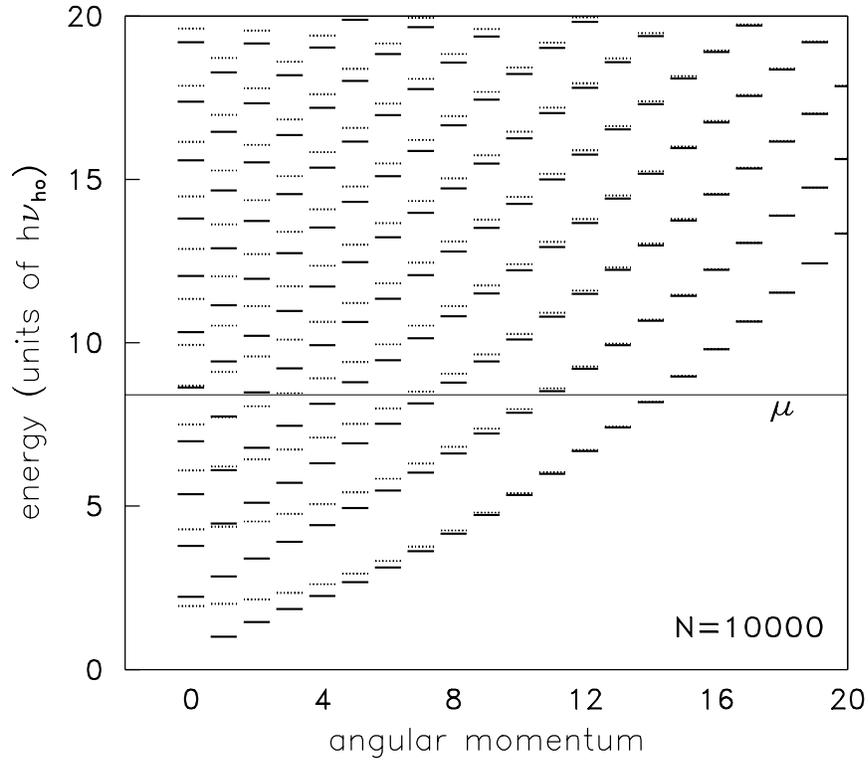}
\caption{Excitation spectrum of $10000$ atoms of $^{87}$Rb in a 
spherical trap with $a_{\rm ho}=0.791$ $\mu$m.  The eigenenergies
of the linearized time dependent GP equations 
(\protect\ref{eq:coupled1})-(\protect\ref{eq:coupled2}) are 
represented by thick solid bars.  Dashed bars correspond to the 
single-particle spectrum of Hamiltonian (\protect\ref{eq:HFhamiltonian}). 
The thin orizontal line is the chemical potential, $\mu=8.41$ in units 
of $\hbar \omega_{\rm ho}$, which is fixed by the solution of the 
stationary GP equation (\protect\ref{eq:GP}).  } 
\label{fig:spectrum}
\end{figure}

The simplest way to understand the role of these single-particle
excitations is to look at the spectrum  obtained by solving numerically 
the two Bogoliubov-like equations  (\ref{eq:coupled1})-(\ref{eq:coupled2}). 
In Fig.~\ref{fig:spectrum} we show the eigenstates evaluated for a 
condensate of $10^4$ atoms of $^{87}$Rb in a spherical trap (Dalfovo 
{\it et al.}, 1997a). Each state, having energy $\varepsilon$ and angular 
momentum $\ell$, is represented by a thick solid bar. For a given 
angular momentum, the number of radial nodes, i.e., the quantum number
$n_r$, increases with energy. 

By looking at the eigenstates at high-energy and multipolarity in the 
spectrum of Fig.~\ref{fig:spectrum},  one notes that the splitting 
between odd and even states is approximately $\hbar \omega_{\rm ho}$ 
and the spectrum resembles the one of a 3D harmonic oscillator. Actually, 
the states with the same value of $(2 n_r +\ell)$ would be degenerate 
in the harmonic oscillator case, while here they have slightly different
energies,  the  states with lowest angular momentum being shifted upwards 
as a results of the mean-field produced by the condensate in the central 
region of the trap. Indeed the high-energy part of the spectrum is expected 
to be well reproduced by a single-particle description in mean-field 
approximation. The single-particle picture is obtained by neglecting
the coupling between the positive ($u$) and negative ($v$) frequency
components of the order parameter (\ref{eq:linearized}) in the 
Bogoliubov-type equations (\ref{eq:coupled1})-(\ref{eq:coupled2}),
which is responsible for the collectivity of the solutions. This
corresponds to setting $v=0$ in Eq.~(\ref{eq:coupled1}), which then
reduces to the eigenvalue problem $(H_{\rm sp}-\mu)u=\hbar\omega u$,  
for the single-particle (sp) Hamiltonian
\begin{equation}
H_{\rm sp} = - (\hbar^2/2m) \nabla^2 +  V_{\rm ext}({\bf r}) +
2 g n({\bf r}) \; .
\label{eq:HFhamiltonian}
\end{equation}
In this case, the eigenfunctions $u({\bf r})$
satisfy the normalization condition $\int  d{\bf r} \ u_i^\ast({\bf r})
u_j({\bf r})=\delta_{ij}$.  This approximation is directly related to
Hartree-Fock theory, as we will discuss  in Sec.~\ref{sec:therm2}.

Once the condensate density and the chemical potential
are calculated from the stationary GP equation (\ref{eq:GP}), the 
single-particle excitation spectrum of the Hamiltonian 
(\ref{eq:HFhamiltonian}) can be easily calculated. The eigenstates 
are shown as dashed orizontal bars in Fig.~\ref{fig:spectrum}. One 
sees that the general structure of the spectrum is very similar to 
the one obtained with the Bogoliubov-type equations 
(\ref{eq:coupled1})-(\ref{eq:coupled2}) apart from the states with 
low energy and multipolarity.  The lowest levels,  with energy
well below $\mu$ and small angular momentum, are in fact the collective 
modes discussed in the previous sections (for instance, the lowest 
states with $n_r=0$ are the monopole, dipole, and quadrupole modes 
for which the theory, in the limit $Na/a_{\rm ho} \gg 1$, predicts 
$\varepsilon= \sqrt{5}$, $1$, and $\sqrt{2}$, respectively, in units
of $\omega_{\rm ho}$). The single-particle spectrum, which does 
not account for collective motion of the condensate, fails to describe 
these states. It is worth noticing, however, that even below $\mu$ there 
are many states, with relatively high $\ell$, which are well approximated 
by  the single-particle Hamiltonian (\ref{eq:HFhamiltonian}).  
Actually, the  numerical analysis reveals that, for these states,
the condition $|v| \ll |u|$ is well satisfied (You, Hoston and 
Lewenstein, 1997; Dalfovo {\it et al.}, 1997a).  These 
excitations are mainly located near the surface of the condensate, 
where $H_{\rm sp}$ has a minimum. The existence of such
a minimum is evident in the large $N$ limit, where the Thomas-Fermi
approximation for the condensate density is accurate. In this case, one 
has
\begin{equation}
H_{\rm sp}-\mu =  - (\hbar^2/2m) \nabla^2  + {1\over 2} m \omega^2_{\rm ho}
|r^2 -R^2| \; ,
\label{eq:largeNHF}
\end{equation}
where $R=[2 \mu/(m\omega^2_{\rm ho})]^{1/2}$ is the classical radius
of the condensate and we have taken, for simplicity, a spherical trap.
Of course, for finite values of $N$ the minimum of the single-particle
potential is rounded. 

The fact that the Bogoliubov-type  spectrum exhibits states of 
single-particle nature localized near the surface represents an important 
difference with respect to the uniform Bose gas, where no single-particle 
states are present at energy lower than the chemical potential. The 
transition between the collective and single-particle character can be 
understood in terms of length scales. In fact, an excitation inside the
condensate can no longer be phonon-like when its wavelength is of the
order of, or shorter than, the healing length $\xi$ [see
Sec.~\ref{sec:stationaryGP}  and Eq.~(\ref{eq:healinglength})]. This
happens for states with a large number  of radial nodes and energy
larger than $\mu$. Conversely, for states localized mainly at the
surface, the appropriate length scale is the surface thickness $d$,
introduced in Sec~\ref{sec:TF} [see Eq.~(\ref{eq:surfacethickness})].
In this case, excitations cannot be collective, and hence cannot be 
described by the equations of collisionless hydrodynamics, if their 
wavelength is smaller than $d$. This happens when their angular momentum 
is larger than $\ell \sim R/d\sim N^{4/15}$. This critical value of $\ell$ 
corresponds to an energy of the order of $\mu (a_{ho}/R)^{4/3}$, 
so that the transition from the collective to the single-particle behavior 
occurs at energies smaller than $\mu$ in states with high multipolarity.
These states can be viewed as atoms rotating in the outer part of the 
condensate (Lundh, Pethick and Smith, 1997; Dalfovo {\it et al.}, 1997a).

\begin{figure}[t]
\epsfysize=8cm
\hspace{3cm}
\epsffile{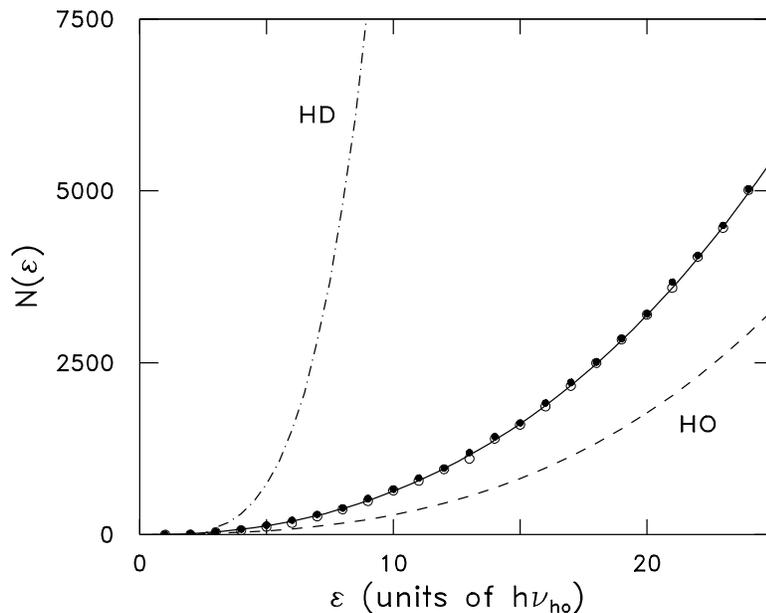}
\caption{Number of states $N(\varepsilon)$ {\it vs.} energy. 
Solid circles are obtained by counting the Bogoliubov-type 
states in the spectrum shown in Fig.~\protect\ref{fig:spectrum}
(thick solid bars). Open circles corresponds to counting the 
single-particle states in the same figure (dashed bars). 
Both calculations are compared with the predictions of the
semiclassical approximation (\protect\ref{eq:semiclnstates}) 
(solid line), as well as with  the ones of the noninteracting 
harmonic oscillator (dashed line) and of the collisionless 
hydrodynamic equations in the Thomas-Fermi regime (dot-dashed 
line). Chemical potential is $\mu=8.41$ in this scale. }
\label{fig:ne}
\end{figure}

In order to discuss the relevance of single-particle excitations in 
the statistical behavior of these trapped Bose gases it is useful to
evaluate the density of states.  For a finite system one can easily
count the number of available states, with energy $\varepsilon'$ and
angular momentum $\ell$, each one multiplied by its degeneracy 
$(2\ell +1)$, up to a given energy $\varepsilon$:
\begin{equation}
N (\varepsilon) = \sum_{\varepsilon' < \varepsilon } (2 \ell +1 ) \; .
\label{eq:numberofstates}
\end{equation}
The density of states is the derivative of (\ref{eq:numberofstates}). 
In Fig.~\ref{fig:ne} we show the quantity  $N(\varepsilon)$  obtained
by summing the levels of the two spectra of Fig.~\ref{fig:spectrum}. The
agreement between the results of the Bogoliubov-type equations (solid 
circles) and of the single-particle theory (open circles) is remarkable 
even at low energy, indicating that the effects of collectivity are not 
relevant in  the sum (\ref{eq:numberofstates}). Indeed, the number of states
which are badly reproduced by the single-particle Hamiltonian is small
and their degeneracy factor $(2\ell +1)$ is also small, so that their
contribution to the sum (\ref{eq:numberofstates}) is negligible. 
The effects of two-body forces on the density of states are nevertheless 
sizable, as emerges from the comparison with the prediction of the  
noninteracting model (dashed line).  We also report the results obtained 
using the dispersion relation (\ref{eq:hdsdispersion}) for the excitations
in the $Na/a_{\rm ho}\gg 1$ limit. This gives a poor approximation for 
$N(\varepsilon)$, revealing that the hydrodynamic picture becomes completely 
inadequate for excitation energies of the order of $\mu$.

The number of states $N(\varepsilon)$ associated with the single-particle
Hamiltonian (\ref{eq:HFhamiltonian}) can be also calculated using the
semiclassical approximation. In this case one counts the available states
through a simple integration over phase space
\begin{equation}
N(\varepsilon) = \int_0^\varepsilon d\varepsilon^\prime
\; \int \frac{d{\bf r}d{\bf p}}{(2\pi\hbar)^3} \delta (\varepsilon^\prime
-\varepsilon^{\rm sp}({\bf p},{\bf r})) \;,
\label{eq:semiclnstates}
\end{equation}
where $\varepsilon^{\rm sp}({\bf p},{\bf r})=p^2/2m 
+V_{\rm ext}({\bf r})+2gn({\bf r})
-\mu$ is the semiclassical energy corresponding to the Hamiltonian
$H_{\rm sp}-\mu$. In Fig.~\ref{fig:ne} the prediction of
Eq.~(\ref{eq:semiclnstates}) is shown as a solid line. The semiclassical
approximation is expected to be valid only for
$\varepsilon\gg\hbar\omega_{\rm ho}$.
However, the low energy states which are not reproduced by this
approximation give a negligible contribution to $N(\varepsilon)$ and
the semiclassical prediction is practically
indistinguishable from the Bogoliubov spectrum in the whole range of 
energies. 

As we will see in Sec.~\ref{sec:thermodynamics}, the relevance 
of single-particle excitations in determining the density of states 
makes Hartree-Fock theory and the semiclassical approximation very 
effective tools for the investigation of the thermodynamic properties 
of these trapped gases as well as their dynamic behavior at finite 
temperature.

\section{Effects of interactions: thermodynamics}
\label{sec:thermodynamics}

\subsection{Relevant energy scales}
\label{sec:therm1}

\begin{figure}[t]
\epsfysize=10cm
\hspace{3cm}
\epsffile{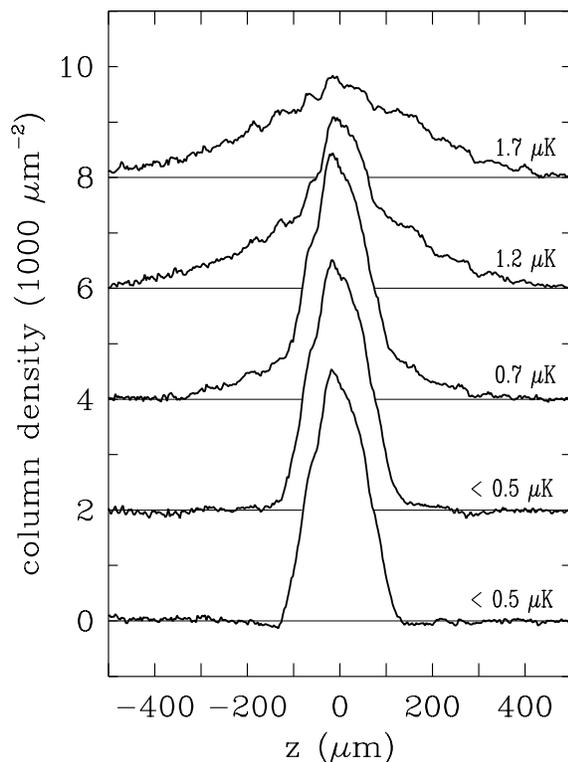}
\caption{Column density of atomic clouds from phase-contrast images at
several values of temperature in the MIT trap. From Stamper-Kurn
and Ketterle (1998).} 
\label{fig:profiles-T}
\end{figure}

The occurrence of Bose-Einstein condensation is revealed  by an  abrupt
change  in the thermodynamic properties of the system below the critical
temperature. In the presence of harmonic trapping a sharp peak 
appears in both the density and velocity  distributions superimposed
on the broader distribution of the thermal component. By further lowering
the temperature the height of the condensate peak increases, while the 
tails of the thermal component are reduced, until they completely 
disappear at very low temperatures, as shown in Fig.~\ref{fig:profiles-T}.  
At the transition, the temperature dependence of the energy shows a 
sudden change in slope which reflects the occurrence of a maximum in the 
specific heat.   

In Sect.~\ref{sec:theidealgas} we have discussed the thermodynamic behavior
of the noninteracting gas. In this model BEC takes
place below the critical temperature $k_BT_c^0 = \hbar\omega_{\rm ho}
(N/\zeta(3))^{1/3}$.  The fraction of atoms in the condensate and their
energy obey the simple laws $N_0/N=1-(T/T_c^0)^3$ and
$E\propto T^4$, respectively [see Eqs.~(\ref{eq:condfractioho}) and
(\ref{eq:Etothoperpart})]. A major question is to understand
whether the predictions of the ideal gas are adequate and under which
conditions the effects of interactions become sizable. This is
the main purpose of the present section.

The effects of two-body interactions in a dilute Bose gas are
expected to be significant only in the presence of the condensate,
since only in this case can the density become relatively high due
to the occurrence of the peak  in the center of the trap. A first
important consequence of repulsive forces is the broadening
of the condensate peak. This effect, already discussed in Sect.~III
at zero temperature, provides a dramatic change in the density distribution
also at finite $T$ and its  experimental observation is an important
evidence of the role played by two-body forces. The opposite happens in the
presence of attractive forces which produce a further narrowing
of the  peak and a consequent increase of the peak density. 
In the following we will mainly discuss the case of systems composed by
a large number of particles interacting with repulsive forces.

Let us discuss the effects of a repulsive interaction by estimating
the relevant energies of the system. At zero temperature the
interaction energy per particle can be simply estimated using the
Thomas-Fermi approximation $E_{\rm int}/N=(2/7)\mu$ where $\mu =(1/2) 
\hbar \omega_{\rm ho} (15Na/a_{\rm ho})^{2/5}$  is the value of the 
chemical potential [see Eq.~(\ref{eq:muTF})]. It is useful to 
compare $E_{\rm int}/N$, or equivalently $\mu$, with the thermal 
energy $k_BT$. If $k_BT$ is 
smaller than $\mu$, then one expects to observe important effects in the
thermodynamic behavior due to interactions. If instead $k_BT$ is larger
than $\mu$, interactions will provide only perturbative corrections.
Thus for repulsive forces the chemical potential provides an
important scale of energy lying  between the oscillator energy and
the critical temperature: $\hbar\omega_{\rm ho} < \mu < k_BT_c^0$.
A useful parameter is the ratio
\begin{equation}
\eta = \frac{\mu}{k_BT_c^0} = \alpha\left( N^{1/6}\frac{a}{a_{\rm ho}}
\right)^{2/5}
\label{eq:eta}
\end{equation}
between the chemical potential calculated at $T=0$ in Thomas-Fermi
approximation and the critical temperature for noninteracting particles in
the same trap. Here $\alpha=15^{2/5}(\zeta(3))^{1/3}/2\simeq 1.57$ is
a numerical coefficient.  If one uses the typical values for the
parameters of current experiments, one finds that $\eta$ ranges
from $0.35$ to $0.40$.  Thus,
one expects that interaction effects will be
visible also at values of $T$ of the order of $T_c^0$.

It is worth discussing the dependence of the parameter $\eta$ on the
relevant parameters of the system. First one should point out that this
dependence is different from that of the interaction
parameter $Na/a_{\rm ho}$ already introduced in Sect.~\ref{sec:groundstate}
to account for  the effects of two-body interactions in the GP equation
for the condensate. The parameter
$Na/a_{\rm ho}$ determines the value of the chemical potential in units
of the oscillator energy, while $\eta$ fixes it in units of
the critical temperature. This brings a different dependence of $\eta$ on
$N$ which turns out to be very smooth ($\eta \sim N^{1/15}$).
Thus, in order to change the value of this parameter, and consequently
the effects of interactions on the thermodynamic behavior, it is
much more effective to modify the ratio $a/a_{\rm ho}$ between the
scattering and  oscillator lengths rather than the value of $N$.

Another important feature of the parameter $\eta$ is that it can
be expressed explicitly in terms of the traditional ``gas parameter"
$a^3n$, through the relation $\eta =2.24 [a^3 n_{T=0}(0)]^{1/6}$
[see Eq.~(\ref{eq:gasparameter})]. Notice that in this formula
$n_{T=0}(0)$ is the density at  the center of the trap evaluated at
zero temperature. Due to the $1/6$-th power entering this relation,
the value of $\eta$  can be easily of the order of $1$ even if the
gas parameter is very small. For example, taking $a^3n = 10^{-5}$ one 
finds $\eta=0.33$. Equation (\ref{eq:eta})  can be also written in 
terms of the ratio between the transition temperature $k_BT_c^0$ and 
the energy  $\hbar^2/ma^2$; one has, in fact, $\eta=1.59 (k_BT_c^0)^{1/5}
(\hbar^2/ma^2)^{-1/5}$. This  expression reveals that in the 
thermodynamic  limit, where $N\to\infty$ and $\omega_{\rm ho}\to 0$ 
with $N\omega_{\rm ho}^3$ kept fixed, the parameter $\eta$ has a 
well defined value.

In the absence of the condensate ($T>T_c$) interaction effects
are less important  because the system is
very dilute. In this case one can  estimate the interaction energy
using the expression $E_{\rm int}/N\simeq gN/R^3_T$ where $R_T =(2k_BT/m
\omega_{\rm ho}^2)^{1/2}$ is the classical radius  of the thermal cloud.  
For temperatures of the order of $T_c$ one finds,
\begin{equation}
\frac{E_{\rm int}}{N k_BT_c^0} \sim N^{1/6}
\frac{a}{a_{\rm ho}} \sim \eta^{5/2} \; .
\label{eq:ratNT}
\end{equation}
This ratio depends on the interaction parameter $\eta$ through a higher
power law as compared to the analogous ratio for the energy of the
condensate, which is linear in $\eta$ [see Eq.~(\ref{eq:eta})], and 
the effect of $E_{\rm int}$ is hence much smaller for noncondensed atoms.

The above discussion emphasizes the importance of the dimensionless
parameter (\ref{eq:eta}) which can be used to discuss the effects of
interactions on the thermodynamic behavior of the system both at low
and high temperatures. Actually, in Sec.~\ref{sec:therm4} we will
show  that  in the thermodynamic limit the  system exhibits a scaling
behavior on this parameter. 

\subsection{Critical temperature}
\label{sec:therm2}

The first quantity we discuss is the critical temperature. As anticipated 
in the previous section, at the onset of BEC the system is very dilute 
and one does not expect atom interactions to give large corrections 
to thermodynamics.
Nevertheless the role of interactions on critical phenomena is an
important question from a conceptual viewpoint. It is interesting to
understand, in particular, the differences between the behavior of
uniform and nonuniform Bose gases. As concerns the comparison with
experiments, one should however note that finite size corrections to
$T_c$ [see Eq.~(\ref{eq:dtc})] cannot be in general ignored,
being in many cases of the same order as the ones due to interactions.

In the noninteracting model the system can be cooled, remaining in
the normal phase, down to the temperature $T_c^0$ which satisfies the
condition $n(0) \lambda_T^3 = \zeta(3/2) \simeq 2.61$. Here 
$\lambda_T=[2\pi\hbar^2/ (mk_BT)]^{1/2}$ is the thermal wavelength and
$n(0)$ is the density at the center of the trap which, at the critical
temperature, is given by the thermal density (\ref{eq:nThor}).
The presence of repulsive interactions has  the  effect of expanding
the atomic cloud, with a consequent decrease of density. The opposite
happens for attractive forces, which tend to compress the system.  
Lowering (increasing) the  peak  density has  then the consequence of
lowering  (increasing) the value of the critical temperature. This
effect is absent in the case of a uniform gas where
the density is kept fixed in the thermodynamic limit. It consequently
represents a typical feature of trapped Bose gases that is worth
discussing in some detail.

The shift in the critical temperature due to the mechanism described
above can be easily estimated by treating the interaction in mean-field
approximation. The simplest scheme is Hartree-Fock (HF) theory, which
consists in assuming the atoms to behave as ``noninteracting"  bosons 
moving in a self-consistent mean-field: 
\begin{eqnarray}
H_{HF} &=& -\frac{\hbar^2\nabla^2}{2m} + V_{\rm eff}({\bf r})
\nonumber \\
&=& -\frac{\hbar^2\nabla^2}{2m} + V_{\rm ext}({\bf r}) + 2gn({\bf r}) \; ,
\label{eq:Ueff}
\end{eqnarray}
where the last term, $2gn({\bf r})$ is a mean-field generated by the
interactions with the other atoms.  This method has been first applied 
to the study of trapped Bose gases by Goldman, Silvera and Leggett (1981) 
and Huse and Siggia (1982) and it has since been adopted in many papers 
(Bagnato, Pritchard and Kleppner, 1987; Oliva, 1989;
Giorgini, Pitaevskii and Stringari, 1996, 1997a and 1997b ;  Chou, 
Yang and Yu, 1996; Minguzzi, Conti and Tosi, 1997; Shi and Zheng, 1997b).

In Eq.~(\ref{eq:Ueff}) the quantity $n({\bf r})$ is the total density of 
the system, the sum of the density of both the condensate and thermal 
components. The single-particle energies and the density $n({\bf r})$ are 
obtained by solving a Schr\"odinger equation with a density-dependent
effective potential.  In the presence of  Bose-Einstein  condensation, the
equations for the single-particle excitations are coupled to the equation
for the order parameter and the whole set of equations must be solved using
a self-consistent procedure.  At zero temperature the Hamiltonian
$H_{HF}$ coincides with the  single-particle Hamiltonian
(\ref{eq:HFhamiltonian}), which describes the excitations of time dependent
Gross-Pitaevskii theory after neglecting the quasi-particle amplitude
$v_j$ in the equations of motion (see discussion in
Sec.~\ref{sec:collectivevssingleparticle} ).  

In the semiclassical approximation (see Sec.~\ref{sec:thermodynamiclimit}) 
one can easily calculate the thermal averages over the eigenstates of the 
Hamiltonian (\ref{eq:Ueff}). The thermal density of the system
is given by the ideal gas formula (\ref{eq:nThor})
\begin{equation}
n_T({\bf r}) =
\lambda_T^{-3} g_{3/2}\left(e^{-[V_{\rm eff}({\bf r}) - \mu]
/k_BT}\right) \;,
\label{eq:nrhf}
\end{equation}
where we have replaced $V_{\rm ext}$ with $[V_{\rm eff}-\mu]$.
Bose-Einstein condensation starts at the
temperature for which the normalization condition
\begin{equation}
N = \int d{\bf r} \; n_T({\bf r},T_c,\mu_c)
\label{eq:defTc}
\end{equation}
can be satisfied with the value of the chemical potential $\mu_c$
corresponding to the minimal eigenvalue of the Hamiltonian (\ref{eq:Ueff}).
For large systems the leading contribution arises from interaction effects
\begin{equation}
\mu_c =  2gn(0) \;,
\label{eq:muc2}
\end{equation}
where, working to the lowest order in $g$, one can calculate the central
density $n(0)$ using the noninteracting model.  Equation (\ref{eq:muc2})
ignores finite size effects, given by (\ref{eq:mucritical}) for the ideal 
gas.

By expanding the right hand side of (\ref{eq:defTc}) around 
$\mu_c = 0$ and $T_c=T_c^0$  one obtains the following result for
the shift  $\delta T_c=T_c-T_c^0$ of the critical temperature
(Giorgini, Pitaevskii and Stringari, 1996)
\begin{equation}
\frac{\delta T_c}{T_c^0}  =
-1.3 \frac{a}{a_{\rm ho}} N^{1/6} \;.
\label{eq:deltaTc}
\end{equation}
Equation (\ref{eq:deltaTc})
shows that, to lowest order in the coupling constant, the shift of
$T_c$  is linear in the scattering  length and is negative for repulsive
interactions  ($a>0$). In this case, the ratio (\ref{eq:deltaTc})
can be expressed in
terms of the parameter $\eta$ defined in (\ref{eq:eta}), and one has
$\delta T_c/T_c^0=-0.43\ \eta^{5/2}$. For a typical configuration with
$\eta=0.4$, the shift is $\sim 4$\%;   this can be compared with the 
shift (\ref{eq:dtc}) arising from the finite size correction.  Unlike the 
shift (\ref{eq:deltaTc}) due to interactions, the finite size 
effect (\ref{eq:dtc}) depends on the anisotropy of the trap and decreases
with $N$.  Taking, for example, $N=10^5$ and $\lambda=\sqrt8$ one finds
that finite size effects provide a negative correction of  $\sim 2$\%. For
larger values of $N$ these corrections become negligible and one
can safely  use prediction (\ref{eq:deltaTc}). For attractive
interactions ($a<0$), equation (\ref{eq:deltaTc}) predicts instead a
positive shift.  However, in this case, finite size effects are always
important because the value of $N$ cannot be large.

First measurements of the critical temperature (Ensher {\it et al.}, 1996),
as shown in Fig.~\ref{fig:condfrac-exp}, indicate the occurrence of a 
negative shift with respect to $T_c^0$ by about 6\%,  in agreement with 
the theoretical predictions. However the experimental uncertainties are 
at present too large to draw definitive conclusions from this
comparison. 

Let us conclude this section by recalling that in the mean-field approach
discussed above the relation  between $T_c$ and the critical density in the
center of the trap  remains the same as for the  noninteracting model:
$n(0) \lambda^3_{T_{c}}=\zeta(3/2)=2.61$ [see Eq.~(\ref{eq:nrhf})] and it
is   interesting to  look for effects which violate this relation.
These can be either finite size or  many-body effects  beyond mean-field
theory. These latter effects have been recently calculated  in the uniform gas
 through a Path Integral Monte  Carlo simulation of the homogeneous 
hard-sphere Bose  gas (Gr\"uter, Ceperley and Lalo\"e, 1997). 
This work  has shown that the critical
temperature $T_c$ as a function of the gas parameter $na^3$ first  increases
from the noninteracting value 
$T_c^0=(2\pi\hbar^2/mk_B) [n/\zeta(3/2)]^{2/3}$,
reaches a maximum for $na^3\simeq 0.01$  where  $T_c/T_c^0\simeq1.06$ and
finally decreases for larger values of $na^3$.  For the densities relevant 
for the experiments in traps  the effects on $T_c$ calculated by these 
authors are much smaller than the mean-field correction (\ref{eq:deltaTc}).

\subsection{Below $T_c$}
\label{sec:therm3}

Below the critical temperature $T_c$, Bose-Einstein condensation 
results in a sharp enhancement of the density in the central region 
of the trap. This makes interaction effects much more important than 
above $T_c$, as discussed in Sec.~\ref{sec:therm1}.  In this section we
will be dealing only
with systems interacting with repulsive forces and we will  consider
the limit of large $N$ where finite size effects can be ignored.
The main purpose is to develop a perturbative scheme which permits
one to obtain simple analytic formulas for the temperature dependence
of the condensate fraction and of the energy of the system, providing 
a useful guide to understand the role of two-body interactions. 
We will use the finite temperature Hartree-Fock scheme 
already presented in the previous section. An important result emerging 
from this analysis is that, to lowest order in the
coupling constant, the corrections to the thermodynamic quantities due to
interaction effects are linear in the parameter $\eta$ defined in
(\ref{eq:eta}). A more complete analysis of the thermodynamic behavior,
based on  self-consistent numerical calculations will be presented in
Sec.~\ref{sec:therm4}.

A first important  problem concerns the  temperature dependence
of the order parameter and of the chemical potential. As long as
$N_0(T)a/a_{\rm ho} \gg 1$ and one ignores the interaction with the thermal
component, the Thomas-Fermi approximation (\ref{eq:rhoTF})
to the GP  equation provides  a good description for the condensate also at
$T>0$.  Equation (\ref{eq:muTF}) then permits one to estimate the temperature
dependence of the chemical potential whose value is fixed by the
number of atoms in the condensate. One can write
\begin{equation}
\frac{\mu(N_0,T)}{k_BT_c^0}\simeq\frac{\mu(N,T=0)}{k_BT_c^0}
\left(\frac{N_0}{N}\right)^{2/5}=\eta (1-t^3)^{2/5} \; .
\label{eq:chpot}
\end{equation}
In order to express the condensate fraction in terms of the reduced
temperature $t= T/T_c^0$ we have used the noninteracting  prediction
$N_0 = N(1-t^3)$. Inclusion of corrections to this law would yield
higher order effects in the interaction parameter. Equation 
(\ref{eq:chpot}) provides a  useful estimate of $\mu$, which is
expected to be accurate in the range $\mu<T<T_c^0$.
For smaller temperatures,  Eq.~(\ref{eq:chpot}) misses the
thermal contributions arising from collective excitations. These
effects represent however very small corrections and will be
ignored in the present discussion.

Concerning the uncondensed atoms, at high temperature 
they can be treated  as free particles governed by the effective 
mean-field potential $V_{\rm eff}({\bf r})$ given by (\ref{eq:Ueff}).
The form of this potential can be simplified by ignoring the
contribution  to the density $n({\bf r})$ due to the dilute thermal
component and by evaluating the condensate density
in the Thomas-Fermi approximation. This yields the simple result
$V_{\rm eff}({\bf r})-\mu=
|V_{\rm ext}({\bf r})-\mu|$ [see also Eq.~(\ref{eq:Ueff})]. 
In practice most of the thermal atoms occupy
regions of space lying outside the condensate where $V_{\rm ext}>\mu$ and
$V_{\rm eff}=V_{\rm ext}$.   As a consequence, to a first approximation
the effective potential felt by thermal atoms is the same as without
interaction. However, this does not mean that interaction effects are
negligible. In fact, these atoms have a chemical potential
(\ref{eq:chpot})  quite different from the noninteracting value
and the corresponding contribution to  the thermodynamic averages
is modified.

Let us first  discuss the problem of  thermal depletion. Using the
semiclassical picture one can write
\begin{equation}
N_T = \int \frac{d{\bf r} d{\bf p}}{(2\pi\hbar)^3}
\left\{ \exp[ (p^2/2m +
V_{\rm eff}({\bf r}) - \mu)/k_BT ] -1 \right\}^{-1} \;.
\label{eq:NT}
\end{equation}
Explicit integration of (\ref{eq:NT}), using the Thomas-Fermi 
approximation for the effective mean-field potential, 
$V_{\rm eff}({\bf r})-\mu=|V_{\rm ext}({\bf r})-\mu|$, 
leads to the result 
\begin{equation}
\frac{N_0}{N}
= 1 - t^3 - \frac{\zeta(2)}{\zeta(3)} \eta t^2 (1-t^3)^{2/5} \; ,
\label{eq:cdfr}
\end{equation}
valid to the lowest order in the interaction parameter $\eta$.

Equation (\ref{eq:cdfr}) shows that the effects of the interaction
 depend linearly  on $\eta$
and are consequently expected to be much larger  than the ones in the
shift of the critical temperature  (\ref{eq:deltaTc}) which behave
like $\eta^{5/2}$.  For example, taking $\eta=0.4$ and $t=0.6$ one finds
that interactions reduce the value of $N_0$ by 20\% as compared to the
prediction of the noninteracting model.  It is worth noting that  the
quenching of the condensate  represents a peculiar behavior of
trapped Bose gases and takes place because the thermal component of the 
gas is, in large part, spatially
separated from the condensate. In a uniform  system one has an opposite
mechanism. In fact, in this case, the condensate  and the thermal components
completely overlap and the effective  potential is enhanced due to the
interaction  term $2gn$ [see Eq.~(\ref{eq:Ueff})].
This effect is only partially compensated by the presence of the
chemical potential and the final result is a suppression of the
thermal component.

In a similar way one can calculate the energy of the system. The main
effects of temperature and interactions are twofold: on the one hand the 
number of atoms in the condensate is reduced at finite temperature and
the density in the central region of the trap decreases. As a consequence
the atom cloud becomes larger but more dilute and the interaction energy
is reduced as compared to the zero temperature case. On the other hand 
the particles out of the condensate are
thermally distributed with a modified Bose factor  as in (\ref{eq:NT}).
By explicitly calculating the two contributions, one finds that the total
energy of the system  exhibits  the following temperature dependence:
\begin{equation}
\frac{E}{Nk_BT_c^0} = \frac{3\zeta(4)}{\zeta(3)} t^4 +
\frac{1}{7} \eta (1-t^3)^{2/5}
(5 + 16 t^3) \;.
\label{eq:entot}
\end{equation}
Notice that the contribution of the interaction, which is again linear
in $\eta$,  can be obtained directly starting from result (\ref{eq:chpot})
for the chemical potential, through the use of general relations of
thermodynamics. Analogously to Eqs.~(\ref{eq:chpot}) and (\ref{eq:cdfr}),
expression (\ref{eq:entot}) is valid in the temperature regime $\mu<T<T_c$
and to the lowest order in the parameter $\eta$.  

Another useful quantity  is the release energy, which
corresponds to the energy of the system after switching off the trap.
Using the same approximations as discussed above,  one can easily calculate
also this quantity, for which one finds the result
\begin{equation}
\frac{E_{\rm rel}}{Nk_BT_c^0} \equiv \frac{E-E_{\rm ho}}{Nk_BT_c^0}
= \frac{3\zeta(4)}{2\zeta(3)} t^4 + \frac{1}{7} \eta (1-t^3)^{2/5}
\left( 2+\frac{17}{2}t^3 \right) \;.
\label{eq:enrel}
\end{equation}
The release energy can be extracted from time of flight measurements and, 
consequently, equation (\ref{eq:enrel}) provides a useful
formula to check the  effects of two-body interactions
directly from experiments.

\begin{figure}[t]
\epsfysize=8cm
\hspace{3cm}
\epsffile{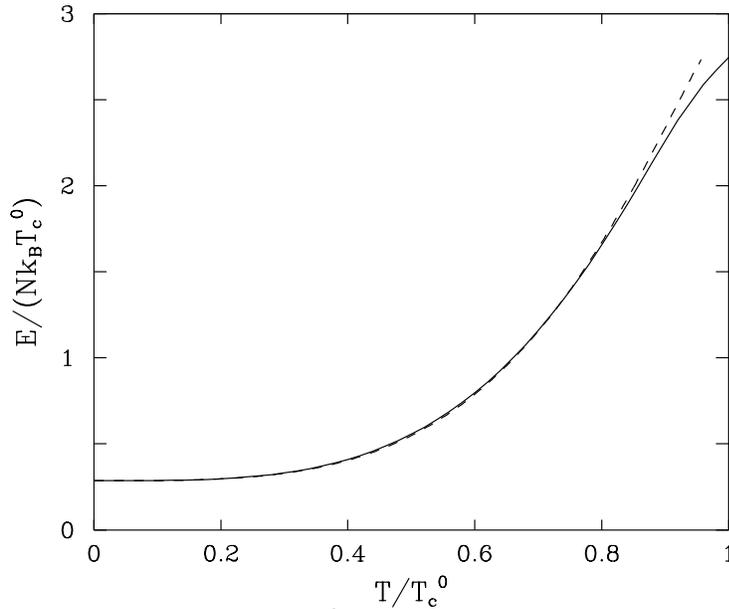}
\caption{ Energy per particle as a function of $T/T_c^0$ for $\eta=0.4$. 
The solid line refers to the perturbative expansion (\protect\ref{eq:entot}); 
the dashed line is the result of the self-consistent calculation based on 
the Popov approximation 
[see Eqs.~(\protect\ref{eq:popov})-(\protect\ref{eq:popov2})].} 
\label{fig:s-energy}
\end{figure}

The formulas presented in this section account for first order effects
in the coupling constant $\eta$. Their validity is ensured only for relatively
high  temperatures and weakly interacting gases. In order to appreciate the
accuracy of these predictions, in Fig.~\ref{fig:s-energy} we compare
the energy predicted in (\ref{eq:entot}) with the one obtained by means of
a self-consistent calculation based on the Popov approximation (see
Sec.~\ref{sec:therm5}).  The agreement is excellent over a wide range of
temperatures except, of course, very close to $T_c^0$ where higher orders 
in $\eta$ give the leading contribution of two-body interactions to 
thermodynamics. 

Expansions similar to the ones discussed in this section can be carried out 
also in the opposite limit of low temperature $t<\eta$, which is the 
analog of the phonon regime of uniform superfluids. Though this 
regime is not easily reachable in current experiments, its theoretical
investigation is rather interesting. The low temperature properties of 
trapped Bose gases are deeply influenced by the thermal excitation of 
the single-particle states localized near the surface of the condensate,
already discussed in Section \ref{sec:collectivevssingleparticle} 
(see also, Giorgini, Pitaevskii and Stringari, 1997b).

\subsection{Thermodynamic limit and scaling}
\label{sec:therm4}

The thermodynamic limit for the noninteracting gas confined in
harmonic traps has been discussed in Sec.~\ref{sec:thermodynamiclimit}.
This limit is reached by letting the total number of particles $N$ increase 
to infinity and the  oscillator frequency $\omega_{\rm ho}$ decrease to
zero, with the product $\omega_{\rm ho}N^{1/3}$  kept fixed. Here
$\omega_{\rm ho}$ is the geometrical average of the three frequencies.
This procedure provides  a natural extension of the usual
thermodynamic limit used in uniform systems where one takes
$N\to\infty$, $V\to\infty$, and keeps the density $n=N/V$  fixed.
In harmonic traps the quantity $\omega_{\rm ho}N^{1/3}$  represents,
together with $T$, the relevant thermodynamic parameter of the
system and replaces the role played by the density in  uniform
systems. In particular it fixes the value of the critical temperature
$k_BT_c^0=\hbar\omega_{\rm ho} N^{1/3}/(\zeta(3))^{1/3}$.
In the thermodynamic limit   all the thermodynamic properties of
the noninteracting model can be expressed  in terms of
the critical temperature $T_c^0$ and the reduced temperature
$t=T/T_c^0$.  Of course dimensionless quantities, like the condensate
fraction or the entropy per particle, will depend only on $t$.

The thermodynamic limit discussed above applies also in the presence
of repulsive interactions. As discussed in Sec.~\ref{sec:therm1}
the parameter $\eta$ depends on $N$ and
$\omega_{\rm ho}$ only through the transition temperature $T_c^0$  and
is consequently well defined in the
thermodynamic limit. 
It is also worth noticing that the dimensionless parameter
$Na/a_{\rm ho}$, which characterizes the effects of two-body interactions
in the Gross-Pitaevskii equation for the ground state behaves as
$Na/a_{\rm ho} \sim N^{5/6} \eta^{5/2}$.
Thus, in  the   thermodynamic limit,  the condition 
$N_0 a/a_{\rm ho}\gg 1$ which ensures the validity of the  Thomas-Fermi 
approximation for the condensate, is always guaranteed below $T_c$.

The above discussion suggests that in the thermodynamic limit the
relevant functions of the system will depend on $T_c^0$, $t$ and
$\eta$ (Giorgini, Pitaevskii and Stringari, 1997a). In the previous section
we have already anticipated  such a behavior by calculating some
relevant thermodynamic functions  to the lowest order  in $\eta$,
as done in Eqs.~(\ref{eq:chpot}), (\ref{eq:cdfr})-(\ref{eq:enrel}).
This points out a scaling behavior exhibited by these systems.
Different configurations, corresponding to different values of
$N$, $m$, trapping frequencies and scattering length, will be
characterized by the same thermodynamic behavior provided they
correspond to the same value of $\eta$.

The scaling behavior can
be  proved in a  general way by noting that in  the
limit $\omega_{\rm ho}\to 0$ the size of the system increases and the
density $n({\bf r})$ changes very slowly. As a consequence in the
thermodynamic limit the density is fixed by the condition (\ref{eq:local})
of local equilibrium, $\mu(T)=\mu_{\rm local}(\bar{n},T)+
V_{\rm ext}({\bf r})$, where
$\mu_{\rm local}(\bar{n},T)$ is the chemical potential of an interacting
uniform system at  density $\bar{n}=n({\bf r})$.  By inverting the
above condition one can write  the density of the gas in the form
 $n({\bf r})=\bar{n}(\mu-V_{\rm ext}({\bf r}),T)$, where $\bar{n}(\mu,T)$ is
the density of the uniform gas as a function of chemical potential 
and temperature. Notice that the inversion of the function
$\mu_{\rm local}(\bar{n})$ requires that $\mu_{\rm local}$ be a 
monotonous function
of the density. This condition is satified by interacting systems
where the stability condition implies $\partial \mu/\partial n >0$. 
The total number of atoms $N$ is obtained by integrating the density over 
space  co-ordinates.   
Introducing the new variable $\xi=V_{\rm ext}({\bf r})$,
one  can write  
\begin{equation}
N=2\pi
\left(\frac{2}{m\omega_{\rm ho}^2}\right)^{3/2} \int_0^\infty d\xi \
\bar{n}(\mu-\xi,T)\sqrt{\xi} \;.
\label{eq:scal}
\end{equation}
The parameters of the trap enter  the above equation through the
combination  $\omega_{\rm ho}N^{1/3}\propto T_c^0$. On the other hand
the integral on the right hand side  requires the knowledge of the
density of the uniform system as a function of chemical
potential and $T$.  Equation (\ref{eq:scal}) shows that the
knowledge of the thermodynamic behavior of the interacting uniform
system would permit one, in principle, to determine 
the thermodynamics of the
trapped  gas (Damle {\it et al.}, 1996). For a dilute Bose gas, where the
interaction  is accounted for by a single parameter (the scattering length
$a$), the integral (\ref{eq:scal}) depends on the quantities $\mu$,
$T$ and   $\hbar^2/ma^2$, the latter being the only energy  one
can construct  with the mass $m$ and the scattering length $a$. As
a consequence, inversion of (\ref{eq:scal}) yields the following
general dependence for the chemical potential $\mu = \mu(T, T_c^0,
\hbar^2/ma^2)$. Due to dimensionality arguments the above expression
can be always written in the form
\begin{equation}
\mu=k_BT_c^0
f(t,\eta) \; .
\label{eq:scal1}
\end{equation}
Here $f$ is a dimensionless function depending on the reduced
temperature $t=T/T_c^0$ and on the scaling parameter $\eta$, fixed by
the ratio between $k_BT_c^0$ and $\hbar^2/ ma^2$  [see discussion
after Eq.~(\ref{eq:eta})].  A similar scaling behavior applies to
the other thermodynamic functions.

\begin{figure}[t]
\epsfysize=7cm
\hspace{3cm}
\epsffile{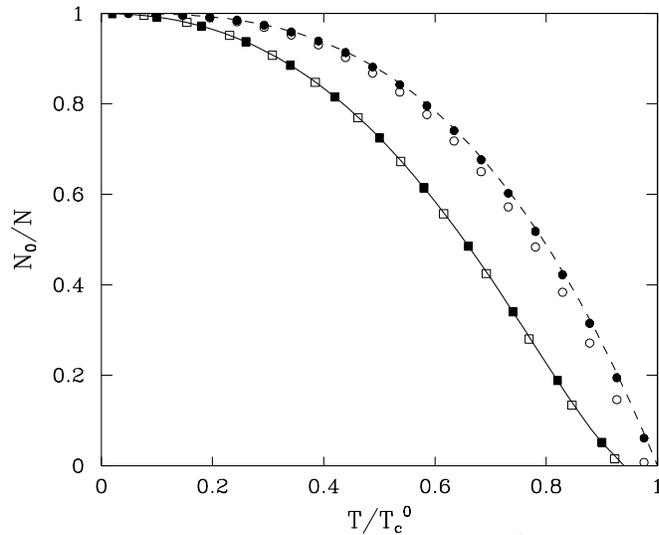}
\caption{Theoretical predictions for the condensate fraction {\it vs.} 
$T/T_c^0$ for interacting (squares) and noninteracting (circles) 
particles in two different traps. Concerning interacting particles, we
show here the results obtained from a self-consistent mean-field 
calculation, within Popov approximation, for $N=5\times 10^4$ rubidium 
atoms in a trap with $a/a_{\rm ho}= 5.4\times 10^{-3}$ and $\lambda=
\protect \sqrt{8}$ (open squares) and for $N=5\times 10^7$ sodium atoms 
in a trap with $a/a_{\rm ho}=1.7\times 10^{-3}$ and $\lambda=0.05$ (solid
squares). The numerical results are compared with the prediction 
of the scaling limit for $\eta=0.4$ (solid line). Open and solid 
circles correspond to $N=5\times 10^4$ and $N=5\times 10^7$ 
noninteracting particles, respectively, in the same two traps as
the corresponding open and solid squares.  The dashed 
line is the $1-t^3$ curve of the noninteracting model in the 
thermodynamic limit.  }
\label{fig:condfracscaling}
\end{figure}

The above discussion applies to the thermodynamic limit where
$N\to\infty$. An important question is to understand whether in the
available experimental conditions, where $N$ ranges between $10^4$
and $10^7$, this limit is reached in practice or finite size
effects are still significant. In Fig.~\ref{fig:condfracscaling}
we show, as an example, the behavior of the condensate fraction.
This quantity depends, in the thermodynamic limit, only on the
variables $t$ and $\eta$. In the figure we plot the numerical results
obtained   from a self-consistent mean-field calculation based on the
Popov approximation  (see Sec.~\ref{sec:therm5}) for two very different
configurations,  both corresponding to the same value of $\eta =0.4$.
Open squares correspond to  $N=5\times 10^4$ rubidium 
atoms in a trap with $a/a_{\rm ho}= 5.4\times 10^{-3}$ and $\lambda=
\protect \sqrt{8}$, while solid squares correspond to $N=5\times 10^7$ 
sodium atoms in a trap with $a/a_{\rm ho}=1.7\times 10^{-3}$ and 
$\lambda=0.05$.  One sees that both set of data 
coincide with the asymptotic scaling function (solid line), calculated 
with the same value of $\eta$, by taking the thermodynamic limit 
in the equations of the Popov approximation 
(Giorgini, Pitaevskii and Stringari, 
1997b). The figure points out, in an explicit way, how very different
configurations can  give rise to the same thermodynamic behavior,
if the corresponding scaling parameter $\eta$ is the same.  It is also
interesting to notice that the scaling behavior is reached faster in the
presence of two-body  interactions than for noninteracting particles. In
the latter case, in fact, finite size  effects, which are responsible for
the deviations from the scaling law $(1-t^3)$ (dashed line), are  
more visible (open and solid circles). 

The scaling behavior is very well reproduced also by the other
thermodynamic quantities. For this reason, in the next section, we
will  discuss the  behavior of interacting Bose gases confined in
harmonic traps calculating directly the various physical quantities
in  the thermodynamic  limit.

\subsection{Results for the thermodynamic functions}
\label{sec:therm5}

In Sec.~\ref{sec:therm3} we have used Hartree-Fock theory to  estimate the
temperature dependence of the chemical potential, condensate fraction,
release energy to lowest order in the scaling parameter $\eta$. The full
equations of  Hartree-Fock theory can be  solved numerically in a
self-consistent way (Minguzzi, Conti and Tosi, 1997; Shi and Zheng  1997b;
Giorgini, Pitaevskii and Stringari, 1997b) going beyond the perturbative 
scheme. Hartree-Fock theory is expected to be quite accurate at high 
temperature except, of course,  very close to $T_c$, where mean-field 
theories are inadequate (Shi, 1997; Shi and Griffin, 1998). The accuracy 
of  Hartree-Fock at high $T$ is justified by the
crucial role played by single-particle excitations, as we have already
seen in Sec.~\ref{sec:collectivevssingleparticle} for the density
of states.  This theory is instead less accurate at low $T$, since it 
ignores the effects of collectivity which characterize the low energy
part of the excitation spectrum. Such collective effects are instead
properly included in time dependent Gross-Pitaevskii theory, as
discussed in Sec.~\ref{sec:TDGP}.  A mean-field scheme, describing
correctly  both the high and low temperature regimes, is provided by the
so called Popov approximation (Popov, 1965 and 1987; Griffin, 1996; Shi, 
1997) whose  application to interacting
bosons in harmonic traps  has been considered by several authors in the
last few years.  This mean-field scheme is based,  on the one hand, on a
finite $T$ extension of the Gross-Pitaevskii equation, in which the
interaction between condensed and noncondensed atoms is explicitly
accounted for, and, on the other hand, on Bogoliubov-type equations
for the excitations of the system. The corresponding
equations have the form
\begin{equation}
\left(-{\hbar^2\nabla^2 \over 2m} + V_{\rm ext}({\bf r}) +
g[n_0({\bf r})+2n_T({\bf r})]\right)\phi=\mu\phi
\label{eq:popov}
\end{equation}
and
\begin{eqnarray}
\varepsilon_i u_i({\bf r}) &=& \left( -\frac{\hbar^2\nabla^2}{2m}
+V_{\rm ext}({\bf r})
-\mu+2gn({\bf r})\right) u_i({\bf r}) + gn_0({\bf r})v_i({\bf r})
\label{eq:popov1}
 \\
- \varepsilon_i v_i({\bf r}) &=& \left( -\frac{\hbar^2\nabla^2}{2m}
+V_{\rm ext}({\bf r})
+\mu+2gn({\bf r})\right) v_i({\bf r}) + gn_0({\bf r})u_i({\bf r})
\label{eq:popov2}
\end{eqnarray}
where $n_0({\bf r})=|\phi({\bf r})|^2$ is the condensate density, while
the thermal density $n_T$ is calculated  through the relation
$n_T = \sum_j(|u|_j^2 +|v|_j^2) [\exp(\beta\varepsilon_j)-1]^{-1}$
with $u_j$, $v_j$ and $\varepsilon_j$ solutions of 
(\ref{eq:popov1})-(\ref{eq:popov2}). These quantities are now
temperature dependent. The sum  $n({\bf r})=n_0({\bf r})+n_T({\bf r})$ 
is the total density. The functions $u_i$, $v_i$ entering 
Eqs.~(\ref{eq:popov1})-(\ref{eq:popov2}) are normalized according 
to condition (\ref{eq:normalizationuandv}). Notice that in this 
approximation the thermal component is  treated as a
thermal bath generating an additional static external field in the
equation for the condensate. One also ignores here the $T=0$ quantum 
depletion $n_{\rm out}({\bf r}) = \sum_j |v_j({\bf r})|^2$ which 
has been  shown to be very small in these trapped gases 
(see Sec.~\ref{sec:corrections}). Finally, at low temperature, 
when $n_T$ is negligible compared with $n_0$, equation 
(\ref{eq:popov}) coincides with the stationary GP equation 
(\ref{eq:GP}), while Eqs.~(\ref{eq:popov1})-(\ref{eq:popov2})
reduce to (\ref{eq:coupled1})-(\ref{eq:coupled2}). 

\begin{figure}[t]
\epsfysize=9cm
\hspace{3cm}
\epsffile{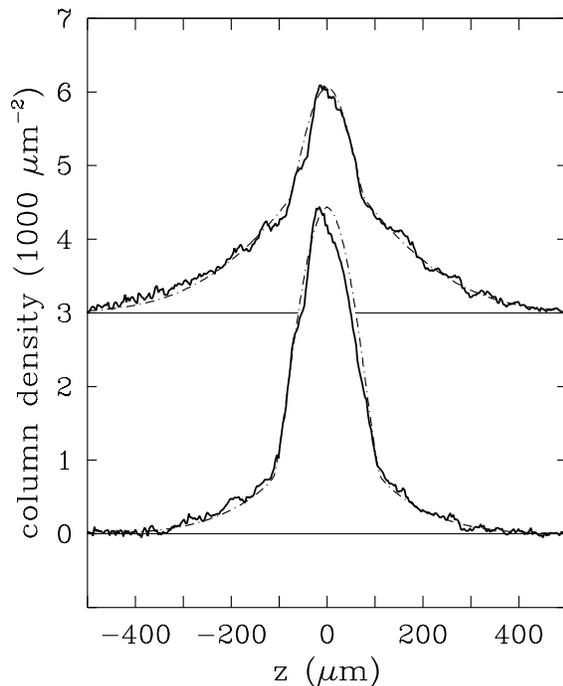}
\caption{ Axial profiles of a cloud of sodium atoms. The thick solid 
lines are two of the profiles already shown in 
Fig.~\protect\ref{fig:profiles-T}, obtained at MIT (Stamper-Kurn and 
Ketterle, 1998) from phase-contrast images at different temperatures: 
$T=0.7 \mu$K (lowermost), $T=1.2 \mu$K (uppermost). Dot-dashed lines 
are theoretical predictions obtained from 
Eqs.~(\protect\ref{eq:popov})-(\protect\ref{eq:popov2}), using 
$N$ and $T$ as fitting parameters: the lower curve corresponds to 
$N=1.4 \times 10^7$ and $T=0.8$ $\mu$K, and the upper one to  
$N=2.3 \times 10^7$ and $T=1.1$ $\mu$K. In both cases,  a difference 
in temperature of about $10$ \% \ between the experimental estimate 
and the result of the fit is consistent with the experimental 
uncertainty.  } 
\label{fig:profiles-T-theo}
\end{figure}

From the solution of 
Eqs.~(\ref{eq:popov})-(\ref{eq:popov2}) one obtains density profiles
in good agreement with experimental data. As an example, we
take two of the density profiles already shown in 
Fig.~\ref{fig:profiles-T} and we plot them again in 
Fig.~\ref{fig:profiles-T-theo} together with the theoretical 
prediction from Eqs.~(\ref{eq:popov})-(\ref{eq:popov2}), using the
number of particles, $N$, and the temperature, $T$, as fitting parameters.  
The same equations have been used also for fitting the experimental data 
by Hau {\it et al.} (1998).  

Using the distribution function of the excited states,
$f_j=[\exp(\beta\varepsilon_j)-1]^{-1}$, and the combinatorial
expression for the entropy, $S=k_B\sum_j\{\beta \varepsilon_jf_j-
\ln [1-\exp(-\beta\varepsilon_j)]\}$, one can work out all
the thermodynamic quantities (Giorgini, Pitaevskii and Stringari, 1997b).
The  comparison between the predictions of Hartree-Fock and Popov
theories has revealed that there are no significant differences
between the two approaches for most thermodynamic  quantities. 
Only  increasing the value of the  interaction parameter  $\eta$ one
can observe some differences. This is a further evidence of the
negligible role played in thermodynamics by the collective modes
of the condensate, which are ignored in the Hartree-Fock scheme. 
This behavior, in accordance with the analysis of the density of
states made in Sec.~\ref{sec:collectivevssingleparticle},  
represents a significant difference with respect to the case
of  uniform superfluids.

In the following we present  results for various thermodynamic
quantities (chemical potential, condensate fraction and release
energy) obtained using the Popov approximation in the thermodynamic
limit.  As discussed in the preceeding section, this limit is well
achieved   in the configurations realized in present experiments.
The results  are consequently presented as a function of the reduced
temperature  $t$ for different values of the scaling parameter $\eta$. We
have considered  the value $\eta=0.4$, which corresponds to the typical
configurations  realized in actual experiments, and the value $\eta=0.6$,
which would  correspond to a more correlated gas.

\begin{figure}[t]
\epsfysize=8cm
\hspace{3cm}
\epsffile{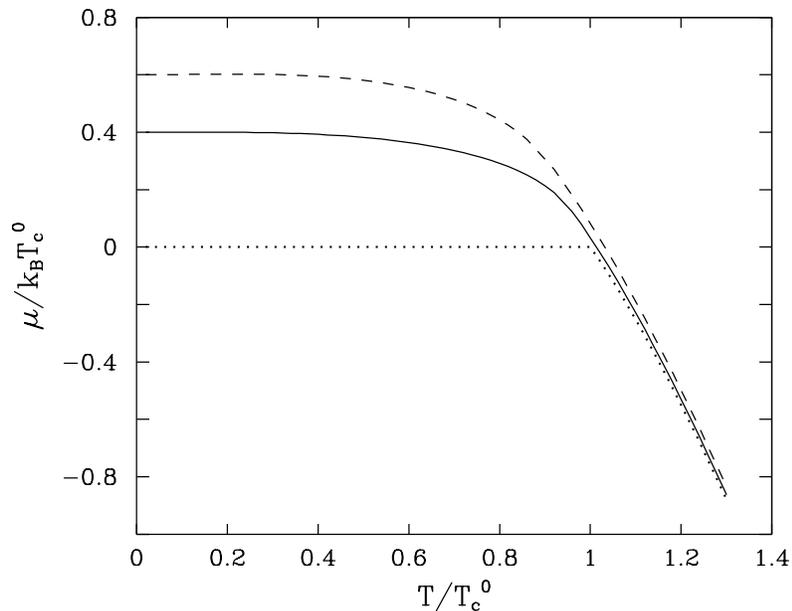}
\caption{Chemical potential as a function of $T/T_c^0$ in the 
thermodynamic  limit.  Solid line: $\eta=0.4$, dashed line: $\eta=0.6$.
The dotted line refers to the non-interacting model ($\eta=0$).} 
\label{fig:s-mu}
\end{figure}

First, in Fig.~\ref{fig:s-mu}, we show the chemical potential in units
of $k_BT_c^0$ as a function of the reduced temperature $t$. Notice that 
for  $t\to 0$ the plotted quantity coincides, by definition,
with the  parameter $\eta$ [see definition (\ref{eq:eta})].  In the
classical limit, $T\gg T_c^0$, the chemical potential approaches the
ideal gas value  $\mu/k_BT_c^0 = t\ln (\zeta(3)/t^3)$.

\begin{figure}[t]
\epsfysize=8cm
\hspace{3cm}
\epsffile{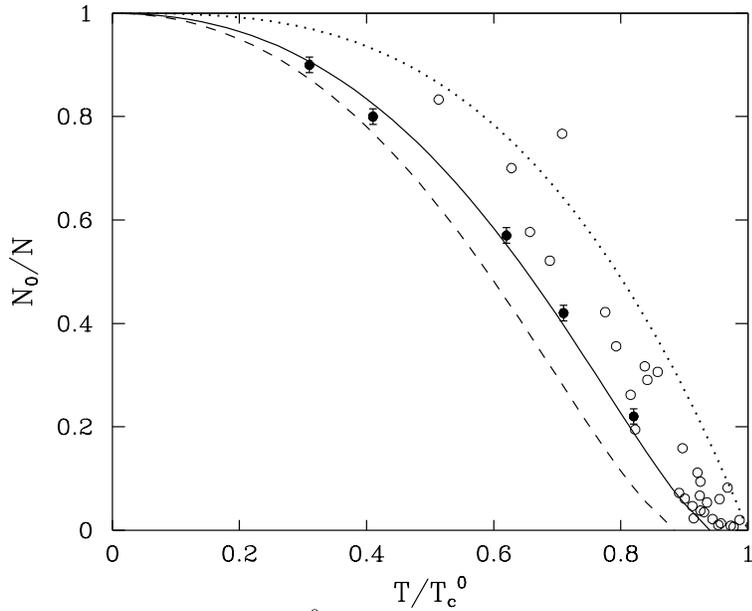}
\caption{Condensate fraction as a function of $T/T_c^0$ in the 
thermodynamic limit. As in Fig.~\protect\ref{fig:s-mu}, the curves 
are the theoretical predictions for $\eta=0.4$ (solid line), 
$\eta=0.6$ (dashed line) and the noninteracting case $\eta=0$
(dotted line).  Open circles are the experimental data by Ensher 
{\it et al.} (1996), corresponding to $\eta$ in the range 
$0.39-0.45$.  Solid circles with error bars are the path
integral Monte Carlo results by Krauth (1997), with $\eta=0.35$. } 
\label{fig:s-condfrac2}
\end{figure}

The results for the condensate fraction $N_0/N$ are given in 
Fig.~\ref{fig:s-condfrac2}. The open circles are the experimental 
points (Ensher {\it et al.}, 1996). In the experiment the number of 
atoms $N$ varies with $t$ and the corresponding value of $\eta$ ranges 
from 0.39 to 0.45.  The data are compared with the predictions of the 
mean-field theory for $\eta=0.4$ (solid line) and the noninteracting gas 
model (dotted line). The experimental points are shifted from the
noninteracting case to lower temperature, but not as much as predicted by 
mean-field theory. However, the experimental uncertainties are 
still too large to draw any definitive conclusion. 

\begin{figure}[t]
\epsfysize=8cm
\hspace{3cm}
\epsffile{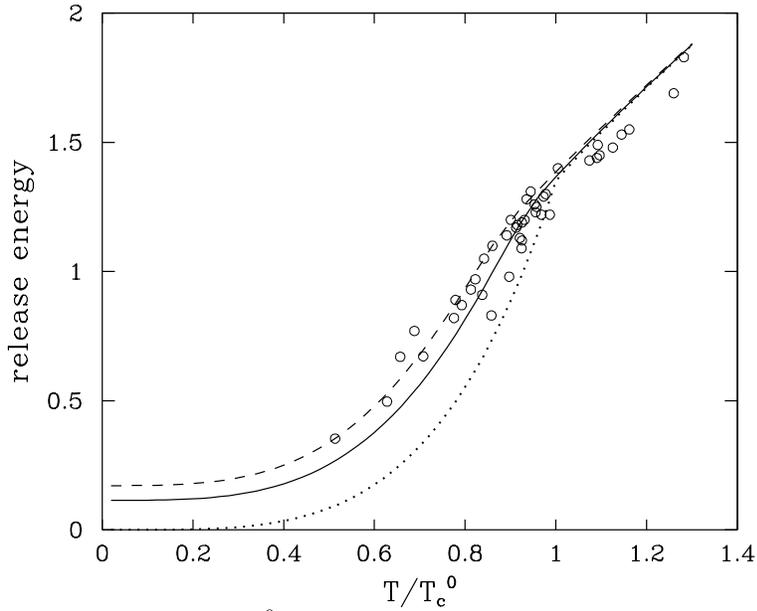}
\caption{Release energy as a function of $T/T_c^0$ in the 
thermodynamic limit. The curves refer to the same values of $\eta$ 
as in  Fig.~\protect\ref{fig:s-mu}.  Circles are the experimental 
data by Ensher {\it et al.} (1996).} 
\label{fig:s-release}
\end{figure}

In Fig.~\ref{fig:s-release} we show the results for the release energy. 
The dots are the experimental data (Ensher {\it et al.}, 1996) which, 
below $T_c$, lie well above the noninteracting curve, showing again 
a clear evidence for the effects of two-body interactions. We notice 
that, by differentiating the total energy with respect to $T$, one 
could calculate the specific heat. In the present experiments however, 
this quantity  is not directly available because only the release 
energy is measured.  What one can see from a fit to the experimental 
data of the release energy is the occurrence of a characteristic  
bump in the derivative near the transition temperature  (Ensher 
{\it et al.}, 1996); this behavior is in good agreement with the 
prediction obtained by taking the derivative of the theoretical 
curves in Fig.~\ref{fig:s-release}.

Finally, the above results of the mean-field theory at finite 
temperature can be also compared with the ones of Quantum Monte 
Carlo calculations (Krauth, 1996; Holzmann, Krauth and Naraschewski, 
1998). It is worth noticing that the possibility of  making a close 
comparison between exact Monte Carlo simulations, experimental data 
and mean-field calculations is a rather rare event in the context 
of interacting many-body systems 
and represents a further nice feature of BEC in traps. In 
Fig.~\ref{fig:s-condfrac2}, the condensate fraction obtained with 
path integral Monte Carlo calculations by Krauth (1996) is represented 
by solid circles with error bars. The simulation has been done with 
$10000$ atoms interacting through an hard-core potential, and corresponds
to $\eta=0.35$. The results are very close to the mean-field prediction.
The value of $\eta$ used for the solid curve is actually $0.4$, since 
this value is closer to the experimental situation, but the same 
calculation for $\eta=0.35$ gives an even better agreement, crossing 
precisely the three Monte Carlo data at high temperature. A detailed 
comparison between Monte Carlo results and mean-field theory has been 
recently performed by Holzmann, Krauth and Naraschewski (1998), including 
the analysis of the density profiles of the gas at different temperatures.

\subsection{Collective modes at finite temperature}
\label{sec:finiteTexcitations}

In Sec.~\ref{sec:dynamics} we have studied the collective excitations
of a trapped Bose gas at zero temperature. In this case, all the atoms
are in the condensate and there are no collisions. In the collisionless
regime the force acting on a given particle comes from the mean-field
created by the other particles; this field generates a collective
oscillation of the system, which is sometimes called Bogoliubov sound 
and is the analog of {\it zero sound} for normal Fermi liquids. 

At finite temperature the situation is more complicated. On the one
hand both the condensate and the thermal cloud can oscillate.
On the other hand collisions between excitations can play an
important role and one must distinguish between a collisional  and a
collisionless  regime. 

So far the temperature dependence of these 
oscillations has been analysed experimentally both at JILA and MIT. 
In the first case, Jin {\it et al.} (1997) investigated the $m=0$ 
and $m=2$ modes for a system of $\sim 10^4$ Rb atoms in a trap with 
asymmetry parameter $\lambda=\sqrt{8}$. At low temperatures 
($T\stackrel{<}{\sim}0.4 T_c^0$) they found that the oscillations of 
the condensate have frequencies in good agreement with the predictions 
of the  $T=0$ Gross-Pitaevskii equation (see Sec.~\ref{sec:dynamics}), 
while at higher temperatures the frequency exhibits an unexpected 
temperature dependence with very different 
behavior for the two modes. In the second case, Stamper-Kurn {\it et al.} 
(1998c) studied the low-lying $m=0$ mode for a much larger system ($N\sim 
10^7$ Na atoms) in a cigar-shaped trap with $\lambda \ll 1$. Similarly
to the JILA group, they observed a shift of the collective frequency 
with respect to the $T=0$ value. Both groups measured also  the
damping of the collective modes of the condensate, finding a rather 
strong $T$-dependence. They also observed the oscillations of the 
thermal cloud below and above the critical temperature. 

Mean-field approaches have been used to predict the properties of the 
collective excitations in the collisionless regime. This regime is achieved 
at low temperatures and for low densities of the thermal cloud. In this case, 
the oscillations of the condensate behave similarly to the $T=0$ case and can 
still be called Bogoliubov's sound modes.  Several authors have used 
finite $T$ extensions of the time-dependent Gross-Pitaevskii equation
in the  Popov approximation [see Eqs.~(\ref{eq:popov})-(\ref{eq:popov2})]
(Hutchinson, Zaremba and Griffin, 1997; Dodd {\it et al.}, 1998a and 1998b).
This approach, in which  the thermal component  is treated as a static
thermal bath,  does not account for any damping mechanism. In order to
include damping, a dynamic description of the oscillations of both the 
condensate and the thermal cloud is needed (Minguzzi and Tosi, 
1997; Giorgini, 1998). The dynamic coupling between the motion of the two
components might also be important in the determination of the temperature
dependence of the frequency shift. There are still open questions on 
this problem and several aspects of the theory are expected to be 
clarified in the next future.  

\begin{figure}[t]
\epsfysize=8cm
\hspace{3cm}
\epsffile{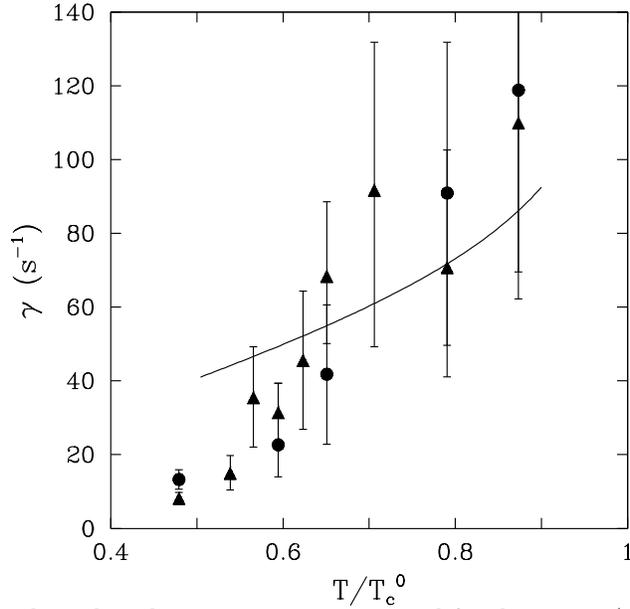}
\caption{Temperature dependent damping rates, $\gamma$, measured 
for the $m=0$ (triangles) and $m=2$ (circles) modes (Jin {\it et al.}, 
1997). The solid line is the theoretical estimate (\protect\ref{eq:damp})
where we have used the Thomas-Fermi approximation for the local sound 
velocity, $c(T)=(\mu(T)/m)^{1/2}$, calculated in the center 
of the trap,  and expression (\protect\ref{eq:chpot}) for the temperature 
dependence  of the chemical potential.  } 
\label{fig:damping}
\end{figure}

Damping processes in the collisionless regime have been investigated 
using  perturbative approaches (Liu
and Shieve, 1997; Pitaevskii and Stringari, 1997; Liu, 1997; Fedichev, 
Shlyapnikov and Walraven, 1998). An important mechanism of collisionless 
damping is provided by the Landau damping. In Fermi liquids, Landau damping 
originates from the coupling between single-particle excitations and 
zero sound at $T=0$ (Lifshitz and Pitaevskii, 1981, \S 30; Pines and 
Nozi\`eres, 1966, Vol.I). In a Bose gas, an analogous damping occurs because 
the thermal bath  of elementary excitations can absorb quanta of the 
collective oscillation. Landau damping  increases with temperature,
because of the larger number of elementary excitations  available at
thermal equilibrium. For temperatures larger than the chemical 
potential  it increases linearly with $T$ and for a uniform system the 
ratio between the imaginary, $\gamma= - {\rm Im}(\omega)$, and real, 
${\rm Re}(\omega)$, parts of the collective frequency takes the analytic 
form (Sz\'{e}pfalusy and Kondor, 1974; Shi, 1997; Pitaevskii and 
Stringari, 1997)
\begin{equation}
\frac{\gamma}{{\rm Re}(\omega)} = \frac{3\pi}{8}\frac{k_BTa}{\hbar c} \; .
\label{eq:damp}
\end{equation}
This equation is valid for excitations with energy $\hbar {\rm Re}
(\omega) \ll \mu \ll k_BT$, which in a uniform system are phonons 
propagating with the sound  velocity $c$. Equation (\ref{eq:damp}) can 
be used for a rough estimate of the damping of collective excitations 
in traps by taking for $c$ the value of the sound velocity in the center 
of the trap. As an example, in Fig.~\ref{fig:damping} we show the damping 
rate, $\gamma$, measured for the $m=0$ and $m=2$ modes by Jin {\it et al.}
(1997), compared with the theoretical estimate (\ref{eq:damp}). Taking 
into account that this estimate, which is expected to apply at high 
temperatures, is very rough and that the effects  of geometry and 
multipolarity are completely neglected, the agreement between theory 
and experiment can be considered reasonable. Recently Fedichev, 
Shlyapnikov and Walraven (1998) have indeed argued that Landau damping 
is strongly influenced by the geometry of the traps, being particularly 
effective in anisotropic traps, due to the randomness of  the excitation 
spectrum. 

At high temperature and/or high density, collisions are more important 
and can affect the nature of collective excitations. In Bose superfluids 
the collisional regime is described by the equations of two-fluid 
hydrodynamics and is characterized by the occurrence of two distinct 
oscillations: first and second sound. In liquid $^4$He, first sound is 
a density wave with {\it in phase} oscillations of the superfluid and 
normal components, while second sound is an almost pure entropy wave 
with {\it opposite phase} motion of the two components. For a dilute Bose 
gas the situation is different because the interaction between the condensate 
and the thermal cloud is very weak. In particular, except at very low 
temperature, first sound mainly involves the thermal cloud and reduces 
to the usual hydrodynamic sound above $T_c$; conversely, second sound is
essentially the oscillation of the condensate and disappears above $T_c$ 
(Lee and Yang, 1959;   Griffin and Zaremba, 1997).  Similar features 
have been pointed out also in the  presence of harmonic trapping (Zaremba,
Griffin and Nikuni, 1997; Shenoy and Ho, 1998). 
 
An important limiting case is represented by the collective motion of the 
gas above the critical temperature, where the gas exhibits an almost 
classical behavior. In the collisional regime, one can then use the 
equations of hydrodynamics for classical gases in order to obtain 
explicit results for the collective frequencies. For example the 
coupled quadrupole and monopole modes with $m=0$ obey the following 
dispersion law (Griffin, Wu and Stringari, 1997; Kagan, Surkov and 
Shlyapnikov, 1997a):
\begin{equation}
\omega^2 = {1\over 3}[5\omega_{\perp}^2 + 4 \omega_z^2
\pm (25 \omega_{\perp}^4 + 16 \omega_z^4 -32
\omega_{\perp}^2\omega_z^2)^{1/2}]  \; .
\label{eq:GriffinS}
\end{equation}
If the trap is spherical ($\omega_{\perp}=\omega_z=\omega_{\rm ho}$),
the two solutions have frequency $\omega =\sqrt{2} \ \omega_{\rm ho}$ 
(quadrupole) and $\omega=2\omega_{\rm ho}$ (monopole). Notice that 
the frequency of the surface (quadrupole) mode is equal to the one of 
the oscillation of the condensate at zero temperature, given by 
Eq.~(\ref{eq:pml}). In the limit of highly deformed cigar-shaped traps 
($\omega_z\ll\omega_{\perp}$), the lowest frequency of (\ref{eq:GriffinS}) 
becomes $\omega= \sqrt{12/5}\ \omega_z$. 

This collisional regime, yielding the dispersion law (\ref{eq:GriffinS}),
is achieved if $\omega \tau \ll 1$, where $\tau$ is a typical collision 
time. In the opposite limit ($\omega \tau \gg 1$) one finds the 
collisionless regime, where the gas oscillates with frequencies fixed by 
the trapping potential, corresponding to the predictions of the 
noninteracting model. Oscillations of 
this type have been observed by Jin {\it et al.} (1997) for both the 
gas above $T_c$ and the thermal component below $T_c$; the frequency 
of the $m=0$ and $m=2$ modes was found to be roughly twice the trap 
frequency, in agreement with the prediction for the ideal gas. 

In the intermediate regime where $\omega\tau\sim 1$ one expects a smooth
cross-over from collisionless to collisional hydrodynamic modes. A useful 
interpolation  formula is provided by  the law (Kavoulakis, Pethick 
and Smith, 1998)
\begin{equation}
\omega^2 = \omega_C^2 + \frac{\omega_{HD}^2-\omega_C^2}{1-i\omega\tau} \;,
\label{eq:freqint}
\end{equation}
predicted by general theory of relaxation phenomena (Landau and Lifshitz, 
1987, \S 81). Here $\omega_C$ and $\omega_{HD}$ are the frequencies of the 
mode in the collisionless and collisional hydrodynamic regimes respectively. 

An estimate of the collisional time  $\tau$ can be obtained by
considering again  a classical picture of the system. One finds $\tau
\simeq l_{\rm mfp}/v_T$, where $v_T=\sqrt{2k_BT/m}$ is the thermal 
velocity and $l_{\rm mfp}=(n_T\sigma)^{-1}$ is the mean free path
which is fixed by the $s$-wave cross section $\sigma=8\pi a^2$ and by
the density. For frequencies of the order of the oscillator frequency
the condition $\omega\tau\ll 1$ is equivalent to requiring that the
mean free path be much  smaller than the thermal radius $R_T=
\sqrt{2k_BT/m\omega_{\rm ho}^2}$. Near $T_c$ one finds that the 
collisional frequency $1/\tau$ behaves like $\eta^5k_BT_c/\hbar$ 
where $\eta$ is the scaling parameter defined in (\ref{eq:eta}). 
Despite the smallness of the factor $\eta^5$, the collisionless 
frequency can be of the same order of the collective frequencies 
of the system because of the factor $N^{1/3}$ contained in $T_c$, 
so that increasing $N$ favours the achievement of the collisional 
hydrodynamic regime. Notice that the
multipolarity of the excitation can play an important role in
characterizing the relaxation of the collective oscillation. For
example the dipole mode cannot have any relaxation mechanism in the
presence of harmonic trapping. The same happens in the case of the
monopole excitation if the harmonic trap is isotropic. In both
cases the collisionless ($\omega_C$) and hydrodynamic ($\omega_{HD}$)
frequencies coincide.

\begin{figure}[t]
\epsfysize=8cm
\hspace{3cm}
\epsffile{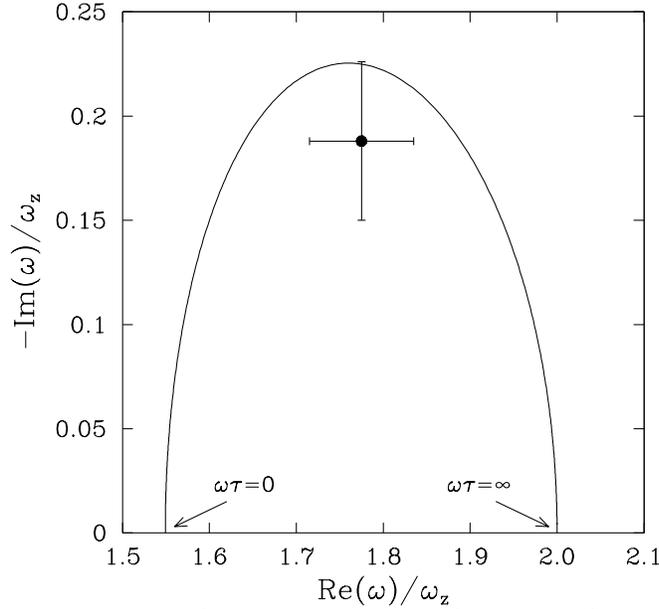}
\caption{The imaginary part of $\omega$ against the real part, as given 
by the interpolation formula (\protect\ref{eq:freqint}), in the  case of 
the low-lying $m=0$ mode observed at MIT (Stamper-Kurn {\it et al.}, 1998c). 
In the collisional hydrodynamic 
regime the frequency of the mode is given by $\omega_{HD}=(12/5)^{1/2}
\omega_z$  [see  Eq.~(\protect\ref{eq:GriffinS})], while in the 
collisionless regime it is given by the noninteracting value 
$\omega_C=2\omega_z$. In both cases the motion is undamped (${\rm Im}
(\omega)=0$).  Stamper-Kurn {\it et al.} (1998c) measured a 
frequency of about $30$ Hz with a damping rate of about $20$ s$^{-1}$; 
the corresponding values ${\rm Re}(\omega) \sim 1.78 \omega_z$ and 
$-{\rm Im}(\omega) \sim 0.19 \omega_z$ are represented by the solid circle.
The theoretical curve near this point corresponds to collision time 
such that ${\rm Re} (\omega) \tau \sim 1$.   } 
\label{fig:s-omega}
\end{figure}

In Fig.~\ref{fig:s-omega} we plot the imaginary part of $\omega$ 
against the real part, as given in (\ref{eq:freqint}), for the 
case of the low-lying $m=0$ mode observed in the cigar-shaped trap
at MIT (Stamper-Kurn {\it et al.}, 1998c). In this experiment, the
thermal cloud is found to oscillate with a frequency of 
$1.78 \omega_z$, which is larger than the hydrodynamic prediction 
$\sqrt{12/5}\ \omega_z$, but lower  than the noninteracting value 
$2\omega_z$. A damping rate of about $20$ s$^{-1}$ was also 
observed, corresponding to $-{\rm Im}(\omega) = 0.19 \omega_z$.  
In Fig.~\ref{fig:s-omega} the experimental results are represented
by the solid circle with error bars, which turns out to be reasonably 
close to the theoretical curve, the difference being of the order of the 
experimental uncertainty. The part of the curve near the maximum 
corresponds to values of collision time such that ${\rm Re}(\omega)
\tau \sim 1$ and this suggests that, differently from the JILA 
experiment (Jin {\it et al.}, 1997), the motion of the thermal 
cloud in the MIT experiment is affected by collisions. 

Finally, we mention that Stamper-Kurn {\it et al.} (1998c) observed also 
the {\it opposite phase} dipolar oscillation of the thermal cloud 
and the condensate, occurring below $T_c$ (Zaremba, Griffin and 
Nikuni, 1998). This mode exhibits strong damping.


\section{Superfluidity and coherence phenomena}
\label{sec:superfluidity}

Superfluidity is one of the most spectacular consequences
of Bose-Einstein condensation. However, the explicit connection between
superfluidity and BEC is not trivial and has been the object of a
longstanding and deep investigation in the last decades, mainly for
its importance in  understanding the physics  of liquid helium.
In macroscopic bodies superfluidity shows up with many peculiar
features: absence of viscosity, reduction of the moment of inertia, 
occurrence of persistent currents, new collective phenomena (second
sound, third sound, etc.), quantized vortices, and others. Several 
properties are usually interpreted as coherence effects associated 
with the phase, $S$, of the order parameter whose gradient fixes 
the velocity of the superfluid through ${\bf v}_s = (\hbar/ m) 
\mbox{\boldmath$\nabla$} S$. A major question is to understand whether 
some of these effects can be observed also in trapped gases. Of course 
in a mesoscopic system one expects the manifestations of superfluidity to
be different from the ones exhibited by macroscopic bodies.
In particular, traditional experiments based on the study of transport
phenomena  are not easily feasible in trapped gases. On the other hand,
interference patterns, associated with phase coherence, have been
already observed (Andrews, Townsend {\it et al.}, 1997) and successfully
compared with theory.  This opens a promising field of research
based on the investigation of coherence phenomena,
including the realization of the so-called ``atom laser" (Ketterle, 1998).

\subsection{Rotational properties: vortices and moment of inertia}
\label{sec:rotational}

Among the several properties exhibited by superfluids,  the occurrence
of quantized vortices and the strong reduction of the moment of inertia
represent effects of primary importance.

In a dilute Bose gas the  structure of  quantized  vortices can be
investigated  starting from the Gross-Pitaevskii equation. Indeed one of
the primary motivations of the GP theory was the study of vortex states
in weakly interacting bosons (Gross, 1961 and 1963; Pitaevskii, 1961). 
These studies were further developed by Fetter (1972)  including higher
order effects in the interaction.

\begin{figure}[t]
\epsfysize=8cm
\hspace{3cm}
\epsffile{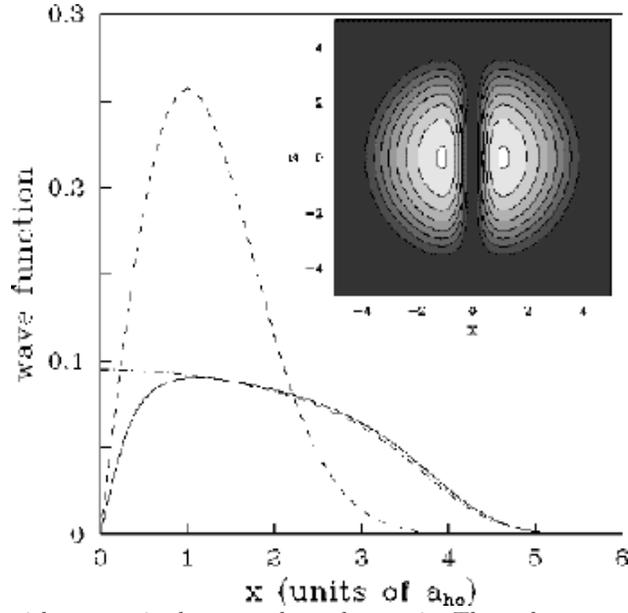}
\caption{Condensate with a quantized vortex along the $z$-axis. The order
parameter, $\phi_v(x,0,0)$, is plotted in the case of $10^4$ rubidium  
atoms confined in a spherical trap with $a_{\rm ho}=0.791$ $\mu$m.  
Distances are in units of the oscillator length $a_{\rm ho}$ and
the curves correspond  to $(a_{\rm ho}^3/N)^{1/2} \phi_v(r)$, so 
that they are normalized to $1$ when $\phi_v$ in normalized to $N$.   
The dot-dashed line is the solution of the stationary GP equation 
(\protect\ref{eq:GP}), or equivalently of Eq.~(\protect\ref{eq:vorgpe})
with $\kappa=0$; the solid line is the profile of a vortex with $\kappa=1$,
from (\protect\ref{eq:vorgpe});  the dashed line is the noninteracting 
wave function (\protect\ref{eq:vorsol}). In the inset, the contour plot 
for the density in the $xz$-plane, $n(x,0,z)=|\phi_v(x,0,z)|^2$, is
given. Luminosity is proportional to density, the white area being the
most dense. [note: this is version of reduced quality for e-archive 
only]  }
\label{fig:vortex}
\end{figure}

A quantized vortex along the $z$-axis can be described by writing the
order parameter in the form
\begin{equation}
\phi({\bf r})=\phi_v(r_\perp,z) \exp[i\kappa \varphi]
\label{eq:vortex}
\end{equation}
where $\varphi$ is the angle around the $z$-axis, $\kappa$ is an integer,
and $\phi_v(r_\perp,z)=\sqrt{n(r_\perp,z)}$.  This vortex state has
tangential velocity
\begin{equation}
v = \frac{\hbar}{mr_{\perp}}\kappa \; .
\label{eq:vorvel}
\end{equation}
The number $\kappa$ is the quantum of circulation and the angular momentum
along $z$ is $N\kappa\hbar$. The equation for the modulus  of the order
parameter is obtained from the GP equation (\ref{eq:GP}).  The kinetic
energy brings a new centrifugal term arising from the velocity flow which
pushes the atoms away from the $z$-axis. 
The GP equation then takes the form
\begin{equation}
\left[ - {\hbar^2 \nabla^2 \over 2m}  + { \hbar^2\kappa^2
\over 2m r_{\perp}^2} + {m \over 2} (\omega_{\perp}^2 r_{\perp}^2
+ \omega_z^2  z^2  ) +  g \phi^2_v(r_\perp,z) \right] \phi_v(r_\perp,z)
= \mu \phi_v(r_\perp,z) \;.
\label{eq:vorgpe}
\end{equation}
Due to the presence of the centrifugal term, the solution of this equation
for $\kappa\ne 0$ has to vanish on the $z$-axis. An example is shown in 
Fig.~\ref{fig:vortex}, where the solid line represents the condensate wave
function, $\phi_v(x,0,0)$, for a gas of $10^4$ rubidium atoms in a spherical 
trap and with vorticity $\kappa=1$. In the inset, we give the contour plot 
for the density in the $xz$-plane, $n(x,0,z)=|\phi_v(x,0,z)|^2$. 

For noninteracting systems the solution of Eq.~(\ref{eq:vorgpe}) is
analytic and, for $\kappa=1$, has the form
\begin{equation}
\phi_v(r_\perp,z) \ \propto \  r_{\perp} \exp\left[-\frac{m}{2\hbar}\left(
\omega_{\perp}r_{\perp}^2 + \omega_z z^2 \right)\right] \; .
\label{eq:vorsol}
\end{equation}
In this case, the vortex state corresponds to putting all the atoms  in
the $m=1$ single-particle state. Its energy is then $N 
\hbar\omega_{\perp}$ plus the ground state energy. In Fig.~\ref{fig:vortex},
the corresponding wave function is shown as a dashed line. Similarly to what 
happens for the ground state without vortices,  the presence of repulsive 
forces reduces dramatically the density with respect to the
noninteracting gas, the condensate wave function becoming much broader. 

The structure of the core of the vortex is fixed
by the balance between the kinetic energy  and  the two-body interaction
term. For a uniform  Bose gas the size of the core is of the order of
the healing length $\xi =  (8\pi n a)^{-1/2}$, already introduced in
Sec.~III.B, where $n$ is the density of the system. For the trapped gas,
an  estimate of the core size can be obtained using for $n$ the central
value of the density in the absence of vortices. If the trap is spherical,
as in the case of  Fig.~\ref{fig:vortex}, the ratio between $\xi$ and
the radius $R$ of the condensate takes the form (Baym and Pethick, 1996)
\begin{equation}
\frac{\xi}{R} = \left(\frac{a_{\rm ho}}{R}\right)^2 \; .
\label{eq:vorhl1}
\end{equation}
where we have used the Thomas-Fermi approximation for the central density
and the radius $R$. For the condensate in the figure, the radius is
about $4.1$ in units of $a_{\rm ho}$ and the ratio $\xi/R$ is then
$\sim 0.06$. The actual core size depends, obviously, also on the
position on the $z$-axis and becomes larger when the vortex
line reaches the outer part of the condensate, where the density 
decreases. This can be clearly seen in the inset of Fig.~\ref{fig:vortex}. 

\begin{figure}[t]
\epsfysize=8cm
\hspace{3cm}
\epsffile{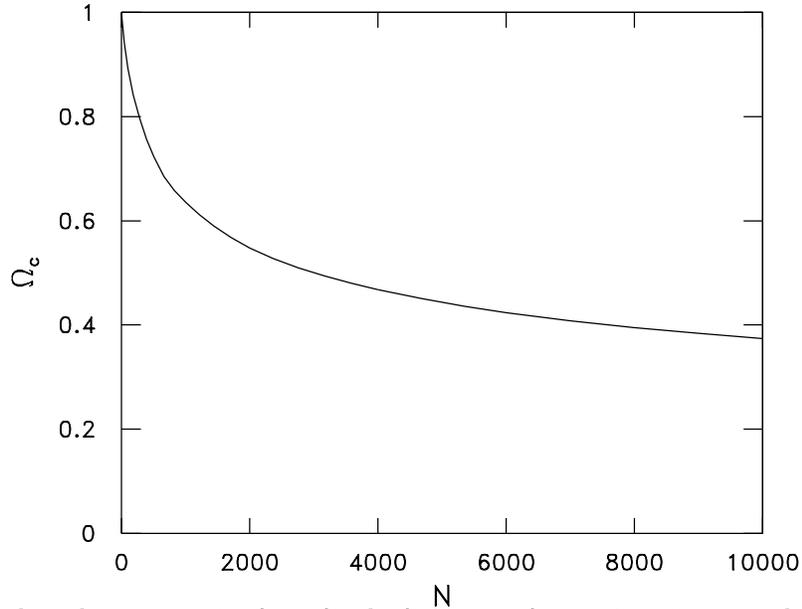}
\caption{Critical angular velocity, in units of $\omega_\perp$,  for the
formation of a $\kappa=1$ vortex in a spherical trap with $N$ atoms
of $^{87}$Rb and $a_{\rm ho}=0.791$ $\mu$m.  } 
\label{fig:omegac}
\end{figure}

The energy of the vortex can be evaluated through the energy functional
(\ref{eq:energy}). The difference between the energy of the vortex state
and the one of the ground state allows one to calculate the critical
frequency needed to create a vortex. In fact, in a frame rotating with
angular frequency $\Omega$, the energy of a system carrying angular
momentum $L_z$ is given by $(E-\Omega L_z)$, where $E$ and $L_z$ are
defined in the laboratory frame. At low rotational frequencies this
energy is minimal without the vortex. If $\Omega$ is large enough the
creation of a vortex can become favourable due to the term  $-\Omega L_z$.
This happens at the critical frequency
\begin{equation}
\Omega_c = (\hbar\kappa)^{-1} [(E/N)_{\kappa}-(E/N)_0] \;.
\label{eq:vorcf}
\end{equation}
where $E_{\kappa}$ is the energy of the system in the presence of a
vortex with angular momentum $N\hbar{\kappa}$. In Fig.~\ref{fig:omegac}
we plot the critical frequency for the creation of a vortex with 
$\kappa=1$ as a function of the number of atoms, for rubidium in
a spherical trap. As shown in the figure, the predicted critical 
frequency is a fraction of the oscillator frequency and, in typical 
experimental conditions, corresponds to a few Hz. It decreases when 
$N$ increases, because for large systems the energy cost associated 
with the occurrence of a vortex increases as $\ln N$, while the gain 
in $\Omega L_z$ is always linear in $N$. This behavior is similar 
to the one exhibited by uniform systems where, approximating  the 
vortex core  as a cylindrical hole of radius  $\xi$, one can  
calculate explicitly the critical frequency; one finds
$\Omega_c = (\hbar/mR^2)\ln (R/\xi)$, where $R$ is the radius of the
region occupied by the vortex flow. Analogous expressions can be derived 
also in the presence of harmonic trapping for large $N$.
Baym and Pethick (1996) and Sinha (1997) have shown that the critical
frequency, in units of $\omega_{\rm ho}$, goes as $\sim (a_{\rm ho}/R)^2
\ln(R/\xi)$. Using the asymptotic solution of the GP equation in the
large $N$ limit, Lundh, Pethick and Smith (1997) have found a useful 
analytic expression for the critical velocity in the case of axially 
symmetric traps:
\begin{equation}
\Omega_c = {5 \hbar \over 2 m R_\perp^2 } \ln {0.671 R_\perp \over \xi}
\label{eq:lundh}
\end{equation}
where $R_\perp$ is the Thomas-Fermi radius of the cloud in the $xy$-plane, 
orthogonal to the vortex line, while the healing length is defined by
$\xi = (8\pi n a)^{-1/2}$, with $n$ equal to the central density of the
gas without vortex. This formula gives a critical frequency 
which differs significantly from the numerical result shown in 
Fig.~\ref{fig:omegac} only for $N$ smaller than about $2000$, while
for larger $N$ it becomes more and more accurate. 

The above discussion regards the structure of vortices for repulsive
interactions. An intriguing problem is how the angular momentum 
is distributed in systems with attractive forces. This question was
recently addressed by Wilkin, Gunn and Smith (1998), who showed
that the lowest eigenstates of fixed angular momentum do not exhibit 
vortex configurations. 

A major question concerning vortices in trapped Bose gases is whether
they can be observed in experiments.  So far no evidence about their
existence has been reported. In principle, it should not be difficult
to produce them in a steadily rotating trap. The value of the critical
frequency is in fact easy to achieve in the laboratory.  However, when one
stops the rotation it is not obvious whether the vortex remains stable. 
The problem of stability of vortices is rather complex even in uniform
superfluids, like $^4$He, where it has been the object of much
experimental and theoretical work (Donnelly, 1991). In a recent
paper, Rokhsar (1997) has argued that a vortex placed at the center of 
a nonrotating harmonic trap is unstable. Other discussions about the 
stability of vortex configurations can be found in Fetter (1998), Benakli
{\it et al.} (1997), Isoshima and Machida (1998). 

Creating a vortex is only part of the problem. A second important
problem is its detection. The excitation energy associated with a vortex
is too small to be observed with measurements of the release energy. In
fact the increase in  the energy per particle is $\hbar \Omega_c$, 
a quantity much smaller than the energy per particle in the ground state, 
given by  $(5/7)\mu$. However, imaging the core of the vortex during the
expansion, after switching-off the trap, should be feasible, as recently
suggested by Lundh, Pethick and Smith (1998).   Promising 
perspectives are also given by the effects of vortices on the shift of 
the collective frequencies of the condensate. These  can be measured
with  high precision and the observation of a breaking of degeneracy 
between states of opposite angular momentum would represent a rather 
unambiguous evidence of the presence and the quantization of the 
vortex.  The shift of the collective frequencies has been already
investigated by several authors.  Sinha (1997) used a large $N$
semiclassical expansion in the Gross-Pitaevskii equation, Dodd {\it et
al.} (1997a) carried out a direct numerical  solution of
the same  equation. Very recently Zambelli and Stringari (1998) 
have developed a sum rule approach yielding an explicit analytic 
expression for the splitting of the quadrupole modes, while 
Svidzinsky and Fetter (1998) have developed a perturbative 
solution of the collisionless hydrodynamic equations.  One should also
recall that the existence of vortices gives rise to a new series of
collective excitations localized near the vortex core. Similar modes
are found in uniform superfluids. In that case, the corresponding
dispersion law of the lowest mode is $\omega = (\hbar k_z^2/2m)
\ln (k_z\xi )$, where $k_z$ is the wavevector associated with a 
periodic motion of the vortex line along the $z$-axis.  The effects 
of thermally excited vortex waves in trapped gases have been 
explored by Barenghi (1996). Another explicit proof of the
existence of vortices would be the observation of an asymmetry in
the velocity of sound when one considers wave packets propagating 
in the same, or in the opposite direction with respect to the vortex
flow. Such a test on the quantization of the superfluid flow requires
a ring-type geometry for the confinement of the atomic cloud.
The quantization of the superfluid flow  in the ring could be  also
revealed using interference experiments (see next section). By letting
the condensate expand one should in fact observe  interference
fringes associated with the modulations of the phase produced by
the quantization of the circulation in the ring. In a similar way,
one could detect a quantized vortex by looking at the phase slip in 
the interference fringes produced by two expanding condensates (Bolda 
and Walls, 1998).

We have already pointed out that if we induce at zero temperature
rotations on an axially symmetric system with angular frequency smaller than
$\Omega_c$, then the system remains in its ground state.  In fact only
the normal (nonsuperfluid) part of the system can participate in the
rotational motion and, consequently, axially symmetric Bose systems can 
possess a moment of inertia only at finite temperature. A deviation of
the moment of inertia from the rigid value represents an important
manifestation of superfluidity.  In liquid $^4$He such deviation has
been directly observed below the lambda temperature, where the system
becomes superfluid (Donnelly, 1991). Measuring the moment of inertia of a
trapped gas is actually a challenging problem, because direct measurements
of angular momentum are difficult to obtain. 

The moment of inertia $\Theta$ relative to the $z$ axis can be defined as
the linear response of the system to a rotational field $H_{\rm ext} = -\omega
L_z$, according to the definition
\begin{equation}
\langle L_z\rangle = \omega\Theta \;,
\label{eq:momin}
\end{equation}
where $L_z =\sum_i(x_ip_i^y-y_ip_i^x)$ is the $z$ component of the angular
momentum and the average is taken on the state perturbed by $H_{\rm ext}$. For
a classical system the moment of inertia takes the rigid value
\begin{equation}
\Theta_{\rm rig} = mN\langle r_\perp^2 \rangle \;.
\label{eq:momrig}
\end{equation}
Vice versa, the quantum mechanical determination of $\Theta$ is much less
trivial. It involves a dynamic calculation and, according to perturbation
theory, can be written as
\begin{equation}
\Theta = \frac{2}{{\cal Z}}\sum_{i,j} { |\langle j|L_z|i\rangle|^2 \over
E_i-E_j }  \exp \left( - { E_j \over k_BT } \right) \;,
\label{eq:momin1}
\end{equation}
where $|j\rangle$ and $|i\rangle$ are eigenstates of the unperturbed
Hamiltonian, $E_j$ and $E_i$ are the corresponding eigenvalues,  and
${\cal Z}$ is the partition function.

The moment of inertia can be easily calculated if one considers the
simplest case of an ideal gas trapped by a harmonic potential.
The result is (Stringari, 1996a)
\begin{equation}
\Theta = \epsilon_0^2 m \langle r_\perp^2 \rangle_0 N_0(T)
+ m\langle r_\perp^2 \rangle_T N_T(T) \;,
\label{eq:momin3}
\end{equation}
where the indices $\langle\rangle_0$ and $\langle\rangle_{T}$ mean
average taken over the densities of the Bose condensed and noncondensed
components of the system, respectively. The quantity
\begin{equation}
\epsilon_0 = \frac{\langle x^2 - y^2\rangle_0}{\langle x^2 + y^2\rangle_0}
\label{eq:defor}
\end{equation}
is the deformation parameter of the condensate given by $(\omega_y-
\omega_x)/ (\omega_y+\omega_x)$. This quantity vanishes for an axially
symmetric trap.

The physics contained in (\ref{eq:momin3}) is very clear. In fact, 
the first term in the moment of inertia arises from the atoms in the 
condensate, which contribute with their irrotational flow and can be
hence interpreted as the superfluid component. The second term arises
instead from the particles out of the condensate which rotate in a rigid
way (normal component).  These two distinct contributions are at the 
origin of an interesting $T$-dependence of $\Theta$. In fact, above $T_c$,
where $N_0 = 0$, the moment of inertia takes the classical rigid
value (\ref{eq:momrig}), while at $T=0$, where all the atoms are in
the condensate, it is given by the irrotational value $\Theta_{\rm irrot}
= \epsilon_0^2\Theta_{\rm rig}$. In the limit of small deformation
(i.e., small $\epsilon_0 $), the deviation of the moment of inertia
from its rigid-body value is given by the useful expression:
\begin{equation}
\frac{\Theta}{\Theta_{\rm rig}} = \frac{N_T \langle r_\perp^2\rangle_{T}}
{N_0 \langle r_\perp^2\rangle_0 + N_T \langle r_\perp^2
\rangle_{T}} \;.
\label{eq:ratmi}
\end{equation}

\begin{figure}[t]
\epsfysize=8cm
\hspace{3cm}
\epsffile{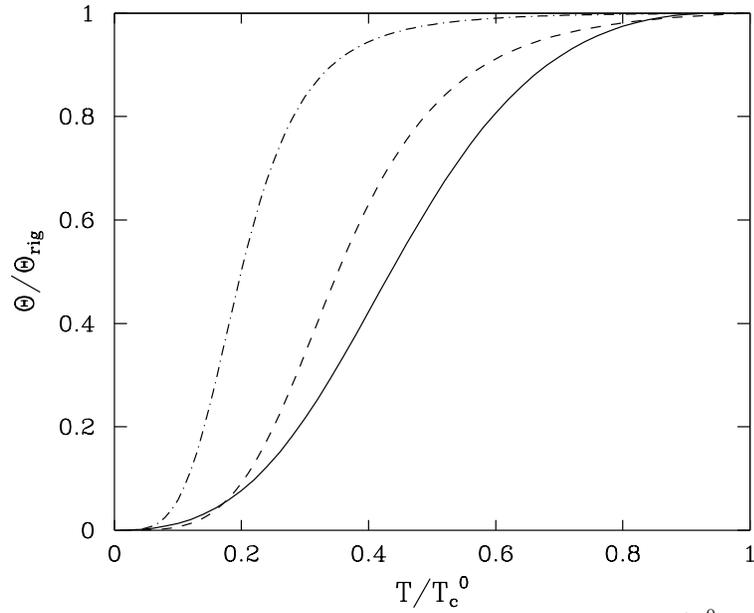}
\caption{Moment of inertia, $\Theta$, divided by its rigid value 
$\Theta_{\rm rig}$,  as a function of $T/T_c^0$. Solid line:
interacting gas in the thermodynamic limit with $\eta=0.4$. The dashed and
dot-dashed lines are the predictions for $5\times 10^7$ and
$5\times 10^4$ noninteracting particles, respectively, in a spherical 
trap.  }
\label{fig:inertia}
\end{figure}

It is worth discussing the behavior of the moment of inertia in the
thermodynamic limit $N\to \infty$, with $N\omega_{\rm ho}^3$ kept constant.
In this limit, the ratio $\langle r_\perp^2\rangle_0/\langle r_\perp^2
\rangle_{T}$ tends to zero.  In fact the square radius of the condensate
increases as $1/\omega_\perp$, while the one of the thermal cloud as
$k_B T/\omega_\perp^2$. In this limit, one then finds $\Theta/
\Theta_{\rm rig}\to 1$ everywhere except at $T\simeq 0$ where $N_T\simeq 0$.
This behavior is not surprising. In  fact if the radius of the
condensate is much smaller than the one of the thermal component then there
is no distinction between $\Theta$ and $\Theta_{\rm rig}$ since in both cases
the leading contribution is given by the thermal component. For finite 
values of $N$ the ratio $\Theta/\Theta_{\rm rig}$ goes smoothly to zero
as temperature decreases. An example is shown in Fig.~\ref{fig:inertia}
for a spherical trap and two different values of $N$ (dashed and
dot-dashed lines).  

How do two-body interactions change the above picture? Result
(\ref{eq:ratmi}) is expected to be valid also in the presence of 
interactions,  to the extent that the relevant excitations are 
well described by a single-particle picture and  the condensate can be
still identified with
the superfluid component.  For example in Hartree-Fock theory one
obtains coupled equations for the condensate wave function and
the single-particle excited states. The irrotational flow for the
condensate follows from the definition of the superfluid velocity
as the gradient of the phase of the order parameter.  On the other
hand, the flow of the atoms out of the condensate is rigid-like if
one treats the thermally excited states in semiclassical approximation,
as done in Sec.~V [see Eq.~(\ref{eq:NT})].  Only at very low temperature,
where the effects of collectivity can be important and  the superfluid
component must be distinguished from the condensate, expression
(\ref{eq:ratmi}) for the moment of inertia is no longer correct.

Interactions can affect the value of the moment of inertia by changing
the temperature dependence of the condensate as well as the value of the
square radii $\langle r_\perp^2 \rangle$. The change in the radii
is particularly significant at large $N$. In fact, unlike for
the noninteracting case, the ratio $\langle r_\perp^2\rangle_0/\langle
r_\perp^2\rangle_{T}$ does not vanish in the thermodynamic limit and is
fixed by the value of the scaling parameter $\eta$. As a consequence
interactions have the important effect of reducing the value of the
moment of inertia with respect to the rigid value in the whole range
of temperatures below $T_c$. This behavior is explicitly shown by the
solid line in Fig.~\ref{fig:inertia}.

\subsection{Interference and Josephson effect}
\label{sec:interference}

\begin{figure}[t]
\epsfysize=10cm
\hspace{3cm}
\epsffile{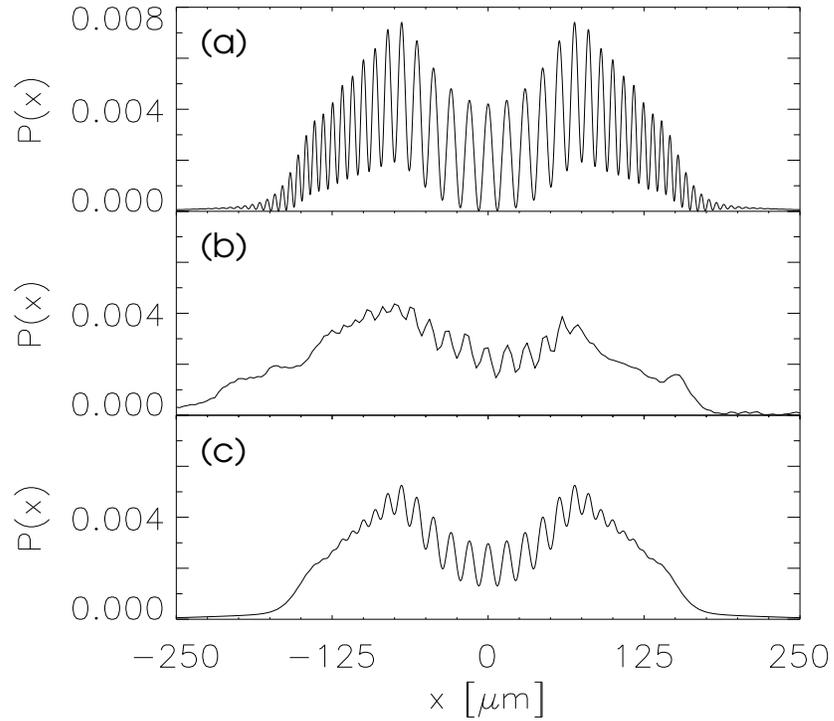}
\caption{Density pattern for the interference of two expanding and 
overlapping condensates. (a) Theory by R\"ohrl {\it et al.} (1997),
based on the solution of the time dependent GP equation. 
(b) Experimental data by Andrews, Townsend {\it et al.}, (1997). 
(c) Theory including the effect of finite experimental resolution. }
\label{fig:interference}
\end{figure}

An important consequence of phase coherence in Bose-Einstein
condensates  is the occurrence of interference phenomena.
A beautiful example is the experiment carried out at MIT (Andrews, 
Townsend {\it et al.}, 1997), where a laser beam was used to
cut a cigar-shaped atomic cloud into two spatially separated parts.
After switching off the confining potential and the laser, the two
independent atomic clouds expand and eventually overlap. Clean
interference patterns have been observed in the overlapping
region (see Fig.~\ref{fig:interference}b).

\par From a qualitative viewpoint, one can imagine the initial
condensates as two point-like pulsed sources placed at distance $d$
on the $z$-axis. Let us consider the interference taking place in the 
region of space where the density of the gas is small enough and hence
the condensate wave function is a linear superposition of two de Broglie 
waves. By using result (\ref{eq:S}) for the phase of each expanding 
condensate, one finds that the relative phase of the two waves
behaves as $[S(x,y,z+d/2)-S(x,y,z-d/2)]=(m/\hbar) \alpha_z(t) zd$. If the 
time delay, $t$, between the switching off of the trap and observation is
large, one has $\alpha_z(t) \to 1/t$. One then finds straight interference
fringes, orthogonal to the $z$-axis, with wavelength given by 
\begin{equation}
\lambda = \frac{h t}{md} \; .
\label{eq:frin}
\end{equation}
Using the spacing between the two  initial condensates as an estimate of 
the distance $d$ one gets  fringe periods in reasonable agreement with the 
observed patterns (Andrews, Townsend {\it et al.}, 1997). Typical
values are $t\simeq 40$ ms, $d \simeq 40$ $\mu$m, and
$\lambda \simeq 20 \mu$m.

Gross-Pitaevskii theory is a natural framework for investigating
interference phenomena in a quantitative way. In this theory, the phase
coherence of the condensate is assumed from the very beginning. The
interference patterns can be obtained by solving numerically the GP equation
(\ref{eq:TDGP}) for two condensates. This has been done for instance
by Hoston and You (1996), Naraschewski {\it et al.} (1996),  
R\"ohrl {\it et al.} (1997), Wallis {\it et al.} (1997a and 1997b). 
Interference phenomena have been investigated also without using 
the concept of  broken gauge symmetry by
Javanainen and Yoo (1996). In Fig.~\ref{fig:interference}
the experimental results are plotted together with the theoretical
calculations by R\"ohrl {\it et al.} (1997). The good agreement between
theory and experiment reveals that the concept of phase coherence,
as assumed in GP theory, works very well.  This was not obvious 
{\it a priori}, since the system is finite-sized and interacting 
and hence phase coherence is expected to be only approximate. 

\begin{figure}[t]
\epsfysize=8cm
\hspace{3cm}
\epsffile{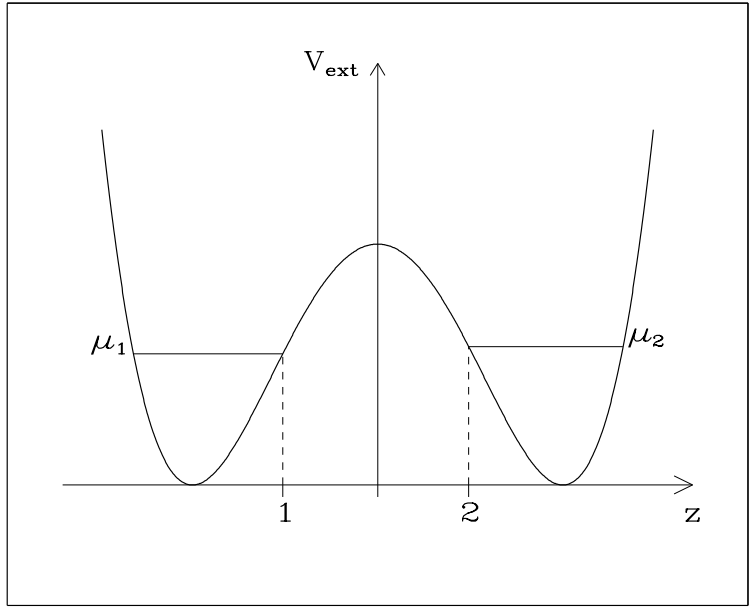}
\caption{Schematic geometry of a double-well trapping potential, $V_{\rm ext}$,
for the Josephson effect.  }
\label{fig:josephson}
\end{figure}

Another interesting manifestation of phase coherence in trapped
condensates is the possible occurrence of Josephson-type effects,
in analogy with well known properties of Josephson junctions in
superconductors and superfluids. The physical idea consists
of considering a double-well trap, with a barrier between the
two condensates. If the chemical  potential in the two traps is
different, a flux of atoms is generated.  In Fig.~\ref{fig:josephson}
we show a simplified scheme. If one assumes  the barrier between
the two wells to be high enough, then Eq.~(\ref{eq:GP})
has two natural solutions, $\phi_1 ({\bf r})$ and $\phi _2 ({\bf r})$,
localized in each potential well, $1$ and $2$, and having chemical
potentials $\mu_1$ and $\mu _2$. A difference between the chemical
potentials in the two traps can be achieved by filling them with a
different number of atoms. The overlap between the condensates occurs
only in the classically forbidden region, where the wave function is
small and nonlinear effects due to interactions can be ignored. Thus
in this region the linear combination
\begin{equation}
\phi({\bf r},t)=\phi_1({\bf r})\exp(-i\frac{\mu_1t}{\hbar})
+\phi_2({\bf r})\exp(-i\frac{\mu _2t}{\hbar})
\label{eq:comb}
\end{equation}
is still a solution of the time dependent equation (\ref{eq:TDGP}). If
the  two condensates are elongated  in the $z$-direction, the current
through the barrier can be written as
\begin{equation}
I(z,t) =\frac{i\hbar}{2m}\  \int\! dxdy \left( \phi ({\bf r},t)
\frac{\partial}{\partial z}\phi^* ({\bf r},t) -\phi^* ({\bf r},t)
\frac{\partial}{\partial z}\phi ({\bf r},t) \right)
 \label{eq:cur}
\end{equation}
Using the wave function (\ref{eq:comb}), the current can be easily
calculated and takes the typical Josephson form
\begin{equation}
I = I_0 \sin [ (\mu_1 - \mu _2 )t/\hbar ]
\label{eq:jos}
\end{equation}
with $I_0 = (\hbar /m) \int dxdy (\phi_1\phi'_2-\phi_2\phi'_1)$. The
calculation of the critical current $I_0$ is a difficult task,  since
it corresponds to a  nonlinear 3D  tunneling problem. If $\mu_1$ and
$\mu_2$ differ from the average value $\mu$ by a small quantity 
$\delta\mu$ and one treats the motion under the barrier in WKB 
semiclassical approximation, the estimated current $I_0$ turns out to 
be proportional to $\exp(- S_0)$, where  $S_0= \int_1^2 dz [ 2m 
(V_{\rm ext} (0,0,z) - \mu )/\hbar^2 ]^{1/2}$ and the integral is 
taken between the points $1$ and $2$ located as in Fig.~\ref{fig:josephson} 
(Dalfovo, Pitaevskii and Stringari, 1996). Zapata,  Sols and Leggett (1998) 
have recently applied the same formalism to realistic 3D configurations,
deriving the Josephson current in the form $I \sim (k_B T_c^0 / \hbar )
\exp(- S_0)$.  Their results suggest that Josephson
effects might be indeed observed in experiments. At finite temperatures
one should also  include possible contributions arising from the incoherent
flux of thermally excited atoms; this ``normal" current is expected to be
proportional to $\delta\mu$. In order to observe the Josephson effect
one must consequently work at low enough temperatures where the system is
fully superfluid. It is also worth noticing that the geometry of the 
trapped gases allows one to realize qualitatively new Josephson-type 
effects, as suggested by Smerzi {\it et al.} (1997), Raghavan {\it et al.}
(1997) and Williams {\it et al.} (1998). The propagation of density
solitons across regions of phase discontinuity in the collision of 
two condensates has been also considered as an analogue of a 
Josephson-like effect (Reinhardt and Clark, 1997). 

An open problem concerns the possible decoherence mechanisms, which 
could affect, or even destroy, the phase coherence in interference and
Josephson-like experiments through phase diffusion processes. The 
fluctuations of the phase can have either a thermal or a 
quantum origin. Actually, even at $T=0$ the phase of the condensate 
must diffuse since having a fixed phase is inconsistent with atom 
number conservation. Many authors have investigated the problem 
of the quantum diffusion of the phase by 
describing the system as a coherent superposition of states with 
different $N$. This yields fluctuations in the chemical potential and 
hence in the phase of the order parameter.  Discussions about this 
effect and on the general problem of phase coherence can be found 
in: Lewenstein and You (1996),  Barnett, Burnett and Vaccarro (1996),  
Wright, Walls and  Garrison (1996), Castin and Dalibard (1997), 
Wallis {\it et al.} (1997b), Imamoglu, Lewenstein and You (1997),
and Dodd {\it et al.} (1997b), Javanainen and Wilkens (1998), 
Leggett and Sols (1998) [see also Parkins and Walls (1998) for 
more discussions and references].

\subsection{Collapse and revival of collective oscillations}
\label{sec:collapse}

\begin{figure}[t]
\epsfysize=8cm
\hspace{3cm}
\epsffile{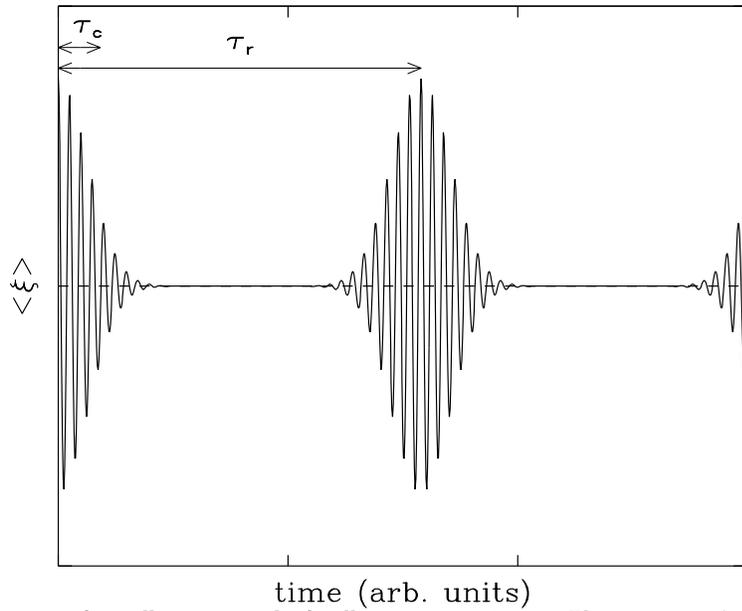}
\caption{Schematic picture for collapse-revival of collective excitations. The
quantity $\xi$ is a generic oscillator co-ordinate and the symbol $\langle \xi
\rangle$ means an average over different replica of the system prepared in the
same conditions. }
\label{fig:collapse}
\end{figure}

In trapped gases one can predict another interesting
``mesoscopic"  phenomenon having no classical analog, namely the
collapse and revival of collective excitations. This process should
not be confused with the decay of coherence in the many-body wave
function, which corresponds to the phase diffusion mentioned at the end
of the previous section and which is also sometimes called ``collapse" of the
condensate. Conversely, the collapse-revival of collective excitations
originates  from a dephasing of an oscillation due to the quantum
fluctuation of the number of quanta. Indeed, a classical oscillation
of the condensate can  be viewed as a coherent superposition of
stationary  states  with different numbers of quanta of the oscillator.
Fluctuations  in the number of quanta cause a dephasing and a consequent
decrease in the amplitude (collapse) of the oscillation.  Since there
is no dissipation of energy in this process, the oscillation can
eventually  reappear (revival) after  a certain time interval.
A schematic picture is shown in
Fig.~\ref{fig:collapse}.  Similar processes of collapse-revival of
coherent quantum states have been already observed in atomic Rydberg
wave packets (Yeazell and Stroud, 1991; Meacher {\it et al.}, 1991),
molecular vibrations (Vrakking {\it et al.}, 1996) as well as atoms and
ions interacting with an electromagnetic field (Meekhof {\it et al.}, 
1996; Brune {\it et al.}, 1996).

Let $E$ be the energy associated with a classical oscillation of the
system induced, for example, by some external sinusoidal drive. By
classical oscillation we mean that the number $n$ of quanta of
oscillation is very large. Let us further suppose  the frequency
$\omega$ to be weakly dependent on the amplitude as in
Eq.~(\ref{eq:quadraticshift}). The energy of the oscillation is
proportional to the square of the amplitude, $E \propto A^2$, and
the coefficient of proportionality can be calculated, for instance, by
solving the time dependent GP equation (\ref{eq:TDGP}). Thus one can
rewrite Eq.~(\ref{eq:quadraticshift}) as
\begin{equation}
\omega = \omega _0 (1+\kappa E) \; ,
\label{eq:cl}
\end{equation}
with $|\kappa | E \ll 1$. Now, one can use the semiclassical
approximation in order to express the energy $E$ in terms of the
number of quanta of oscillation, through $\hbar \omega = (\partial 
E_n/\partial n)$. One finds  $E_n/\hbar= n \omega _0 +n^2(\hbar
\omega _0^2 \kappa/2)$. The wave function describing the coherent
state of the oscillator can be written in the form
$\psi = \sum_n c_n \psi_n \exp(-i E_n t/\hbar)$. The coefficients
\begin{equation}
| c _n |^2 \ \approx \ \frac{1}{\sqrt{2\pi
\bar{n}}} \ \exp \left[ -\frac{(n-\bar{n})^2}{2\bar{n}} \right] \; .
\label{eq:cn2}
\end{equation}
characterize a coherent  gaussian distribution, and $\bar{n}$ is the
average value of quanta,  which is supposed to be much larger than $1$. 
Given a generic oscillator co-ordinate $\xi(t)$, its average over this
Gaussian superposition of states can be easily estimated; it takes
contributions  from $n \rightarrow n \pm 1$ transitions and the result
is $\langle \xi (t) \rangle \propto  \sum _n\mid c _n \mid ^2 \cos
[(\omega _0 + \hbar \omega_0^2 \kappa n)t]$.  For small enough values
of $t$,  one can replace the summation over $n$ by an integral and
one gets a Gaussian damping, or collapse, of the oscillation according
to $\langle \xi \rangle \propto \exp[-(t/\tau _c)^2]$,  where
\begin{equation}
\tau_c ^{-1}  \ = \ \omega _0  (E \hbar \omega _0/2)^{1/2} \mid \kappa \mid 
\label{eq:tauc}
\end{equation}
defines the collapse time.
The periodicity of $\langle \xi (t) \rangle$ gives also the revival
time, $\tau _r = 2\pi/(\hbar \omega_0^2 \kappa)$.  One finds $\tau_c
\approx \sqrt{(1/\bar{n})} \ \tau_r \ll \tau_r$. These expressions
for the time scales were derived by one of us (Pitaevskii, 1997),
and the theory of collapse-revival has been also developed by Kuklov
{\it et al.} (1997) and Graham {\it et al.} (1998).  The revival
time is in agreement with the theory of Averbukh and Perelman
(1989).

An explicit estimate of the collapse time (\ref{eq:tauc}) can be obtained, 
for instance, using Gross-Pitaevskii theory  within the collisionless 
hydrodynamic scheme of Sec.~\ref{sec:nonlinear}. One can solve the
equations of motion for the lowest $m=0$ and $m=2$ modes in an
axially symmetric trap and find the relation between the coefficients
$\kappa$ and $\delta(\lambda)$ entering the expansions (\ref{eq:cl})
and (\ref{eq:quadraticshift}), respectively (Dalfovo {\it et al.},
1997b).  For the $m=2$ mode in the first JILA trap, with about $5000$
rubidium atoms, one finds a collapse time of the order of $5$ s, if the
relative amplitude is about $20$\%.  This time is much larger than the
lifetime reported in the experiment by Jin {\it et al.} (1996, 1997), 
which is of the order of $100$~ms and hence clearly originates from
other damping mechanisms. The collapse time
would be even longer for larger $N$ and this makes the collapse
rather difficult to observe. It is not hopeless, however, to observe
such an effect at very low temperatures where other
damping processes, like Landau damping, become much less
effective, and looking for special values of the anisotropy parameters,
where nonlinear effects and frequency shifts are larger, as suggested in
Sec.~\ref{sec:nonlinear}. 


\section{Conclusions and outlook}
\label{sec:conclusions}

In this paper we have provided an introductory description of the 
properties of Bose condensed gases confined in harmonic traps. The 
main message emerging from our analysis is that, despite the dilute  
nature of these gases, two-body interactions have crucial consequences
on most measurable quantities. This is the combined effect of
Bose-Einstein condensation and of the nonuniform nature  of the
system. In particular the ground state  (Sec.~\ref{sec:groundstate})
and the dynamic  (Sec.~\ref{sec:dynamics}) properties  are affected by
two-body forces in an essential  way. Interactions can be
included using fundamental many-body theories for the
order parameter, depending on a single interaction parameter,
the $s$-wave scattering length. Direct measurements of the density
profiles,  release energy and collective frequencies
have already provided very accurate tests of the theoretical predictions.
Concerning thermodynamics  (Sec.~\ref{sec:thermodynamics}) our analysis
has pointed out the possibility of defining the thermodynamic limit
for these nonuniform systems including the effects of two-body
interactions. Such effects are less important than those 
of the ground state since at finite temperature the system
is more dilute. Nevertheless significant corrections to
the critical temperature and to the $T$-dependence of the
release energy can be predicted and in some cases compared with
experiments.  In Sec.~\ref{sec:superfluidity} we have discussed 
possible superfluid and coherence phenomena exhibited by trapped
Bose gases. This discussion could not be exhaustive
because the evolution of current research in this field
is very rapid.

The mean-field picture of these interacting Bose gases turns out
to be quite accurate in describing most of the available experimental 
results. Deviations from the mean-field predictions are expected to 
arise from ``correlation" effects beyond Bogoliubov theory, when the 
gas parameter $n|a|^3$ is not very small. They can also originate 
from ``mesoscopic" effects associated with the fact that the concepts
of order parameter and gauge symmetry breaking are only approximate
in finite systems and, in particular, the fluctuations of the
phase are not always negligible. There are no experimental evidences 
so far for these effects, but accurate theoretical predictions
concerning both correlation and mesoscopic effects might stimulate
new important experiments in the future.  

In our review we have been able to cover only part of the huge body 
of literature which arose after the experimental discovery of
Bose-Einstein condensation in 1995. We would like to mention here 
some important issues that we have not discussed and
that have been recently at the center of significant theoretical
and/or experimental research.

{\it Kinetics of the condensate:} An important question, not 
yet fully understood,  is the kinetics of the condensate nucleation. 
The process of condensation of a uniform ideal gas was considered 
by Semicoz and Tkachev (1995) on the basis of the Boltzmann equation. 
They assumed that the distribution function depends only on 
the energy of the atoms and found that this function exhibits a 
divergence at zero energy after a finite time interval, corresponding to
the onset of Bose-Einstein condensation. In previous investigations
the mechanism of condensation was predicted to occur only
asymptotically. The next stage of the process is the growth of
the condensate. This was considered recently by Gardiner {\it et al.}
(1997) and Jaksch {\it et al.} (1997) on  the basis of quantum kinetic 
master equations. The kinetics of the Bose gas near critical conditions 
for condensation has also been studied by Monte Carlo simulations 
taking into account the Bose statistics under the random phase 
approximation (Wu, Arimondo and Foot, 1997). In a very recent 
experiment at MIT (Miesner {\it et al.}, 1998) the formation and 
growth of the condensate has been investigated by means of imaging 
techniques. This work has shown clear evidence for a behavior known 
as ``bosonic stimulation", which corresponds to an enhancement of the
condensation rate induced by the condensate itself. Explicitly, if 
$N$ atoms are in the condensate the condensation rate is proportional 
to $(N+1)$. This gain mechanism is familiar in the physics of optical 
lasers and, in the case of trapped atoms, can lead to matter-wave 
amplification. Another important question recently 
investigated experimentally is the decay of the trapped gas due to 
three-body recombination, yielding formation of molecules  and 
loss of atoms from the trap (Burt {\it el al.}, 1997; Stamper-Kurn 
{\it et al.}, 1998a).  The corresponding rate turns out to be
different for a condensate and a thermal cloud, confirming the 
theoretical predictions by Kagan, Svitsunov and Shlyapnikov (1985). 

\begin{figure}[t]
\epsfysize=8cm
\hspace{3cm}
\epsffile{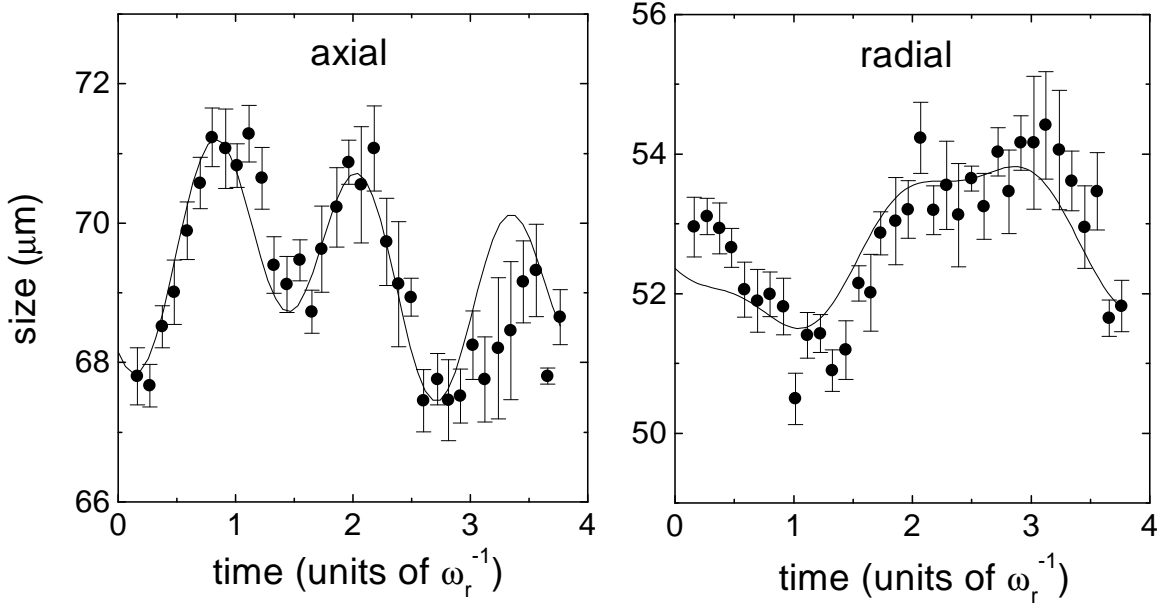}
\caption{Oscillation in the width of the cloud in both the axial and
  radial direction due to the instantaneous change in scattering
  length in the experiment by Matthews {\it et al.} (1998). Time is
  in units of $\omega_r^{-1} \equiv 
  \omega_\perp^{-1} = 9.4$ ms. The solid lines are the time dependent 
  widths calculated using Eqs.~(\protect\ref{eq:ddotb}), with only the
  amplitude of the oscillation and the initial size as free parameters.}
\label{fig:jila-tn}
\end{figure}

{\it Mixtures of condensates:}  Binary  mixtures of condensates can
be obtained experimentally by trapping at the same time two different
atomic species (different isotopes or different alkalis), or two 
different spin states of the same atoms. The confining potential
of the two condensates may be centered at the same point,  or not.
Ground state and dynamic calculations of two interacting condensates
have been carried out by Ho and Shenoy (1996b) using the Thomas-Fermi
approximation and by Esry {\it et al.} (1997) by solving the GP 
equations.  Similar calculations have also been  done by Graham 
and Walls (1998) and by Pu and Bigelow (1998). Interesting behaviors 
can be  predicted depending on the values of the intraspecies and 
interspecies scattering lengths. Hydrodynamic equations have been 
recently derived by Ho and Shenoy (1998). From the experimental 
viewpoint, mixtures of this kind have been created and 
observed at JILA (Myatt {\it et al.}, 1997) with different 
spin states of $^{87}$Rb. In a very recent experiment (Matthews 
{\it et al.}, 1998)  all the atoms  have been converted, via 
two-photon  transitions, into a different hyperfine state.  The 
system in the final configuration is no longer in equilibrium and 
will start oscillating. From the analysis of the subsequent oscillations 
(see Fig.~\ref{fig:jila-tn}) it has been possible to determine with 
high precision the ratio of the intraspecies scattering length relative 
to the final and initial states. Using the same apparatus with a
mixture of the $|F=1, m_F=-1 \rangle$ and $|F=2, m_F=1 \rangle$ spin
states, it was also possible to measure the relative phase of two 
condensates, thus realizing a  ``condensate interferometer" (Hall
{\it et al.}, 1998). Suggestions have been made to use the same 
mixture of states in order to observe nonlinear Josephson-type 
oscillations (Williams {\it et al.}, 1998).

\begin{figure}[t]
\epsfysize=8cm
\hspace{3cm}
\epsffile{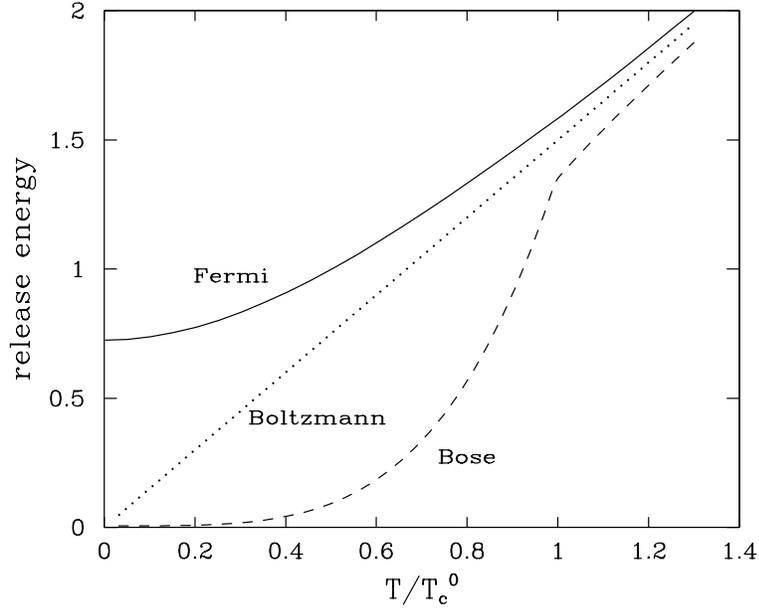}
\caption{ Release energy per particle of an ideal gas in the thermodynamic 
limit, in units of $k_B T_c^0$ with $T_c^0 =  0.94  \hbar \omega_{\rm ho} 
N^{1/3}$. Solid line: Fermi gas. Dashed line: Bose gas. Dotted line: 
Maxwell-Boltzmann gas.  }
\label{fig:fermions}
\end{figure}

{\it Fermions:} The study of degenerate Fermi gases in traps
is expected to be an important issue of future research. Trapping 
of fermionic species has been reported for $^6$Li (McAlexander {\it 
et al.}, 1995) and $^{40}$K (Cataliotti {\it et al.}, 1998). 
Sympathetic cooling of fermions by bosons might yield low temperature
regimes overcoming the problem of the  suppression of collisional
processes exhibited  by polarized Fermi gases at low
temperature. Degenerate Fermi gases behave quite differently from
bosonic sytems. Effects of Fermi statistics can be observed
in the behavior of  the release energy below the Fermi temperature;
for an ideal gas of $N$ fully polarized atoms, the latter is given by
$k_B T_{F}= (6 N \lambda)^{1/3} \hbar \omega_\perp$, with the
usual definition  $\lambda=\omega_z/\omega_\perp$. 
Figure~\ref{fig:fermions} shows  how the release energy of an
ideal Fermi gas compares with the corresponding behavior of an ideal 
Bose gas confined in the same harmonic potential and with the same
number of atoms.  At very low temperatures interacting Fermi gases 
can undergo a phase  transition to a superfluid phase. The resulting
behavior in the presence of a harmonic trap  has been the object
of several studies [see, for instance, Baranov {\it et al.} (1996),
Stoof {\it et al.} (1996), Houbiers {\it et al.} (1997) and 
references therein].

{\it Optical confinement:} The recent realization of Bose-Einstein
condensation in optical traps (Stamper-Kurn {\it et al.}, 1998a) is 
also expected to open important perspectives.
On the one hand one can obtain higher densities, useful, for example,
to study  three-body decay processes and more correlated configurations.
On the other hand different geometrical configurations can be achieved,
like for example quasi 1D structures. Finally, by releasing the
condition of spin polarization imposed by magnetic trapping, this
new method of confinement will permit one to study in a systematic way
the magnetic properties of these gases, including the spinor nature 
of the order parameter (Ho and Shenoy, 1996a; Ho, 1998; Ohmi and 
Machida, 1998). Spin domains in condensates of sodium, made by three
hyperfine states of the $F=1$ multiplet, have been recently observed 
by Stenger {\it et al.} (1998), who have demostrated the 
anti-ferromagnetic character of the spin-dependent interaction. 
A further advantage of the optical traps is that they allow one 
to observe Feshbach resonances for strong-field seeking 
states, as already done by Inouye {\it et al.} (1998) with sodium.
Feshbach resonances are strong variations of the scattering length,
induced by an external field, which occur when a quasibound molecular 
state has nearly zero energy and couples resonantly to the free
states of the colliding atoms. The possibility of tuning the scattering 
length with external magnetic fields provides new perspectives in the
manipulation of Bose condensates.

\vskip 2 cm 

\acknowledgements

An important part of the material presented in this review is the
result of fruitful collaborations with the other members of the Trento
team on Bose-Einstein condensation (M. Guilleumas, C. Minniti, L. Ricci,
L. Vichi and F. Zambelli).  It is a pleasure to thank them. We are indebted to
E.A. Cornell, L.V. Hau, W. Ketterle, D.M. Stamper-Kurn, and C.E. Wieman
for stimulating discussions and for sending us experimental data prior
to publication.   We also gratefully thank E. Arimondo, D. Brink, M. Chiofalo,
A.L. Fetter, A. Griffin, M. Inguscio, F. Lalo\"e, A. Minguzzi, L. Reatto,
G. Tino,  M.P. Tosi, M. Ueda and P. Zoller for many fruitful discussions.  
One of us  (S.G.) would like to thank the European Centre for Theoretical 
Studies in Nuclear Physics and Related Areas (ECT*, Trento) where he 
did part of this work.  This research was supported by the Istituto 
Nazionale per la Fisica della Materia through the Advanced Research 
Project on BEC and, partly, by the National Science Foundation under 
Grant N. PHY94-07194.

\newpage

\bigskip

\bigskip

\bigskip

\bigskip

\bigskip

\bigskip

\bigskip

\bigskip

\bigskip

\bigskip

\bigskip

\bigskip

\bigskip

\bigskip

\bigskip

\bigskip

\bigskip

\bigskip

\bigskip

\bigskip

\bigskip

\bigskip

\bigskip

\bigskip

\bigskip

\bigskip

\bigskip

\bigskip

\bigskip

\bigskip

\bigskip

\bigskip

\bigskip

\bigskip

\bigskip

\vskip 2 cm

\end{document}